\documentclass[10pt,journal,twocolumn]{IEEEtran}

\usepackage{dsfont}
\usepackage{graphicx}
\usepackage{subfigure}
\usepackage{cite}
\usepackage{amsfonts,amssymb}
\usepackage{lipsum}
\usepackage{bm}
\usepackage{subeqnarray}

\usepackage{algorithm}
\usepackage{algorithmic}

\usepackage{caption}
\usepackage{lipsum}
\usepackage{multicol}
\graphicspath{{figures/}}
\usepackage{graphicx}
\usepackage{amsfonts,amssymb}
\usepackage{lipsum}
\usepackage{bm}
\usepackage{subeqnarray}
\usepackage{caption}
\graphicspath{{figures/}}
\usepackage[cmex10]{amsmath}

\usepackage[cmex10]{amsmath}
\usepackage{amsfonts,amssymb}
\usepackage{epstopdf}
\usepackage{amsmath}
\usepackage{bm}
\begin{document}
%
\title{Near Optimal Adaptive Shortest Path Routing with Stochastic  Links  States under   Adversarial Attack}

\author{\normalsize Pan Zhou, \emph{Member, IEEE}, Lin Cheng, \emph{Member, IEEE}, Dapeng Oliver Wu, \emph{Fellow, IEEE}
\thanks{

Parts of this work have been presented at IEEE International Conference on Sensing, Communications and Networking (SECON 2016) \cite{Secon2016}, London,
UK.

Pan Zhou is from Huazhong University of Science \& Technology,
Wuhan, 430074, Hubei, China.

Lin Chen is from Department of Engineering, Trinity College, Hartford,
CT, 06106 USA.

Dapeng Oliver Wu is with Department of Electrical and Computer Engineering, University of Florida, Gainesville, Florida, 32611 USA.
Email: panzhou@hust.edu.cn$^1$, lin.cheng@trincoll.edu$^2$,  wu@ece.ufl.edu$^3$

This work was supported by the National Science Foundation of China under Grant 61401169 and NSF CNS-1116970.
} }



\maketitle
\thispagestyle{empty}
\pagestyle{plain}

\begin{abstract}
We consider the shortest path routing (SPR)  of a network with stochastically time varying link metrics
 under potential adversarial attacks. Due to potential denial of service attacks, the distributions of link
states could be stochastic (benign) or adversarial at different temporal and spatial locations. Without any \emph{a priori},
 designing an  adaptive SPR protocol to cope with all possible situations in practice  optimally is a very challenging issue. In this paper, we  present the first
 solution by formulating it as a multi-armed bandit (MAB) problem. By introducing a novel control
 parameter into the exploration phase for each link, a martingale inequality is applied in the
  our combinatorial adversarial MAB framework.    As such, our proposed algorithms could automatically  detect features of the environment  within a unified framework and find the optimal
SPR strategies with almost optimal learning performance in all possible cases over time. Moreover, we study important issues related to
 the practical implementation, such as decoupling route selection with multi-path route probing, cooperative learning among multiple sources,
the ``cold-start" issue and delayed feedback of our algorithm. Nonetheless, the proposed SPR algorithms can be implemented with low  complexity and they are
proved to scale very well with the network size. Comparing to
  existing approaches in a  typical network scenario under jamming attacks,
    our algorithm has a 65.3\%  improvement of network delay given a learning period and a 81.5\% improvement of learning duration under a specified network delay.
\end{abstract}

\begin{keywords}
Shortest Path Routing, Online learning,  jamming,  stochastic and adversarial multi-armed bandits
\end{keywords}

%
\IEEEpeerreviewmaketitle

\section{Introduction}
Shortest path routing (SPR) is a basic functionality of networks to \emph{route packets} from
sources to destinations. Consider a network with known topology deployed in a wireless
environment, where the link qualities vary stochastically with time.   As security is critical to network performance, it is vulnerable
to a wide variety of attacks. For example, a malicious attacker may perform a denial of service (DoS) attack
by jamming in a selected area of links or creating routing worm \cite{Zou05} to cause severe congestions over the network. As a result, the link metrics for the SPR (e.g., link delays) are hard to predict.  Although the source
can measure links by sending traceroute probing packets along selected paths, it is hard to
obtain an accurate link measurement by a single trial   due to noise, inherent dynamics of links (e.g.,
fading and short-term interference, etc.) and  the unpredictable adversarial behaviors (e.g., DoS attack on
traffic and  jamming attack, etc.). Compared with the classic SPR problem where the assumed average link metrics is a known priori,
the source  is necessary to learn about the link metrics over time.

A fair amount of SPR algorithms have been proposed by considering either the stochastically distributed link metrics
 \cite{Info_Rout13}, \cite{YiToN12},\cite{Asil10} (i.i.d. distributed) or
security issues where \emph{all} link metrics are assumed to be adversarially distributed
\cite{info05, Routing07, Routing04} (non-i.i.d. distributed) that can vary in an arbitrary way by attackers.  In particular, the respective online
learning problems fit into the stochastic Multi-armed bandit (MAB) problem \cite{Robbins1985} and the adversarial MAB problem \cite{non_MAB02} perfectly. The main idea is,  by probing each link along with the balance between ``exploration" and ``exploitation" of sets of routes over time, true average
link metrics will gradually  be learned and the optimal SPR can be found by minimizing the term ``regret" that qualifies  the learning
performance, i.e., the gap between routes selected by the SPR algorithm and the optimal one known in hindsight, accumulated over time.
A known  fact is that stochastic MAB and adversarial MAB have the optimal regrets $O(log(t))$ \cite{Robbins1985}
and $O(\sqrt{t})$ \cite{non_MAB02} over time $t$, respectively.  Obviously, the learning performance of stochastic MAB
is much better than  that of adversarial MAB.

As we know, the assumption of the known nature of the environments, i.e., stochastic or
adversarial, in most existing works is very restrictive in describing practical network environments.  On the one hand, existing SPR protocols
may perform poorly in practice. Consider a  network deployed in a potentially  hostile environment, the mobility pattern,  attacking approaches and strengths, numbers and locations of attackers are often unrevealed. In this case, most likely, certain
 portions of links  may (or may not) suffer from denial of service attackers that are adversarial, while the unaffected others are stochastically  distributed. To design an optimal SPR algorithm, the adoption of
the typical adversarial MAB model \cite{info05,Routing07,Routing04} on all links will lead to
  undesirable  learning performance (large regrets) in finding the SPR, since a great portion of links can be benign as the link states are still
 stochastically distributed. 

 On the other hand,  applying
stochastic MAB model \cite{Info_Rout13},\cite{YiToN12},\cite{Asil10} will face practical implementation issues, even though
 no adversarial behavior exists. In almost all practical networks (e.g., ad hoc and sensor
 networks),  the commonly seen occasionally disturbing events would make the stochastic distributed link
  metrics contaminated. These include the burst traffic injection, the jitter effect of electronmagnetic waves,
  periodic battery replacements, and the unexpected routing table corruptions and reconfigurations, etc.
  In this case, the link metric distributions will not be i.i.d.  for a small portion of time during the whole learning process.
  Thus, it is unclear whether the stochastic MAB theory can still be applied, how it affects the learning performance and to
  what extend the contamination is negligible. Therefore, the design of the SPR protocol without any prior knowledge of the
   operating environment is very challenging.


In this paper, we propose a novel adaptive online-learning based SPR protocol to address this
challenging issue at the first attempt and it achieves near optimal
 learning performance in all different situations within a unified online-learning framework. The proposed algorithm
  neither needs to distinguish the stochastic and adversarial
MAB problems nor needs to know the time horizon of running the
protocol. Our idea is based on the well-known EXP3 algorithm from adversarial MAB \cite{non_MAB02} by introducing a novel control parameter into
the exploration probability to \emph{detect} the metrics evolutions of each link. In contrast to hop-by-hop routing where intermediate node is responsible to decide the
next route, our online routing decision is made at the endhost (i.e, source nodes) that is capable to select a globally optimal path.  Owing to  the lack
of link quality knowledge, the  limited observation of
the network from the endhost makes the online-learning based SPR very challenging.
Therefore, the regret grows not only with time,
but also with the network size.  Moreover, to further
 accelerate the learning process in practical large-scale networks, we need to study the following important issues: each endhost in every time slot decouples route selection and probing by
 sending ``smart"
   probing packets (these packets do not carry any useful data) over multiple paths to measure link metrics
along with the selected path in the network, cooperative learning among multiple endhosts, and the ``cold-start"  and
delayed feedback issues in practical deployments. Our main contributions are summarized as follows:




\emph{ 1)} We design the first adaptive SPR protocol to bring
the \emph{stochastic} and \emph{adversarial} MABs into a unified framework with promising practical applications in unknown network environments. The  environments are  generally categorized into four typical regimes, where our proposed SPR algorithms
 are shown to achieve almost optimal regret performance in all regimes and are resilient to different kinds of attacks.



\emph{ 2)} We extend our algorithm to accelerated learning and see a $1/m$-factor reduction in regret for a probing rate of $m$. We also consider the practical ``cold-start" issue of the SPR algorithms, i.e., when the endhost is unaware of the $m$ and
total number of links $n$ at the beginning and the sensitiveness of the algorithm   to that lacked information, and the delayed feedback issue.

\emph{ 3)} The proposed algorithms can be implemented by dynamic programming, where its time and space complexities is comparable to the classic Dijkstra's algorithm.
 Importantly,  they achieve optimal regret bounds with respect to the network size.

\emph{ 4)} We conduct diversified  experiments on  both real trace-driven and synthetic datasets and  demonstrate that all
advantages of the
  algorithms are real and can be applied in practice.

The rest of this paper is organized as follows.  Section II discusses related works. Section III describes the  problem formulation. Section IV
 studies the single-source adaptive optimal SPR problem  with solid performance analysis. In Section V,
 we study the accelerated learning and practical implementation issues.  Section VI discusses the computationally
  efficient implementation of AOSPR-EXP3++. Section VII conducts numerical experiments. Important
  proofs for single-source and accelerated learning SPR algorithms are put in Section VIII and Section IX, respectively. The paper concludes in Section X.

\vspace{-.1cm}
\section{Related Work}

Online learning-based routing has been proposed to deal with networks in dynamically changing environments, especially in
wireless ad hoc networks with fixed topology. Some existing solutions focus on the hop-by-hop optimization of route selections, e.g., \cite{BToN12},\cite{Asil10},  and references therein.
Meanwhile, most of the other works consider the much more challenging endhost-based routing, e.g.,
 \cite{Info_Rout13},\cite{YiToN12}, \cite{Routing07},\cite{LiuWiOpt12}. In \cite{BToN12},
reinforcement learning (RL)
techniques are used to update the link-level metrics.
It is worth pointing
out that RL is generally targeted at a broader set of learning
problems in Markov Decision Processes (MDPs) \cite{Ref98}. It is well-known that such learning algorithms
 can guarantee optimality only asymptotically (to infinity), which cannot be relied upon in mission-critical applications. MAB  problems constitute a special class of MDPs, for which the regret
learning framework is generally viewed
as more effective both in terms of convergence and computational
complexity for the finite time optimality solutions. Thus, the use of MAB models is highly identified.  If  \emph{path measurements} are available for a set of independent paths, it belongs to the classic MAB problem. If \emph{link
measurements} are available such that the dependent paths can share this information, it is named as the\emph{ combinatorial semi-bandit} \cite{OR2014} problems. Obviously, the exploitation of sharing measurements of overlapping links among different paths can accelerate learning and  result in much lower regrets and better scalability  \cite{OR2014}. 

Importantly, existing works are mainly based on two types of MAB models: \emph{adversarial} MAB\cite{info05,Routing07,Routing04} and \emph{stochastic} MAB
\cite{Info_Rout13,YiToN12,Asil10,LiuWiOpt12}. The work in \cite{Routing04} studied the minimal delay SPR against an oblivious adversary, and the regret is a suboptimal $O(t^{2/3})$.  The throughput-competitive route selection against an \emph{adaptive} adversary was studied in \cite{info05} with regret $O(t^{2/3})$, which yields the
worst routing performance. Gy$\ddot{o}$rgy \emph{et al.} \cite{Routing07} provided a complete routing
framework under the oblivious adversary attack, and it is based on both link and path measurements with
 order-optimal
regrets $O(t^{1/2})$. The works in \cite{Info_Rout13,YiToN12,Asil10,LiuWiOpt12}  considered  benign environments
 to be better modeled by the stochastic setting without adversarial events, where link weights follow
some unknown stochastic  (i.i.d.) distributions. Bhorkar \emph{et al.} \cite{Asil10} consider routing based on each link (hop-by-hop), who has an order-optimal
regret $O(logt)$. The first solution for SPR as the stochastic combinatorial semi-bandit MAB problem was seen
in \cite{YiToN12}, and it indicates  a regret $O(n^4logt)$ given the number of links $n$. As noticed, the regret of
 endhost-based routing greatly increases with the network size.  \cite{Info_Rout13} probed the least measured links for i.i.d. distributed
   links and  it had considered the practical delayed feedback issue, which has improved regrets compared with \cite{YiToN12}.
   Although the algorithm could handle temporally-correlated links, it is not suitable for the adversarial link condition.
   In \cite{LiuWiOpt12}, the author
   proposed  an adaptive  SPR algorithm under stochastically varying link
   states that achieves an $O(k^4logt)$ regret, where $k$ is the dimension of the path set.

%


The stochastic and adversarial MABs have co-existed in parallel for almost two decades. Recently,  \cite{Seldin14}
tried to bring them together in the classic MAB framework.    Our current work is motivated by  \cite{Seldin14} by using a novel exploration parameter over each
channel  to detect its evolving patterns, i.e., stochastic, contaminated, or adversarial, but
they do not generalize their
idea to describe general environmental scenarios (No mixed adversarial and stochastic regime, which is very typical scenario) with potential engineering and security applications.
 Our current work  uses the idea of introducing the novel exploration parameter  \cite{Seldin14} into our special
  \emph{combinatorial semi-bandit } MAB problem by exploiting the link
 dependency among different paths, which is a nontrivial and much harder problem. This new framework avoids the
   computational inefficiency issue for general combinatorial adversary bandit problems as indicated in  \cite{XYICDCS14} \cite{Lugosi12}. It achieves
  a regret bound of order $O({k_r}\sqrt {tn\ln n})$, which only has a factor of $O(\sqrt{k_r})$ factor off when compared to the optimal
  $O(\sqrt {{k_r}tn\ln n})$ bound in the combinatorial adversary bandit setting \cite{OR2014}. However, we do believe that the regret bound
  in our framework is the optimal one for the exponential weight (e.g. EXP3 \cite{non_MAB02}) type of algorithm settings in the sense that
   the algorithm is computationally  efficient. Thus, our work is also a first computationally  efficient combinatorial MAB
    algorithm for general unknown environments\footnote{ As noticed,
    the stochastic combinatorial bandit problem does no have this issue as indicated in \cite{XYICDCS14}\cite{Branislav2015}.}.
 What is more surprising and encouraging, in the stochastic regimes (including the contaminated stochastic regime), our algorithms achieve a regret
 bound of order $\tilde{O}(\frac{{n{k}\log {{(t)}}}}{\Delta })$. In the sense of
  channel numbers $n$ and size of links within each strategy $k$, this is the best result to date for combinatorial
  stochastic bandit problems \cite{Branislav2015}. Please note that
  in \cite{YiToN12}, they have a regret bound of order ${O}(\frac{{n^4\log {{(t)}}}}{\Delta })$;
   in \cite{Liu12}, the
  regret bound is ${O}(\frac{{n^3\log {{(t)}}}}{\Delta })$; in \cite{LiuWiOpt12},  regret bound is
  ${O}(\frac{{k^4\log {{(t)}}}}{\Delta })$  and in \cite{Dani08}, the regret bound is ${O}(\frac{{n^2\log^3 {{(t)}}}}{\Delta })$. Thus, our
  proposed algorithms are order optimal with respect to $n$ and ${k}$ for all different regimes, which indicates the optimal
  scalability for general wireless communication systems or networks.

\section{Problem Formulation}
\subsection{Network Model}
We consider the given network modeled by a directed acyclic graph with a set of vertices connected by edges, and
\emph{sources} vertices have streams of data packets to send to the distinguished \emph{destination} vertices. Formally,
let $V$ denote the set of nodes and $E$ the set of links with $|E|=n$. For any given source-destination pair $(s,d)$, let $\mathcal{P}$
denote the set of all candidate paths as routing strategies belongs to $(s,d)$ with $|\mathcal{P}|=N$. We represent each path \textbf{i}, as a routing
 strategy, has $\textbf{i} \in \mathcal{P} \subset \{0,1\}^n$. Overlaps (sharing links) between different paths are allowed. Let $k_\textbf{i}$ denote the
length of each path $\textbf{i}$ and $k$ denote the maximum length of path(s) within $\mathcal{P}$. Thus, the size of
$N$ is upper bounded by $n^k$, which is exponentially large to the number of edges $n$, and therefore a   computationally
efficient algorithm is desirable.

At each time slot $t$, depending on the traffic and link quality, each edge $e$  may experience a different unknown varying link weight $\ell_t(e)$.
A packet traversed over the chosen path have  a sum of weights $\ell_t(\mathbf{i})$ equals to  $\sum\nolimits_{e \in \mathbf{i}}  \ell_t(e)$ of links composing
 the path. If there are adversary events imposed on a link (or the related routers, which will finally affect the link weight), it is  attacked. We denote the respective
 set and  number of
 these attacked links by  $E_a$ and $k_a$.
We assume the link weights to be additive, where the typical additive metric is link delays (there are others, e.g., log of delivery ratio).  We do not make any assumption on the distribution  of each $\ell_t(e), \forall e \in E$, it can, by default, follow some  unknown stochastic process (i.i.d.), and coud be attacked arbitrarily by diversified
 attackers (non-i.i.d.) that is different across different links. Without loss of generality (W.l.o.g), we transform the link weights such that $\ell_t(e) \in
[0, 1]$
 for all $e$ and $t$,  and there is a single attacker launches all attacks.

\vspace{-.2cm}
\subsection{Problem Description}
The main task for a given source-destination pair is to find a path $\textbf{i} \in \mathcal{P} $ with minimized path weights $\ell_t(\mathbf{i})$
over time. If each link weight $\ell_t(e)$ is known at every time slot, the problem can be efficiently solved by classic routing solutions
(e.g., Dijkstra's algorithm). Otherwise, it necessitates online learning-based approaches.

W.l.o.g, we consider  source routing, where source $s$ periodically sends probes along the potential paths to measure the network and adjust its choices over time. We use the \emph{link-level} measurements to
record the link weights  as in traceroute  on the probed   paths, i.e., if path $\textbf{i}$ is probed at the beginning of time slot $t$, all its link
weights are observed at the end of $t$. If multi-path probing is allowed  with a  budget of $M_t$ paths at time $t$, all the probed path
$\hat{\textbf{i}}_1,...,
\hat {\textbf{i}}_{M_t}$ will be traced out and their link weights are observed. Let  ${L_t}(\textbf{i}) = \sum\nolimits_{s = 1}^t \ell_t(\textbf{i})=
{\sum\nolimits_{s = 1}^t { \sum\nolimits_{e \in \textbf{i}} {{\ell_s(e)}} } }$ be the cumulative weight up to $t$ for a selected path $\textbf{i}$.
 Then,  ${\textbf{i}^*} \buildrel \Delta \over = \mathop {\arg \min }\nolimits_{\textbf{i} \in \mathcal{P}} \left\{ {{L_t}(\textbf{i})} \right\}$
denotes the expected minimum  weight path. Let $\mathbf{I}_t$ denotes a particular path
chosen at time slot $t$ from $\mathcal{P}$, then for a particular SPR algorithm, the cumulative weight up to time slot $t$ is ${{\hat L}_t} \left( {{\textbf{I}_s}} \right)= \sum\nolimits_{s = 1}^t {{\ell_s}\left( {{\textbf{I}_s}} \right)}  = \sum\nolimits_{s = 1}^t {\sum\nolimits_{e \in {\mathbf{I}_s}}^{} {{\ell_s}\left( e \right)} }$. Our goal is to jointly select a path $\textbf{I}_s$ (and a set of probing path
 if allowed, i.e., $\textbf{i} \in M_s$) at each time slot $s$ up to time $t$ ($s=1,2,...,t$) such that $\textbf{I}_s$ converges to
  ${\textbf{i}^*}$ as fast as possible in all different situations. Specifically, the performance of the SPR algorithm is qualified by
  the \emph{regret} $R(t)$, defined as the difference between the selected paths by
the proposed algorithm and the expected minimum  weight path up to $t$ time slots. Note that  $R(t)$ is a random
variable because $\textbf{I}_t$ depends on link measurement. We use $\mathbb{E}_t [\cdot]$ to denote expectations on
 realization of all strategies as random variables up to round $t$.  Therefore, the expected regret can be written as
  \begin{IEEEeqnarray*}{l}
  \begin{array}{l}
\!\!\!\!\!\!\!\!\! R(t)
=\mathbb{E} [ \sum\limits_{s = 1}^t {\mathbb{E}_s [ {\sum\limits_{e \in {\textbf{I}_s}} {{\ell_s(e)}} } ]}]  - \mathop {\min }\limits_{\textbf{i} \in {\mathcal{P}}} ( { \mathbb{E}   [  {\sum\limits_{s = 1}^t
 \mathbb{E}_s [ {\sum\limits_{e \in \textbf{i}} {{\ell_s(e)}}  }] }  ]} ).
 \end{array}\IEEEyesnumber \label{eq:Regrets}
\end{IEEEeqnarray*}
The goal of the algorithm is
to minimize the regret. 



\subsection{The Four Regimes of Network Environments}
Since our algorithm does not need to know the nature of the environments,  different characteristics  of  the environments  will affect its performance differently. We categorize them into four typical regimes as shown in Fig. 1.
\begin{figure}
\vspace{-.3cm}
\centering
\includegraphics[scale=.38]{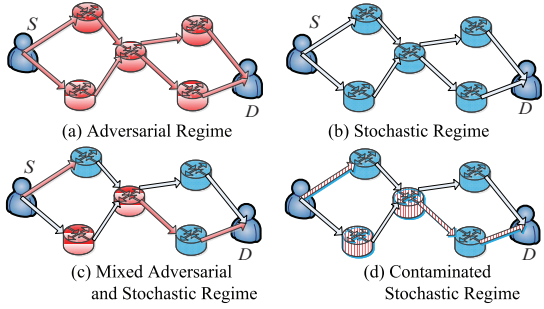}
\caption{SPR in Different Regimes of Unknown Environments}
\label{fig:digraph}
\vspace{-.5cm}
\end{figure}

\subsubsection{Adversarial Regime}
In this regime, there is an attacker attacks (e.g., send interfering power and corrupt routers by worms, etc.)
over all $n$ links such that link weights  suffered completely (See Fig.1 (a)) that lead to
metric value (e.g. link delay) loss.  Note that the adversarial regime
as a classic  model of the well known non-stochastic MAB problem \cite{non_MAB02} implies that the attacker launches attack in every time slot. It is the most general setting and the other three regimes can be regarded as  special cases of the
  adversarial regime.


\textbf{Attack Model:} Different attack philosophies will lead to different level of effectiveness. We focus on the following
two type of jammers in the adversarial regime:¡¡

  a) \emph{Oblivious attacker:} an oblivious attacker attacks different links with different attacking strength as a result of different
data rate reductions, which is independent of the past communication records it might have observed.


 b) \emph{Adaptive attacker:} an adaptive attacker selects its attacking strength on the targeted (sub)set of links by utilizing its
 past experience and observation of the previous communication records. It is  very powerful and can infer the
  SPR protocol and can launch attacks  with different level of strength over a subset of links or routers during a
  single time slot based on the historical monitoring records.  As shown in a recent work \cite{Arora12}, no bandit algorithm can guarantee a sublinear
regret $o(t)$ against an adaptive adversary with unbounded memory, because the adaptive adversary can mimic the
behavior of SPR protocol to attack, which leads to a linear regret (the attack can not be defended). Therefore, we consider a more practical
 \emph{$\theta$-memory-bounded adaptive adversary} \cite{Arora12} model. It is an adversary constrained
to loss functions that depends only on the $\theta+1$ most recent strategies.

\subsubsection{Stochastic Regime}
In this regime, the transceiver is communicating over $n$ stochastic links as shown in Fig.1 (b). The link weights $\ell_t(e), \forall e \in 1,...,n$  of each link $e$ are sampled independently from an unknown distribution that depends on $e$, but not on $t$. We use $\mu_e = \mathbb{E}
\left[ {{\ell_t(e)}} \right]$ to denote the expected loss of link $e$. We define link $e$ as the \emph{best link}
if $\mu (e) = {\min _{e'}}\{ {\mu (e')} \}$ and \emph{suboptimal link} otherwise; let $e^*$ denote some best link. For each link $e$,
define the gap $\Delta (e) = \mu (e) - \mu ({e^*})$;
let $\Delta_e = {\min _{e:\Delta (e) > 0}}\left\{ {\Delta (e)} \right\}$ denote the minimal gap of links.
The regret can be rewritten as
  \begin{IEEEeqnarray*}{l}
R(t) = \sum\nolimits_{e=1}^n {\mathbb{E}\left[ {{N_t}(e)} \right]} \Delta (e). 
\IEEEyesnumber \label{eq:StocR}
\end{IEEEeqnarray*}
Note that we can calculate the regret either from the perspective of links $e \in 1,...,n$ or from the perspective of strategies
$\textbf{i} \in \mathcal{P}$. However, because of the set of strategies (paths) grows exponentially with respect to $n$ and
it does not exploit the link dependency between different strategies, we can calculate the regret from links, where tight regret
bounds are achievable.

\subsubsection{Mixed Adversarial and Stochastic Regime}
This regime assumes that the attacker only attacks $k_a$ out of $k$ active links at each time slot shown in Fig.1 (c). There is always a
$k_a/k$ portion of links under adversarial attack while the other $(k-k_a)/k$ portion is stochastically distributed.


\textbf{Attack Model:} We consider the same attack model as in the adversarial regime. The difference here is that the attacker
only attacks a subset of links of size $k_a$ over the total  $k$ links.


\subsubsection{Contaminated Stochastic Regime}
The definition of the contaminated stochastic regime comes from many practical observations that only a few links (or routers)
 and time slots are exposed to adversary. In this regime, for the oblivious attacker, it selects some slot-link pairs $(t,e)$ as ``locations" to attack before the SPR starts, while the remaining link weights are generated the same as in the stochastic regime. We can introduce and define
the \emph{attacking strength} parameter $\zeta \in [0, 1/2)$. After certain $\tau$ timslots, for all $t> \tau$ the total number of contaminated
locations of each suboptimal link up to time $t$ is $t\Delta(e)\zeta$ and the number of contaminated locations of each best
link is $t\Delta_e \zeta$. We call a contaminated stochastic regime \emph{moderately contaminated}, if $\zeta$ is at most $1/4$, we can prove
that for all $t> \tau$ on the average over the stochasticity of the
loss sequence the adversary can reduce the gap of every link by at most one half.


\begin{algorithm}
\caption{AOSPR-EXP3++: An MAB-based Algorithm for AOSPR}
\begin{algorithmic}
\STATE \textbf{Input}: $n, k, t$,  and See text for definition of $\eta_t$ and $\xi_t(e)$.
\STATE \textbf{Initialization}: Set initial link losses $\forall e \in [1, n], \tilde{\ell}_0(e)= 0$. Then the initial link and
 path weights $\forall e \in [1, n], w_0(e)= 1$ and $\forall \textbf{i} \in [1, N],
W_0(\textbf{i})= k$, respectively. 
\!\STATE \textbf{Set}: \!\! $\beta_t \!\! =\!\! \frac{1}{2}\sqrt {\frac{{\ln n}}{{tn}}}$; ${\varepsilon _t}\left( e \right) \!= \! \min \left\{ {\frac{1}{{2n}},{\beta _t},{\xi _t}\left( e \right)} \right\},\forall e \in \left[ {1,n} \right]$.
\FOR { time slot $t=1,2,...$}
\STATE 1: The source selects a path $\mathbf{I}_t$ at random according to the  probability $\rho _t(\textbf{i}),\forall \mathbf{i} \in \mathcal{P}$, with $\rho _t(\textbf{i})$ computed as follows:
\begin{IEEEeqnarray*}{l}{\rho _t}(\textbf{i}) = \left\{ \begin{array}{l}
\!\!\!  \left( {1 - \sum\nolimits_{e=1}^{n} {{\varepsilon _t}(e)} } \right)\frac{{{w_{t - 1}}\left( \textbf{i} \right)}}{{{W_{t - 1}}}} + \! \sum\limits_{e \in \textbf{i}} {{\varepsilon _t}(e)} \ \emph{if} \ \textbf{i} \in \mathcal{C} \\
\!\!\!  \left( {1 - \sum\nolimits_{e=1}^{n}  {{\varepsilon _t}(e)} } \right)\frac{{{w_{t - 1}}\left( \textbf{i} \right)}}{{{W_{t - 1}}}} \  \quad   \quad \quad \quad \
\text{\emph{if}} \ \textbf{i} \notin \mathcal{C}
\end{array} \right.
\IEEEyesnumber \label{eq:Alg1p1}
\end{IEEEeqnarray*}

\STATE 2: The source computes the probability ${{\tilde \rho }_t}(e),\forall e \in E$, as
\begin{IEEEeqnarray*}{l}
\begin{array}{l}
\!\!\!\!\!{{{\tilde \rho }_t}}(e) = \sum\nolimits_{\textbf{i}:e \in \textbf{i}} {{\rho _t}(\textbf{i})}
 = \left( {1 - \sum\nolimits_{e=1}^{n} {{\varepsilon _t}(e)} } \right)\frac{{\sum\nolimits_{\textbf{i}:e \in \textbf{i}} {{w_{t - 1}}\left( \textbf{i} \right)} }}{{{W_{t - 1}}}} \\
\quad \quad\quad \quad \quad \quad \quad \quad \ \ + \sum\nolimits_{e \in \textbf{i}} {{\varepsilon _t}(e)} \left| {\left\{ {\textbf{i} \in \mathcal{C}:e
\in \textbf{i}} \right\}} \right|.
\end{array}
\IEEEyesnumber \label{eq:Alg1p2}
\end{IEEEeqnarray*}

\STATE 3: Observe the suffered link loss $\ell_{t-1}(e), \forall e \in \textbf{I}_t$, and update its estimated  value by
  ${\tilde{\ell}_t}(e) = \frac{{{\ell_t}(e)}}{{{{{\tilde \rho }_t}}(e)}} , \forall e \in \textbf{I}_t $. Otherwise, ${\tilde{\ell}_t}(e) =0, \forall e \notin \textbf{I}_t$.

\STATE 4: The source updates all the weights as ${w_t}\left( e \right) = {w_{t - 1}}\left( e \right){e^{ - \eta_t { {\tilde \ell}_t}(e)}} = {e^{ - \eta_t {{\tilde L}_t}(e)}}$ and $
{{\bar w}_t}\left( \textbf{i} \right) = \prod\nolimits_{e \in \textbf{i}} {{w_t}(e)}  = {{\bar w}_{t - 1}}\left( \textbf{i} \right){e^{
- \eta_t {{\tilde \ell}_t}(\textbf{i})}}$,
where ${{\tilde L}_t}(e) = {{\tilde L}_{t - 1}}(e) + {{\tilde \ell}_{t - 1}}(e),{{\tilde \ell}_{t - 1}}(e) =
\sum\nolimits_{e \in \textbf{i}} {{{\tilde \ell}_{t - 1}}(e)}$ and ${{\tilde L}_t}(\textbf{i}) = {{\tilde L}_{t - 1}}
(\textbf{i}) + {{\tilde \ell}_{t - 1}}(\textbf{i})$. The sum of weights of all strategies is calculated as
$
{W_t} = \sum\limits_{\textbf{i} \in \mathcal{P}} {{{\bar w}_t}\left( \textbf{i} \right)}.
$
\ENDFOR
\end{algorithmic}
\end{algorithm}
\vspace{-.1cm}

\section{Single-Source Adaptive Optimal SPR }
\subsection{Coupled Probing and Routing}
This section develops an SPR algorithm for a single source. The design philosophy is that the source  collects the link delays of the
previously chosen paths, based on which it can decide the next time slot routing strategy.
 The main difficulty is that it requires the algorithm to appropriately balance between \emph{exploitation} and
\emph{exploration}. On the one hand, such an algorithm needs to keep exploring the best set of paths;
on the other hand, it needs to exploit the already selected best set of paths so that
they are not under utilized. 

We describe Algorithm 1, namely AOSPR-EXP3++,  a variant based on
EXP3 algorithm, whose performance in the four regimes is proved to be asymptotically
optimal. Our new algorithm uses the fact that when link delays of the chosen path are revealed, it
also provides useful information about other paths with shared common links.  During each time slot, we assign
a link weight that is dynamically adjusted based on the link delays revealed to the source. The weight of a path is determined
by the product of weights of all links. Our algorithm has two control parameters: the \emph{learning rate} $\eta_t$ and the exploration
parameter $\varepsilon_t(e)$ for each link $e$. To facilitate the adaptive and optimal SPR
 without the knowledge about the nature of the environments, the crucial innovation is the introduction
of exploration parameter $\xi_t(e)$ into $\varepsilon_t(e)$ for each link $e$, which is tuned individually for each arm depending on the past observations.

Let $N$ denote the total number of strategies at the source side. A set of \emph{covering strategy} is defined to ensure that
each link is sampled sufficiently often. It has the property that for each link $e$, there is a strategy $\textbf{i} \in \mathcal{C}$ such that
$e \in \textbf{i}$. Since there are only $n$ links and each strategy includes $k$ links, we set $|\mathcal{C}| = \lceil {\frac{n}{{{k}}}} \rceil$.
As such, there is no-overlapping among different paths in the set of the covering strategy to maximize the covering range.
 The value $ \sum\nolimits_{e \in \textbf{i}} {{\varepsilon _t}(e)}$ means the randomized
 exploration probability for each strategy $\textbf{i} \in \mathcal{C}$, which is the summation of each link $e$'s exploration probability ${\varepsilon _t}\left( e \right)$ that belongs to the strategy $\textbf{i}$. The introduction of $\sum\nolimits_{e \in \textbf{i}} {{\varepsilon _t}\left( e \right)}$ ensures $\rho _t(\textbf{i}) \ge  \sum\nolimits_{e \in \textbf{i}} {{\varepsilon _t}(e)}$
 so that a mixture of exponential
 weight distribution and uniform distribution  \cite{Auer95}.

In the following discussion, we show that tuning only the learning rate $\eta_t$ is sufficient to control and obtain the
regret of the AOSPR-EXP3++ in the adversarial regime, regardless of the choice of exploration parameter $\xi_t(e)$. Then we show that
tuning only the exploration parameter $\xi_t(e)$ is sufficient to control the regret of AOSPR-EXP3++ in the stochastic regimes regardless of the choice of
 $\eta_t$, as long as $\eta_t \ge \beta_t$. To facilitate the AOSPR-EXP3++ algorithm without knowing about the nature of environments, we can apply the
 two control parameters simultaneously by setting $\eta_t = \beta_t$ and use the control parameter $\xi_t(e)$ in the stochastic regimes such that it can achieve
 the optimal ``root-t" regret in the adversarial regime and almost optimal ``logarithmic-t" regret in the stochastic regime (though with a suboptimal
 power in the logarithm).

 \subsection{Performance Results in Different Regimes}
We present the regret performance of our proposed AOSPR-EXP3++ algorithm in different regimes as  follows. The analysis
involves with martingale theory and some special concentration inequalities, which are put in Section VII.


\subsubsection{Adversarial Regime}
We first show that tuning $\eta_t$ is sufficient to control the regret of AOSPR-EXP3++ in the adversarial regime, which is a
general result that holds for all other regimes.

\textbf{Theorem 1.}  Under the \emph{oblivious} adversary, no matter how the status of the links change (potentially in an adversarial manner), for $\eta_t = \beta_t$ and any
$\xi_t(e) \ge 0$, the regret of the AOSPR-EXP3++ algorithm for any $t$ satisfies 
\begin{IEEEeqnarray*}{l}
R(t) \le 4{k}\sqrt {tn\ln n}.
\end{IEEEeqnarray*}

Note that Theorem 1 attains the same result as in \cite{Routing07} in the adversarial regime for oblivious adversary, based on which we get result
for the adaptive adversary in the following.

\textbf{Theorem 2.}  Under the \emph{$\theta$-memory-bounded adaptive} adversary, no matter how the status of the links change (potentially in an adversarial manner), for $\eta_t = \beta_t$ and any
$\xi_t(e) \ge 0$, the regret of the AOSPR-EXP3++ algorithm for any $t$ satisfies
\begin{IEEEeqnarray*}{l}
R(t) \le (\theta +1){(4{k}\sqrt {n\ln n} )^{\frac{2}{3}}}{t^{\frac{2}{3}}} + o({t^{\frac{2}{3}}}).
\end{IEEEeqnarray*}

\subsubsection{Stochastic Regime}
Now we show that for any $\eta_t \ge \beta_t$, tuning the exploration parameters $\xi_t(e)$ is sufficient to control the regret of the algorithm
in the stochastic regime. We also consider a different  way of tuning the exploration parameters $\xi_t(e)$ for  practical
 implementation considerations. We begin with an idealistic assumption that the gaps $\Delta(e), \forall e \in n$ is known, just to
give an idea of what is the best result we can have and our general idea for all our proofs.


\textbf{Theorem 3}. Assume that the gaps $\Delta(e), \forall e \in n,$  are known. Let $t^*$ be the
minimal integer that satisfies ${t^*(e)} \ge \frac{{4{c^2}n\ln {{({t^*(e)}\Delta {{(e)}^2})}^2}}}{{\Delta {{(e)}^4}\ln (n)}}$.
For any choice of $\eta_t \ge {\beta _t}$
 and any $c \ge 18$, the regret of the AOSPR-EXP3++ algorithm with $\xi_t(a)=\frac{{c\ln (t\Delta {{(e)}^2})}}{{t\Delta {{(e)}^2}}}$ in the
 stochastic regime satisfies 
   \begin{displaymath}
   \begin{array}{l}
 R(t) \le \sum\limits_{e = 1, \Delta {(e)} >0  }^n {O\left( {\frac{{k\ln {{(t)}^2}}}
{{\Delta {{(e)}}}}} \right)}  + \sum\limits_{e = 1, \Delta {(e)} >0  } \Delta {{(e)}} t^* {{(e)}}\\
\hspace{.7cm} =
{O\left( {\frac{{k n\ln {{(t)}^2}}} {{\Delta_{e} {{}}}}} \right)}
+ \sum\limits_{e = 1, \Delta {(e)} >0  }\tilde O\left( {\frac{n}{{\Delta {{(e)}^3}}}} \right).
 \end{array}
   \end{displaymath}

From the upper bound results, we note that the leading constants $k$  and $n$ are optimal and tight as indicated in CombUCB1 \cite{Branislav2015} algorithm. However, we have a factor of $\ln(t)$ worse of the regret performance than
the optimal ``logarithmic-t" regret as in \cite{Info_Rout13,YiToN12,Asil10},  \cite{Robbins1985},\cite{Branislav2015}, \cite{LiuWiOpt12}, where the
performance gap is
trivially negligible (See numerical results in Section VII).


{\emph{A Practical Implementation by Estimating the Gap}}:
Because of the gaps $\Delta(e), \forall e \in n$ can not be known in advance before running the algorithm.
 Next, we show a more practical result that uses the empirical gap as an estimate
of the true gap. The estimation process can be performed in background for each link $e$ that
starts from the running of the algorithm, i.e.,
\begin{IEEEeqnarray*}{l}
{{\hat \Delta }_t}(e) = \min \left\{ {1,\frac{1}{t}\left({{\tilde L}_t}(e) -\mathop {\min }\limits_{e'} ({{\tilde L}_t}(e'))\right )} \right\}.
\IEEEyesnumber \label{eq:EstD1}
\end{IEEEeqnarray*}
 This is a first algorithm that can be used in many real-world applications.

\textbf{Theorem 4.} Let $c \ge 18$ and $\eta_t \ge \beta_t$. Let $t^*$ be the minimal integer that satisfies $t^*
 \ge \frac{{4{c^2}\ln {{(t^*)}^4} n}}{{\ln (n)}}$, and let ${t^*}(e) = \max \left\{ {{t^*},\left\lceil {{e^{1/\Delta {{(e)}^2}}}} \right\rceil } \right\}$ and
$t^*=max_{\{e \in n\}} t^* {{(e)}}$. The regret of the AOSPR-EXP3++ algorithm
with ${\xi _t}(e) = \frac{{c{{\left( {\ln t} \right)}^2}}}{{t{{\hat \Delta }_{t-1}}{{(e)}^2}}}$, termed as
AOSPR-EXP3++$^\emph{AVG}$, in the stochastic regime satisfies
   \begin{displaymath}
      \begin{array}{l}
 R(t) \le \sum\limits_{e = 1, \Delta {(e)} >0  }^n {O\left( {\frac{{k\ln {{(t)}^3}}}
 {{\Delta {{(e)}}}}} \right)}  + \sum\limits_{e = 1, \Delta {(e)} >0  }^n \Delta {{(e)}} t^* {{(e)}}\\
 \hspace{.7cm}=
 O\left( {\frac{{n k\ln {{(t)}^3}}}
 {{\Delta_{e} }}} \right) + n t^*.
    \end{array}
   \end{displaymath}

From the theorem, we observe that factor of another  $ln(t)$ worse of the regret performance when compared to the idealistic
case. Also, the additive constant  $t^*$ in this theorem can be  very large. However, our experimental results show that
a minor modification of this algorithm achieves a comparable performance with ComUCB1 \cite{Branislav2015} in the stochastic regime.

\subsubsection{Mixed Adversarial and  Stochastic Regime}
The mixed adversarial and stochastic regime can be regarded as a special case of mixing
adversarial and stochastic regimes. Since there is  always a jammer randomly attacking $k_a$  links out of the total $n$ links
  out of the total $k$ links
constantly over time,  we will have the following theorem for the AOSPR-EXP3++$^\emph{AVG}$ algorithm, which is
a much more refined regret performance bound than the general regret bound in the adversarial regime.

\textbf{Theorem 5.} Let $c \ge 18$ and $\eta_t \ge \beta_t$. Let $t^*$ be the minimal integer that satisfies $t^*
 \ge \frac{{4{c^2}\ln {{(t^*)}^4} n}}{{\ln (n)}}$, and Let ${t^*}(e) = \max \left\{ {{t^*},\left\lceil {{e^{1/\Delta {{(e)}^2}}}} \right\rceil } \right\}$ and
$t^*=max_{\{e \in n\}} t^* {{(e)}}$. The regret of the AOSPR-EXP3++ algorithm
with ${\xi _t}(e) = \frac{{c{{\left( {\ln t} \right)}^2}}}{{t{{\hat \Delta }_{t-1}}{{(e)}^2}}}$, termed as
AOSPR-EXP3++$^\emph{AVG}$ under \emph{oblivious jamming} attack, in the mixed stochastic and adversarial regime satisfies
   \begin{displaymath}
      \begin{array}{l}
 R(t) \le \sum\limits_{e = 1, \Delta {(e)} >0  }^{n} {O\left( {\frac{{(k-k_a)\ln {{(t)}^3}}}
 {{\Delta {{(e)}}}}} \right)}  + \sum\limits_{e = 1, \Delta {(e)} >0  }^{n} \Delta {{(e)}} t^* {{(e)}} \\
\hspace{1.0cm} + 4{k_a}\sqrt {tn\ln n} \\
 \hspace{.7cm}=
 O\left( {\frac{{{n} (k-k_a)\ln {{(t)}^3}}}
 {{\Delta_{e} }}} \right) + n t^* + O\left({k_a}\sqrt {tn\ln n}\right).
    \end{array}
   \end{displaymath}

Note that the results in Theorem 5 have better regret performance than the results obtained by adversarial MAB as shown
 in Theorem 1 and the adaptive SPR algorithm in \cite{info05}. Similarly, we have the following result under adaptive adversarial attack.

\textbf{Theorem 6.} Let $c \ge 18$ and $\eta_t \ge \beta_t$. Let $t^*$ be the minimal integer that satisfies $t^*
 \ge \frac{{4{c^2}\ln {{(t^*)}^4} n}}{{\ln (n)}}$, and Let ${t^*}(e) = \max \left\{ {{t^*},\left\lceil {{e^{1/\Delta {{(e)}^2}}}} \right\rceil } \right\}$ and
$t^*=max_{\{e \in n\}} t^* {{(e)}}$. The regret of the AOSPR-EXP3++ algorithm
with ${\xi _t}(e) = \frac{{c{{\left( {\ln t} \right)}^2}}}{{t{{\hat \Delta }_{t-1}}{{(e)}^2}}}$, termed as
AOSPR-EXP3++$^\emph{AVG}$ \emph{$\theta$-memory-bounded adaptive} adversarial attack, in the mixed stochastic and adversarial regime satisfies
   \begin{displaymath}
      \begin{array}{l}
  \!\!R(t) \le \sum\limits_{e = 1, \Delta {(e)} >0  }^{n} {O\left( {\frac{{(k-k_a)\ln {{(t)}^3}}}
 {{\Delta {{(e)}}}}} \right)}  + \sum\limits_{e = 1, \Delta {(e)} >0  }^{n} \Delta {{(e)}} t^* {{(e)}} \\
   \quad\quad   + (\theta +1){(4{k_a}\sqrt {n\ln n} )^{\frac{2}{3}}}{t^{\frac{2}{3}}} + o({t^{\frac{2}{3}}})\\
 \hspace{-.1cm}=
 O\left( {\frac{{{(k-k_a)} n\ln {{(t)}^3}}}
 {{\Delta_{e} }}} \right) + n t^* + O\left((\theta +1){({k_a}\sqrt {n\ln n} )^{\frac{2}{3}}}{t^{\frac{2}{3}}}\right).
    \end{array}
   \end{displaymath}


\subsubsection{Contaminated Stochastic Regime}
We  show that the algorithm AOSPR-EXP3++$^\emph{AVG}$ can still retain ``polylogarithmic-t" regret in the  contaminated stochastic
regime. The following is the result for the \emph{moderately contaminated stochastic regime}.



\textbf{Theorem 7.} Under the setting of all parameters given in Theorem 3, for ${t^*}(e) = \max \left\{ {{t^*},\left\lceil {{e^{4
/\Delta {{(e)}^2}}}} \right\rceil } \right\}$, where $t^*$ is defined as before  and
$t_3^*=max_{\{e \in n\}} t^* {{(e)}}$, and the attacking strength parameter $\zeta \in [0,1/2)$ the regret of the \rm{AOSPR-EXP3++} algorithm in the
contaminated stochastic regime that is contaminated after $\tau$ steps satisfies
   \begin{displaymath}
         \begin{array}{l}
\!\! R(t) \le \!\! \sum\limits_{e = 1, \Delta {(e)} >0  }^n \!\!\!\! {O\left( {\frac{{k\ln {{(t)}^3}}}
 {{(1-2\zeta)\Delta {{(e)}}}}} \right)} \!\! + \!\! \sum\limits_{e = 1, \Delta {(e)} >0  }^n \!\!\!\!  \Delta {{(e)}} \max\{t^* {{(e)}}, \tau\}. \\
 \! \hspace{.7cm}=
  {O\left( {\frac{{n k\ln {{(t)}^3}}}
 {{(1-2\zeta)\Delta_{e}}}} \right)} + n t_3^*.
       \end{array}
   \end{displaymath}



If $\zeta \in (1/4, 1/2)$, we find that the leading factor $1/(1-2\zeta)$ is very large, which is \emph{severely} contaminated. Now, the obtained regret bound is not quite meaningful, which could be much worse than the regret performance in the adversarial regime for
both oblivious and adaptive adversary.

\section{Accelerated AOSPR Algorithm}
This section focuses on the accelerated learning by  multi-path probing, cooperative learning between multiple source-destination
pairs and other practical issues. All important proofs are put in Section VIII.

\subsection{Multi-Path Probing for Adaptive Online SPR}
Intuitively, probing multiple paths simultaneously would offer
the source more available information to make decisions,  which results in faster learning and smaller regret value. At each time slot $t$, the source
gets a budget $1 \le {M_t} \le N$ and picks a subsect ${\mathcal{O}_t} \subseteq \left\{ {1,...,N} \right\}$ of $M_t$ paths to probe and
observe the link weights of these routes. Note that the links weights that belong to the un-probed set of
paths $\mathcal{P} \setminus \mathcal{O}_t$ are still unrevealed. Accordingly, we have the probed and
observed set of links $\tilde{\mathcal{O}}_t$ with the simple property $ e \in \tilde{\mathcal{O}}_t,  \forall e \in \textbf{i} \in \mathcal{O}_t$.
 The proposed algorithm 2 is based on Algorithm 1 with
 $\omega_{t-1}(\textbf{i})= w_{t-1}(\textbf{i})/W_{t-1}, \forall \textbf{i} \in \mathcal{P}$ and
$\omega_{t-1}(e)= \sum\nolimits_{\textbf{i}: e \in \textbf{i}} w_{t-1}(e)/W_{t-1}, \forall e \in E$.
The probability
$\varrho_t=(\varrho_t(\mathbf{1}),...,
\varrho_t(\mathbf{N}))$ of each observed path is computed as
 \begin{IEEEeqnarray*}{l}
\!\!{\varrho _t}(\mathbf{i}) =
{\rho _t}(\textbf{i}) +  \left( {1 - {\rho _t}(\textbf{i})} \right)\frac{{{M_t} - 1}}{{{N} - 1}},
  \ if \ \textbf{i} \ \in
 \mathcal{O}_t,
\IEEEyesnumber \label{eq:Lim1}
 \end{IEEEeqnarray*}
 where a mixture of  the new exploration probability $(M_t-1)/(N-1)$ is introduced and $\rho _t(\textbf{i})$ is defined in (\ref{eq:Alg1p1}).
  Similarly,
the link probability $ { \tilde\varrho}_t=({ \tilde\varrho}_t({1})...,{ \tilde\varrho}_t({n}))$ is computed as
 \begin{IEEEeqnarray*}{l}
\!\!{\tilde \varrho _t}(e) =
{\tilde\rho _t}(e) +  \left( {1 - {\tilde\rho _t}(e)} \right)\frac{{{m_t} - 1}}{{{n} - 1}},
  \ if \ e \ \in
  \mathcal{\tilde O}_t.
\IEEEyesnumber \label{eq:Lim2}
 \end{IEEEeqnarray*}
Here, we have a link-level the new mixing exploration probability
$(m_t-1)/(n-1)$ and $\tilde\rho _t(e)$ is defined in (\ref{eq:Alg1p2}). The probing rate $m_t$ denotes the
number of simultaneous probes at time slot $t$. Assume the link weights measured by different probes within
the same time slot also satisfy the assumption in Section II-A.  The mixing probability $(m_t-1)/(n-1)$
 is informed by the source to all links along the probed and observed paths over a total of $n$ links of the
probed path $M_t$ is a constant value for a network with fixed topology.  The source needs to know the number of $n$ and
gradually collect the value of $m_t$ over time. Thus, the algorithm
faces the problems of ``Cold-Start" and delayed feedback. The design of (\ref{eq:Lim1}) and (\ref{eq:Lim2}) and the proof of all results
in this section are non-trivial tasks
in our unified framework .


\begin{algorithm}
\caption{AOSPR-MP-EXP3++: Prediction with Multi-Path Probing}
\begin{algorithmic}
\STATE \textbf{Input}: $M_1, M_2,...,$,  such that  $M_t \in \mathcal{P}$. Set $\beta_t, {\varepsilon _t}\left( e \right), {\xi _t}\left( e \right)$
at in Alg. 1.
  $\forall \textbf{i} \in \mathcal{P}, \tilde{L}_0(\textbf{i})= 0$ and $\forall e \in E, \tilde{\ell}_0(e)= 0$.

\FOR { time slot $t=1,2,...$}
\STATE 1:  Choose one path $H_t$ according to $\rho_t$ (\ref{eq:Alg1p1}). Get advice $\pi_t^{H_t}$ as the selected path.
Sample $M_t-1$ additional paths uniformly over $N$.  Denote the set of sampled paths by $\mathcal{O}_t$, where $H_t \in \mathcal{O}_t$
 and $|\mathcal{O}_t| = M_t$. Let $\mathds{1}_t^h= \mathds{1}_{\{h \in \mathcal{O}_t\}}$.

 \STATE 2: Update the path probabilities $\varrho_t(\mathbf{i})$ according to (\ref{eq:Lim1}). The  loss of the observed path is
  \begin{IEEEeqnarray*}{l}
{{\tilde \ell}_t}(\textbf{i}) = \frac{{{{ \ell}_t}(\mathbf{i})}}{\varrho_t(\mathbf{i}) } \mathds{1}_t^h,
 \forall \textbf{i} \in {\mathcal{O}_t}.
\IEEEyesnumber
 \end{IEEEeqnarray*}

\STATE 3: Compute the probability of choosing each link ${{\tilde \rho}_t}(e)$ that belongs to the selected path according to (\ref{eq:Alg1p2}).

\STATE 4: let $\mathds{1}(e)_t = \mathds{1}(e)_{e \in h \in \mathcal{O}_t}$. Update the link  probabilities $\tilde \varrho_t(e)$
 according to (\ref{eq:Lim2}). The  loss of the observed  links are
   \begin{IEEEeqnarray*}{l}
{{\tilde \ell}_t}(e) = \frac{{{{ \ell}_t}(e)}}{{{\tilde \varrho}_t}(e) } \mathds{1}(e)_t,
\forall {e} \in {\tilde{\mathcal{O}}_t}.
\IEEEyesnumber
 \end{IEEEeqnarray*}

\STATE 5:  Updates all weights ${w_t}\left( e \right), {{\bar w}_t}\left( \textbf{i} \right), W_t$ as in Alg.  1.
\ENDFOR
\end{algorithmic}
\end{algorithm}

\emph{The Performance Results of Multi-path Probing in the Four Regimes}: If $m_t$ is a constant or lower
 bounded by $m$, we have the following results. 

 \textbf{Theorem 8.}  Under the \emph{oblivious} attack with the same setting of Theorem 1,
 the regret of the AOSPR-EXP3++ algorithm in the accelerated learning with probing rate $m$ satisfies
\begin{IEEEeqnarray*}{l}
R(t) \le 4{k}\sqrt {t\frac{n}{m}\ln n}.
\end{IEEEeqnarray*}

\textbf{Theorem 9.}  Under the \emph{$\theta$-memory-bounded adaptive} attack
with the same setting of Theorem 2, the regret of the AOSPR-EXP3++ algorithm  in the accelerated learning with probing rate $m$ satisfies
\begin{IEEEeqnarray*}{l}
R(t) \le (\theta +1){(4{k}\sqrt {\frac{n}{m}\ln n} )^{\frac{2}{3}}}{t^{\frac{2}{3}}} + o({t^{\frac{2}{3}}}).
\end{IEEEeqnarray*}

We consider the practical implementation in the stochastic regime by estimating the gap as in the
 (\ref{eq:EstD1}), and the result under accelerated learning is given as:

\textbf{Theorem 10.} With all other parameters hold as in Theorem 4, the regret of the AOSPR-EXP3++ algorithm
with ${\xi _t}(e) = \frac{{c{{\left( {\ln t} \right)}^2}}}{{mt{{\hat \Delta }_{t-1}}{{(e)}^2}}}$ in the accelerated learning with probing rate $m$,
 in the stochastic regime satisfies
   \begin{displaymath}
      \begin{array}{l}
 R(t) \le \sum\limits_{e = 1, \Delta {(e)} >0  }^n {O\left( {\frac{{k\ln {{(t)}^3}}}
 {{m\Delta {{(e)}}}}} \right)}  + \sum\limits_{e = 1, \Delta {(e)} >0  }^n \Delta {{(e)}} t^* {{(e)}}\\
 \hspace{.7cm}
 = O\left( {\frac{{n k\ln {{(t)}^3}}}
 {{m\Delta_{e} }}} \right) + n t^*.
    \end{array}
   \end{displaymath}
%

\textbf{Theorem 11.} With all other parameters hold as in Theorem 5, the regret of the AOSPR-EXP3++ algorithm
with ${\xi _t}(e) = \frac{{c{{\left( {\ln t} \right)}^2}}}{{mt{{\hat \Delta }_{t-1}}{{(e)}^2}}}$
 under \emph{oblivious jamming} attack in the accelerated learning with probing rate $m$, in the mixed stochastic and adversarial regime
 satisfies
   \begin{displaymath}
      \begin{array}{l}
 R(t) \le \sum\limits_{e = 1, \Delta {(e)} >0  }^{n} {O\left( {\frac{{(k-k_a)\ln {{(t)}^3}}}
 {{m\Delta {{(e)}}}}} \right)}  + \sum\limits_{e = 1, \Delta {(e)} >0  }^{n} \Delta {{(e)}} t^* {{(e)}} \\
  \hspace{.7cm} + 4{k_a}\sqrt {t\frac{n}{m}\ln n}\\
 \hspace{.7cm}=  O\left( {\frac{{{n} (k-k_a)\ln {{(t)}^3}}}
 {{m\Delta_{e} }}} \right) + n t^* + O\left({k_a}\sqrt {t\frac{n}{m}\ln n}\right).
    \end{array}
   \end{displaymath}

\textbf{Theorem 12.} With all other parameters hold as in Theorem 6, the regret of the AOSPR-EXP3++ algorithm
with ${\xi _t}(e) = \frac{{c{{\left( {\ln t} \right)}^2}}}{{mt{{\hat \Delta }_{t-1}}{{(e)}^2}}}$ under the \emph{$\theta$-memory-bounded adaptive} attack
 in the accelerated learning with probing rate $m$, in the mixed stochastic and adversarial regime satisfies
   \begin{displaymath}
      \begin{array}{l}
\!\!\! R(t) \le \sum\limits_{e = 1, \Delta {(e)} >0  }^{n} {O\left( {\frac{{(k-k_a)\ln {{(t)}^3}}}
 {{m\Delta {{(e)}}}}} \right)}  + \sum\limits_{e = 1, \Delta {(e)} >0  }^{n} \Delta {{(e)}} t^* {{(e)}} \\
   \quad\quad   + (\theta +1){(4{k_a}\sqrt {\frac{n}{m}\ln n} )^{\frac{2}{3}}}{t^{\frac{2}{3}}} + o({t^{\frac{2}{3}}})\\
 \hspace{-.1cm} =
  O\left( {\frac{{{n} (k-k_a)\ln {{(t)}^3}}}
 {{m\Delta_{e} }}} \right) + n t^* + O\left({(\theta +1)({k_a}\sqrt {\frac{n}{m}\ln n} )^{\frac{2}{3}}}{t^{\frac{2}{3}}}\right).
    \end{array}
   \end{displaymath}

\textbf{Theorem 13.}  With all other parameters hold as in Theorem 7, the regret of the \rm{AOSPR-EXP3++} algorithm in the accelerated learning with probing rate $m$ in the
contaminated stochastic regime satisfies
   \begin{displaymath}
         \begin{array}{l}
 \!\!\! R(t) \le  \!\! \!\!  \sum\limits_{e = 1, \Delta {(e)} >0  }^n  \!\!\!\! {O\left( {\frac{{k\ln {{(t)}^3}}}
 {{(1-2\zeta)\Delta {{(e)}}}}} \right)} \!\! + \!\! \sum\limits_{e = 1, \Delta {(e)} >0  }^n  \!\!\!\! \Delta {{(e)}} \max\{t^* {{(e)}}, \tau\}. \\
  \hspace{.6cm}=
   {O\left( {\frac{{n k\ln {{(t)}^3}}}
 {{m(1-2\zeta)\Delta_{e}}}} \right)} + n t_3^*.
       \end{array}
   \end{displaymath}



\subsection{Multi-Source Learning for SPR Routing}
So far we have focused on a single source-destination pair. Now, we turn to study the more
practical multi-source learning with multiple source-destination pairs $\{1,...,S\}$ (set $S=m$ to ease comparison), which may
also accelerate learning if sources can share information. Depending on the approach of information sharing,
we consider two typical cases: \emph{coordinated probing} and \emph{uncoordinated probing}. Both cases assume
the sources share link measurements. The difference is the selection of probing path is either \emph{centralized} or
\emph{distributed}.

In the coordinated probing case, the probing path are selected globally either by a cluster head or different sources. We  refer the algorithm
as  AOSPR-CP-EXP3++. Given total of $M_t$ source-destination pairs at time $t$, the $M_t$ probing paths are sequentially chosen which satisfies the
probing rate of one path per source-destination pair. It is identical to the multi-path probing case except that now the candidate paths
are not from one's own path, but from all source-destination pairs. Thus, the \emph{same results hold} as in the \textbf{Theorem 8-Theorem 13}.

\textbf{Theorem 14.} The regret upper bounds in all different regimes for AOSPR-CP-EXP3++ hold the same as in Theorem 8-Theorem 13.

In the uncoordinated probing case, the probing paths are selected in a distributed manner using AOSPR-EXP3++. We  denote the algorithm
as  AOSPR-UP-EXP3++. As such, links are no longer evenly measured since some links may be covered by more source-destination pairs than others. By applying a linear program
to estimate the low bound on the least probed link over time $t$, we have a scale factor $\bar \kappa$ that defines the
 dependency degree of overlapping paths. Then, we obtain the following result.


\textbf{Theorem 15.} The regret upper bounds in all different regimes for AOSPR-UP-EXP3++ are: if $\bar \kappa= 1$, it is equivalent
to the single source-destination pair case, the {same results hold} as in the {Theorem 1-Theorem 7};  if $\bar \kappa=m$,
it is equivalent to the accelerated learning of multi-path probing case where the regret results hold by
the substitution of  $m$  by $\bar \kappa$ in  the {Theorem 8-Theorem 13}. 

\subsection{The Cold-Start and Delayed feedback Issues}
\subsubsection{The Cold-Start Issue}
 Before the initialization of the algorithm, the source does not know
 the number of links $n$ and the simultaneous probed number of links  $m_t$, which is demanded in the probing probability
 calculation in (\ref{eq:Lim2}). Thus, the algorithm faces the ``cold-start" problem.  Note that the $N$ can be
 a complete collection of  paths of the
 source, it must contain a set of covering strategy $\mathcal{C'}$
 where the whole links of the network is covered and the total number of links $n$ is acquirable. Let $\underline{M}=\min_t\{M_1, M_2,...,M_t,...\}$
 denote the minimal probed path from source over time.
 We have the following Corollary 16 that indicates how long
  it takes for the AOSPR-MP-EXP3++ algorithm to work normally.

\textbf{Corollary 16.}
 It takes at most $\frac{N}{\underline{M}}$ timslots for the  AOSPR-MP-EXP3++ algorithm to finish
the ``Cold-Start" phase and start working normally.

\begin{IEEEproof} Denote the event of the probability that source node probes the paths uniformly over the $N$ possible paths at each time slot
as $\rm{X}$ and the event ${Y}$ of the probability that  the number of probed links over total of $n$ links.  Take the following conditional probability
we have $\mathbb{E}[X] = \mathbb{E}[X|Y]\mathbb{E}[Y]$, where $\mathbb{E}[X] = \frac{{\mathbb {E}[{M_t}]}}{N}$ and
$\mathbb{E}[Y] = \frac{{\mathbb {E}[{m_t}]}}{n}$.  Due to the potential dependency among different paths, $\mathbb{E}[X|Y]\le 1$. Thus,
$\mathbb{E}[Y] \le \mathbb{E}[X]$, i.e., $\frac{{\mathbb {E}[{m_t}]}}{n} \le \frac{{\mathbb {E}[{M_t}]}}{N}$,  which indicates $\frac{N}{M_t } \ge \frac{n}{m_t }, \forall t$.
Since each link has probability $p=\frac{m_t}{n}$ to be probed at every time slot. According to the
geometric distribution, the expected time that every link is probed is $\frac{1}{p} = \frac{n}{m_t } \le \frac{N}{M_t }  \le \frac{N}{\underline{M}}$.
This completes the proof.
\end{IEEEproof}

Nevertheless, for  practical implementations, the accurate number of $m$ and $n$ is still hard to  obtain. It often comes with
errors in acquiring these two values. Hence, we need to know the sensitivity of deviations of the two true values on the regret performance.
The result is summarized as follows. 

\textbf{Theorem 17.} Given the deviation of observed values $m$ is $ m_{\Delta}$ in (\ref{eq:Lim2}), the upper bound of the deviated
of regret $ R_{m_{\Delta}}(t)$ with respect to the original $ R(t)$ given its upper bound $ \bar R(t)$  \\
\emph{(a)} in the \emph{adversarial} regime is $-\frac{1}{2}{m_\Delta }\frac{n}{m}\bar R(t)$ and $-\frac{1}{3}{m_\Delta }\frac{n}{m}\bar R(t)$ for
\emph{oblivious} jammer and  for \emph{adaptive} adversary, respectively;
\emph{(b)} in the \emph{stochastic} regime and \emph{contaminated} regime
 are both $  -\frac{1}{2}\frac{m_{\Delta}}{m}\bar R(t)$;
\emph{(c)} in the \emph{ mixed adversarial and stochastic} regime is $ -\frac{1}{2}\frac{m_{\Delta}}{m}\bar R^{k
-k_a}(t) - \frac{1}{2}{m_\Delta }\frac{n}{m}\bar R^{k_a}(t)
$ for oblivious adversary and is $ -\frac{1}{3}\frac{m_{\Delta}}{m}\bar R^{k-k_a}(t) -\frac{1}{2}{m_\Delta }\frac{n}{m}\bar R^{k_a}(t)
$ for adaptive adversary, where $\bar R^{k-k_a}(t)$ and $\bar R^{k_a}(t)$ represents the upper bounds with $k-k_a$ and $k_a$ links in the stochastic
regime and adversarial regime, respectively. \\
Given the deviation of observed  values $n$ is $n_{\Delta}$ in (\ref{eq:Lim2}), the upper bound of deviated
of regret $ R_{n_{\Delta}}(t)$ with respect to the original $ R(t)$ given its upper bound $ \bar R(t)$\\
\emph{(d)} in the \emph{adversarial} regime is $\frac{1}{2}{n_\Delta }
\frac{n}{m}\frac{{m - 1}}{{n - 1}}\bar R(t) \cong  \frac{1}{2}{n_\Delta }\bar R(t)$ and $\frac{1}{3}{n_\Delta }
\frac{n}{m}\frac{{m - 1}}{{n - 1}}\bar R(t) \cong  \frac{1}{3}{n_\Delta }\bar R(t)$ for
\emph{oblivious} jammer and for \emph{adaptive} jammer, respectively;
\emph{(e)} in the \emph{stochastic} regime and \emph{contaminated} regime are both $\bar R_{\Delta}(t)=0$;
\emph{(f)}  in the \emph{mixed adversarial and stochastic} regime is $\frac{1}{2}{n_\Delta }\bar R^{k_a}(t)$  for oblivious adversary
and is $\frac{1}{3}{n_\Delta }\bar R^{k_a}(t)$  for adaptive adversary.

From the theorem 17, we know that the regret in the adversarial regime is more sensitive to the deviation $m_{\Delta}$ than the
deviation $n_{\Delta}$, which guides the design of the network to acquire accurate value of $m_t$ during the probing phase. For the
stochastic regimes, we also see that the regret is more sensitive to the deviation $m_{\Delta}$ than the
deviation $n_{\Delta}$. Moreover, the relative deviations on $R(t)$ stochastic regimes, i.e, $R_{m_{\Delta}}(t)/R(t)=
\Theta(\frac{m_{\Delta}}{m})$ and $R_{n_{\Delta}}(t)/R(t)= 0$ is much less (sensitive) than that in adversarial regimes, i.e.,
 $R_{m_{\Delta}}(t)/R(t)=
\Theta({m_\Delta }\frac{n}{m})$ and $R_{n_{\Delta}}(t)/R(t)= n_{\Delta}$. We see all these phenomena in the simulations.

\subsubsection{Delayed Feedback Issue} In the network with a large number of links, the link delay feedback to the
source node will spend a lot of
 time, which is prohibitive in the realtime process. Therefore, there are variant delayed feedbacks
 of each link to the source. Moreover, if the path is switched in the middle of a long streaming transmission, the network
 SPR protocol needs a while to find the new optimal transmission rate, and the delay of the first few packets after the switch
 can be very large. In a nutshell, the delayed feedback issue is practically important, and we have the following results.

\textbf{Theorem 18.} Given the largest expected deviations of observed link delay $\tau^*$, the expected delayed-feedback
regret $R_{d}(t)$ with respect to the original $ R(t)$
\emph{(a)} Assuming the delays depend only on time but not on links,  in the \emph{oblivious adversarial} regime is upper bounded by
 ${d_t}\mathbb{E}[R( {\frac{t}{{{d_t}}}} )]$, where ${d_t} = \min \{ t,\tau _t^ *  + 1\} $ and $\tau _t^ *$ is the largest link delay
 at time $t$;
\emph{(b)} Assuming  the delays to be independent
 of the rewards of the actions, in the \emph{stochastic} regime  and \emph{contaminated} regime
is upper bounded by $\mathbb{E}[R(t)] + \sum\nolimits_{e = 1}^n {{\Delta _e}\mathbb{E}[\tau _{e,t}^ * ]}$.

%


%
%
%

\section{The Computationally  Efficient Implementation of the AOPSR-EXP3++ Algorithm}
The implementation of algorithm $1$ requires the computation of probability distributions and storage of $N$ strategies, which is
obvious to have a time and space complexity $O(n^{k_\textbf{i}})$ for a given path of length $k_\textbf{i}$. As the number of links increases, the number of path will become
exponentially large, which is very hard to be scalable and results in low efficiency. To address this important problem,
we propose a computationally efficient enhanced algorithm by utilizing the dynamic programming techniques, as shown in Algorithm 3. The key idea
of the enhanced algorithm is to select links in the selected path one by one until $k_{\textbf{i}}$ links are chosen, instead of choosing a path
from the large path space in each time slot.

We use $S\left( {\bar e,\bar k} \right)$ to denote the path set of which each path selects $\bar k$ links from $\bar e, \bar e+1,
\bar e, ..., n$. We also use $\bar S\left( {\bar e,\bar k} \right)$ to denote the path set of which each path selects $\bar k$ links from
 link  $1,2,..., \bar e$. We define $
{W_t}(\bar e,\bar k) = \sum\nolimits_{\textbf{i} \in S(\bar e,\bar k)} {\prod\nolimits_{e \in \textbf{i}} {{w_t}(e)} }$ and $
{W_t}(\bar e,\bar k) = \sum\nolimits_{\textbf{i} \in \bar S(\bar e,\bar k)} {\prod\nolimits_{e \in \textbf{i}} {{w_t}(e)} },
$
Note that they have the following properties:
\begin{IEEEeqnarray*}{l}
{W_t}(\bar e,\bar k) = {W_t}(\bar e + 1,\bar k) + {w_t}(\bar e){W_t}(\bar e + 1,\bar k - 1),
\IEEEyesnumber \label{eq:Bpart1s}
\end{IEEEeqnarray*}
\begin{IEEEeqnarray*}{l}
{W_t}(\bar e,\bar k) = {W_t}(\bar e - 1,\bar k) + {w_t}(\bar e){W_t}(\bar e - 1,\bar k - 1),
\IEEEyesnumber \label{eq:Bpart2s}
\end{IEEEeqnarray*}
which implies both ${W_t}(\bar e,\bar k)$ and ${\bar W_t}(\bar e,\bar k)$ can be calculated in $O(k_rn)$ (Letting ${W_t}(\bar e,0)=1$ and
$W(n + 1,\bar k) = \bar W(0,\bar k) = 0$) by using dynamic programming for all $1 \le \bar e \le n$ and $1 \le \bar k \le k_{\textbf{i}}$.

\newcounter{mytempeqncnt}
\begin{figure*}[!t]
\normalsize
\setcounter{mytempeqncnt}{\value{equation}}
\setcounter{equation}{13}
\begin{equation}
(1 - \sum\nolimits_{e = 1}^n {{\varepsilon _t}(e)} )\frac{{\sum\nolimits_{k' = 0}^{{k_{\textbf{i}}} - 1} {{{\bar W}_{t - 1}}(e - 1,k'){w_{t - 1}}(e){W_{t - 1}}
(e + 1,{k_{\textbf{i}}} - k' - 1)} }}{{{W_{t - 1}}(1,k')}}\\ + \sum\limits_{e \in \textbf{i}} {\varepsilon _t}(e)\left| {\textbf{i} \in C:e \in
\textbf{i}} \right|\IEEEyesnumber \label{eq:Bpart34s}
\end{equation}
\setcounter{equation}{\value{mytempeqncnt}}
\hrulefill
\vspace*{4pt}
\end{figure*}

In step 1, instead of drawing a path, we select links of the path
one by one until a path is found. Here, we select links one by one in the increasing order of channel indices, i.e., we determine
whether the link $1$ should be selected, and the link $2$, and so on. For any link $e$, if $k' \le k_{\textbf{i}}$ links have been chosen in link
$1,..,e-1$, we select link $e$ with probability
\begin{IEEEeqnarray*}{l}
\frac{{{w_{t - 1}}(e){W_t}(e + 1,{k_{\textbf{i}}} - k' - 1)}}{{{W_{t - 1}}(e,{k_{\textbf{i}}} - k')}}\IEEEyesnumber \label{eq:Bpart3s}
\end{IEEEeqnarray*}
and not select $e$ with probability $
\frac{{{W_t}(e + 1,{k_{\textbf{i}}} - k' - 1)}}{{{W_{t - 1}}(e,{k_{\textbf{i}}} - k')}}.
$
Let $w(e) = {w_{t - 1}}(e)$ if link $e$ is selected in the path $\textbf{i}$; $w(e) = 0$ otherwise. Obviously, $w(e)$ is actually the weight
of $e$ in the path weight. In our algorithm, ${w_{t - 1}}(e) = \prod\nolimits_{e = 1}^n {w(e)}$. Let $c(e)=1$ if $e$ is selected in $\textbf{i}$;
$c(e)=0$ otherwise. The term $\sum\nolimits_{e = 1}^{\bar e} {c(e)}$ denotes the number of links chosen among link $1,2,...,\bar e$ in path
$\textbf{i}$. In this implementation, the probability that a path $\textbf{i}$ is selected, i.e.,  $ \frac{{{w_{t - 1}}(\textbf{i})}}{{{W_{t - 1}}}}$,
can be written as
\begin{IEEEeqnarray*}{l}
\!\!\!\!\!\!\prod\limits_{\bar e = 1}^n {\frac{{w(\bar e){W_{t - 1}}(\bar e + 1,{k_{\textbf{i}}} - \sum\nolimits_{e = 1}^{\bar e} {c\left( e \right)} )}}
{{{W_{t - 1}}(\bar e,{k_{\textbf{i}}} - \sum\nolimits_{e = 1}^{\bar e - 1} {c\left( e \right)} )}}}  = \frac{{\prod\limits_{\bar e = 1}^n {w(\bar e)} }}
{{{W_{t - 1}}(1,{k_{\textbf{i}}})}}.\IEEEyesnumber \label{eq:Bpart5s}
\end{IEEEeqnarray*}
This probability is equivalent to that in Algorithm 1, which implies the implementation is correct.
Because we do not maintain $w_t(\textbf{i})$, it is impossible to compute ${{{\tilde \rho }_t}}(e)$ as we have described in Algorithm 1. Then
${{{\tilde \rho }_t}}(e)$ can be computed within $O(nk_r)$ as in Eq.(\ref{eq:Alg1p2}) for each round.

\begin{figure*}
\vspace{-.2cm}
\begin{minipage}[t]{0.32\textwidth}
\includegraphics[scale=.22]{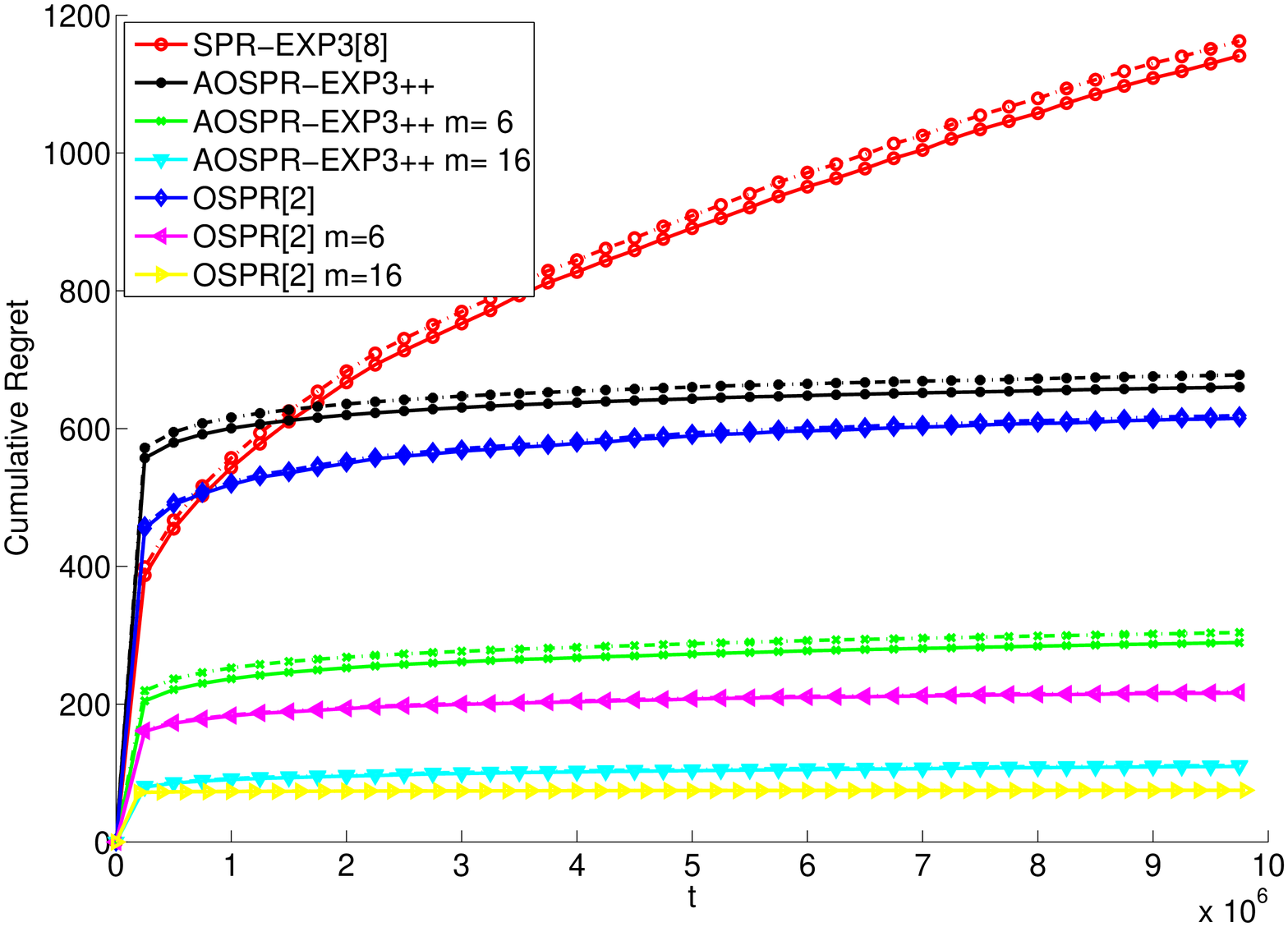}
\caption{Regret in Stochastic Regime}
\end{minipage}
\hspace{.3cm}
\begin{minipage}[t]{0.35\textwidth}
\includegraphics[scale=.24]{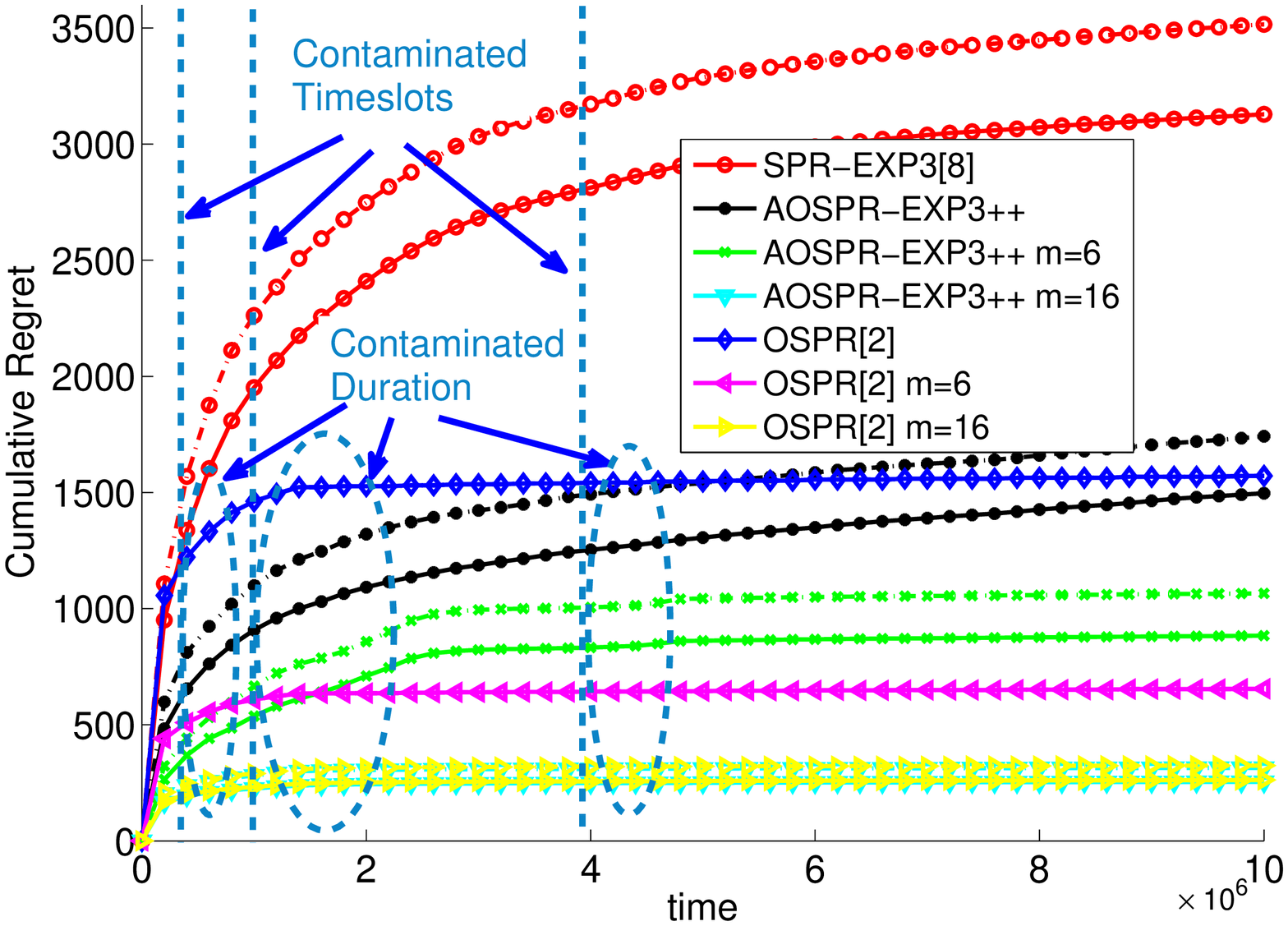}
\caption{Regret in Contaminated Regime}
\end{minipage}
\hspace{-.4cm}
\begin{minipage}[t]{0.32\textwidth}
\includegraphics[scale=.26]{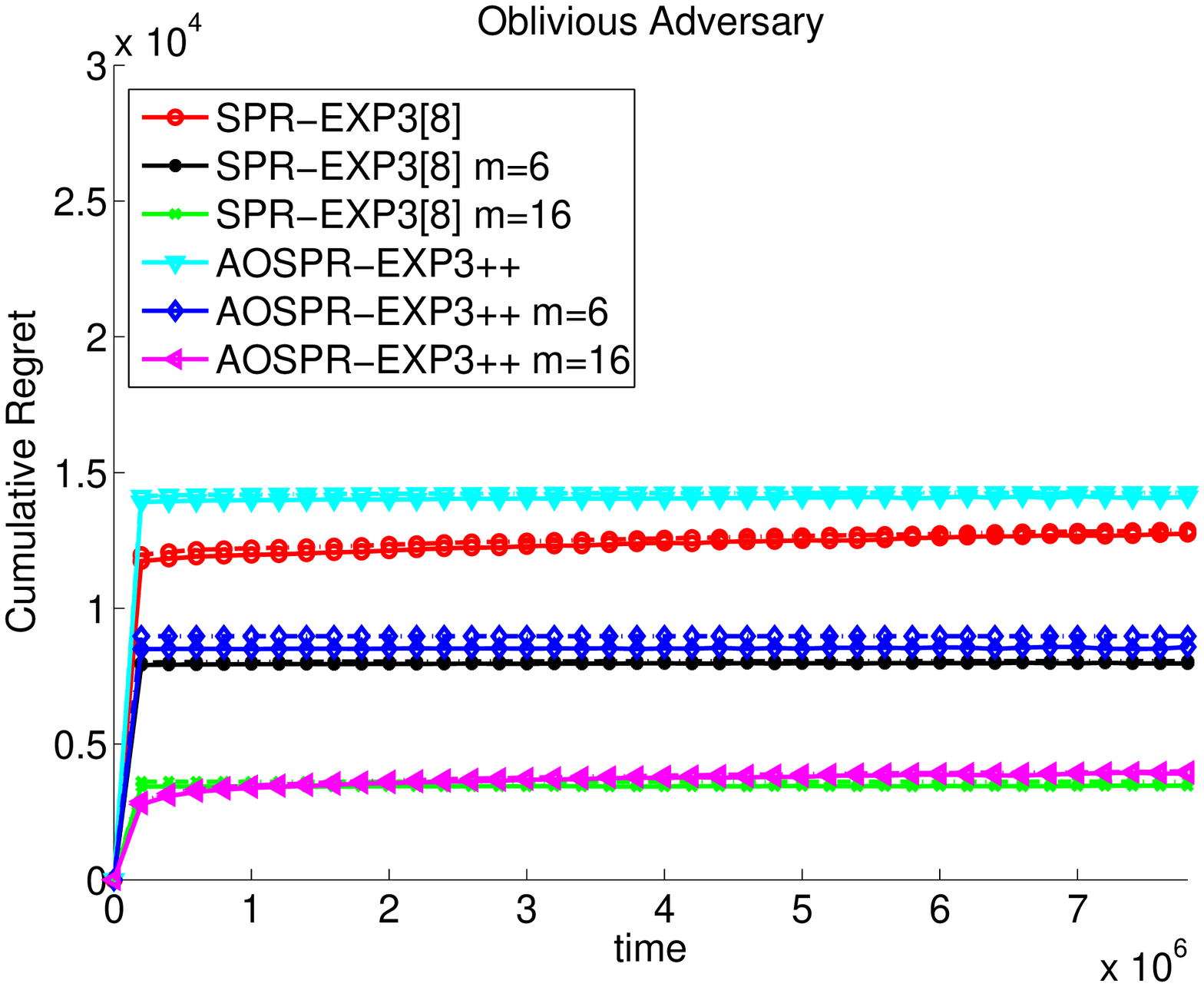}
  \caption{Regret in  Adversarial Regime}
  \end{minipage}
\end{figure*}

For the exploration parameters  $\varepsilon_t(e)$, since there are $k_\textbf{i}$ parameters of $\varepsilon_t(e)$ in the
last term of Eqs. (\ref{eq:Bpart34s}) below and there are $n$ links, the storage complexity is $O(kn)$. Similarly, we have the time complexity
$O(knt)$ for the maintenance of exploration parameters  $\varepsilon_t(e)$. Based on the above analysis, we can summarize the conclusions into the following theorem. Moreover, under delayed feedback, since the base algorithm has a memory requirement of $O(kn)$, the memory
required by the delayed AUFH-EXP3++ by time step $t$ is upper bounded by $O(kn \tau _t^ *)$.

\textbf{Theorem 19.} The Algorithm 2 has polynomial time complexity $O(knt)$,  space complexity $O(kn)$ and space complexity under
 the delayed feedback $O(kn \tau _t^ *)$ with respect to rounds $t$,  parameters $k$ and $n$.

Besides, because of the link selection probability for $q_t(e)$ and the updated weights of Algorithm 2 equals to
Algorithm 1, all the performance results in Section III and IV still hold for Algorithm 2.

\begin{algorithm}
\caption{A  Computational Efficient Implementation of AOPSR-EXP3++}
\begin{algorithmic}
\STATE \textbf{Input}: $n, k_{\textbf{i}}, t$,  and See text for definition of $\eta_t$ and $\xi_t(e)$.
\STATE \textbf{Initialization}: Set initial link weight $w_0(e)=1, \forall e \in [1,n]$. Let $W_t(e,0)=1$ and
$W(n+1,k')= \bar W(0,k')=0$ and compute $W_0(e,k')$ and $\bar W_0(e,k')$ follows Eqs. (\ref{eq:Bpart1s}) and (\ref{eq:Bpart2s}), respectively.
\FOR {time slot $t=1,2,...$}
\STATE 1: The source selects a link $e, \forall e \in [1,n]$ one by one according to the link's probability distribution computed following
 Eq. (\ref{eq:Bpart3s}) until a path with $k_{\textbf{i}}$ chosen links are selected.

\STATE 2: The source computes the probability $q_t(e),\forall e \in \left[ {1,n} \right]$ according to Eq. (\ref{eq:Bpart34s}).

\STATE 3: The source calculates the loss for channel $e$, $\ell_{t-1}(e), \forall e \in \textbf{i}_t$ based on the link
loss $\ell_{t-1}(e)$. Compute the estimated loss $\tilde{\ell}_{t}(e), \forall e \in [1,n]$ as follows:
\begin{IEEEeqnarray*}{l}
{\tilde{\ell}_t}(e) = \left\{ \begin{array}{l}
\frac{{{\ell_t}(e)}}{{{q_t}(e)}} \quad  \text{\emph{if channel}}  e \in {\textbf{i}_t} \\
0         \ \ \quad\quad \emph{otherwise}.
\end{array} \right.
\end{IEEEeqnarray*}

\STATE 4: The source updates all channel weights as ${w_t}\left( e \right) =
{w_{t - 1}}\left( e \right){e^{ - \eta_t { {\tilde \ell}_t}(e)}}
 = {e^{ - \eta_t {{\tilde L}_t}(e)}}, \forall e \in [1,n]$, and computes
  $W_t(e,k')$ and $\bar W_t(e,k')$ follows Eqs. (\ref{eq:Bpart1s}) and (\ref{eq:Bpart2s}), respectively.
\ENDFOR
\end{algorithmic}
\end{algorithm}
\textbf{}

%

\section{Numerical and Simulation Results}

We evaluate the performance of our online adaptive SPR algorithm using a  wireless sensor network (WSN) adopting the IEEE 802.15 standard deployed on a university building. The trace contains QoS metrics of
detailed link quality information, i.e., delay, goodput and packet loss rate, under
an extensive set of parameter configurations are measured. The dataset close to 50 thousand parameter configurations
were experimented and measurement data of more than 200
million packets were collected over a period of 6 months. Each sender-receiver is employed by a pair of
TelosB nodes, each equipped with a TI CC2420 radio using
the IEEE 802.15.4 stack implementation in TinyOS,  which is placed in hallways of a five floor building. The WSN contains $16$ nodes, and there is  line-of-sight path between the
two nodes of a path at a specific distance, which was varied for different
experiments ranging from 10 meters to 35 meters. Each node is forwarding packets under a particular stack parameter configuration, where the
configuration set is finite.

The delay perceived by a packet mainly consists of two parts:
queuing delay and service time delay, which are measured for every
data packet. More specifically, it includes the ACK frame transmission time, retransmission duration, ACK maximal
timeout if damage occurs by the adversarial attack, etc. To quantitatively
answer how all the layer stack parameters contribute to the
delay performance, there are four different types of datasets to emulate the following four typical
 regime of the environments: \emph{1)} the measured link
quality data at night, where the link states distributions are benign and only affected by multi-path reflections from
the walls; \emph{2)} the measured  contaminated link quality data at daytime from $3:00pm-4:00pm$, when university students and employees
walk  most frequently
in the hallway, which is a  particularly harsh wireless environment; \emph{3)} the measured adversarial
 link quality, by the same type of TelosB nodes working under the same  stack parameter configuration but sending garbage data to launch oblivious
   jamming attack during the run of the algorithm. The link delay labeled $N/A$ are replaced
by $1111$ or $999$ in the dataset  to indicate  completed data packet loss;  \emph{4)} the measured adversarial link quality under adaptive jamming attack, where the the algorithm is implemented by a set of $\theta$-memory jammers of our proposed AOSPR-EXP3++ algorithm. We omit the
mixed adversarial and stochastic regime for brevity.   

\begin{figure*}
\begin{minipage}[t]{0.3\textwidth}
\includegraphics[scale=.26]{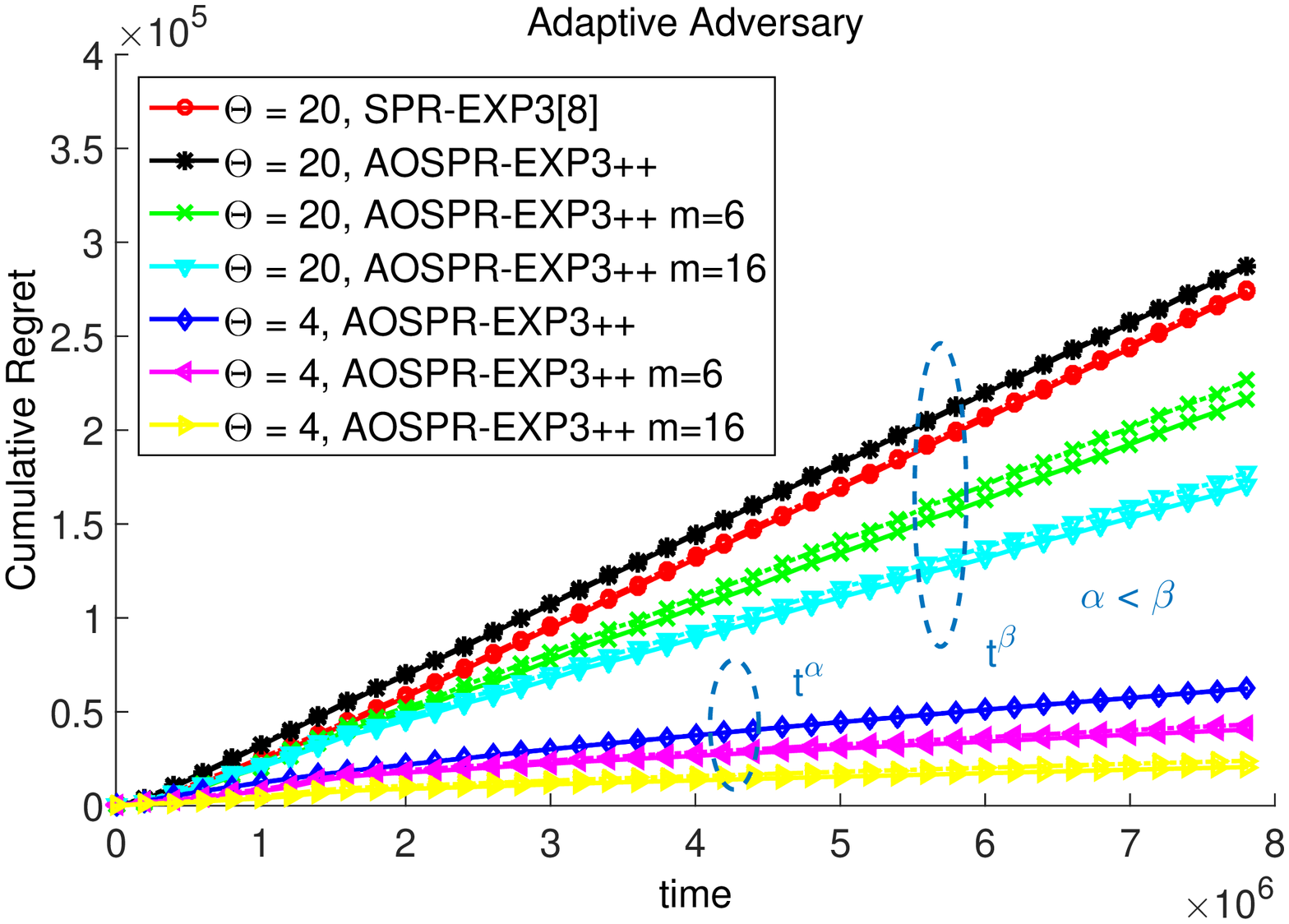}
  \caption{Regret in  Adversarial Regime}
  \end{minipage}
\hspace{.6cm}
  \begin{minipage}[t]{0.33\textwidth}
\includegraphics[scale=.223]{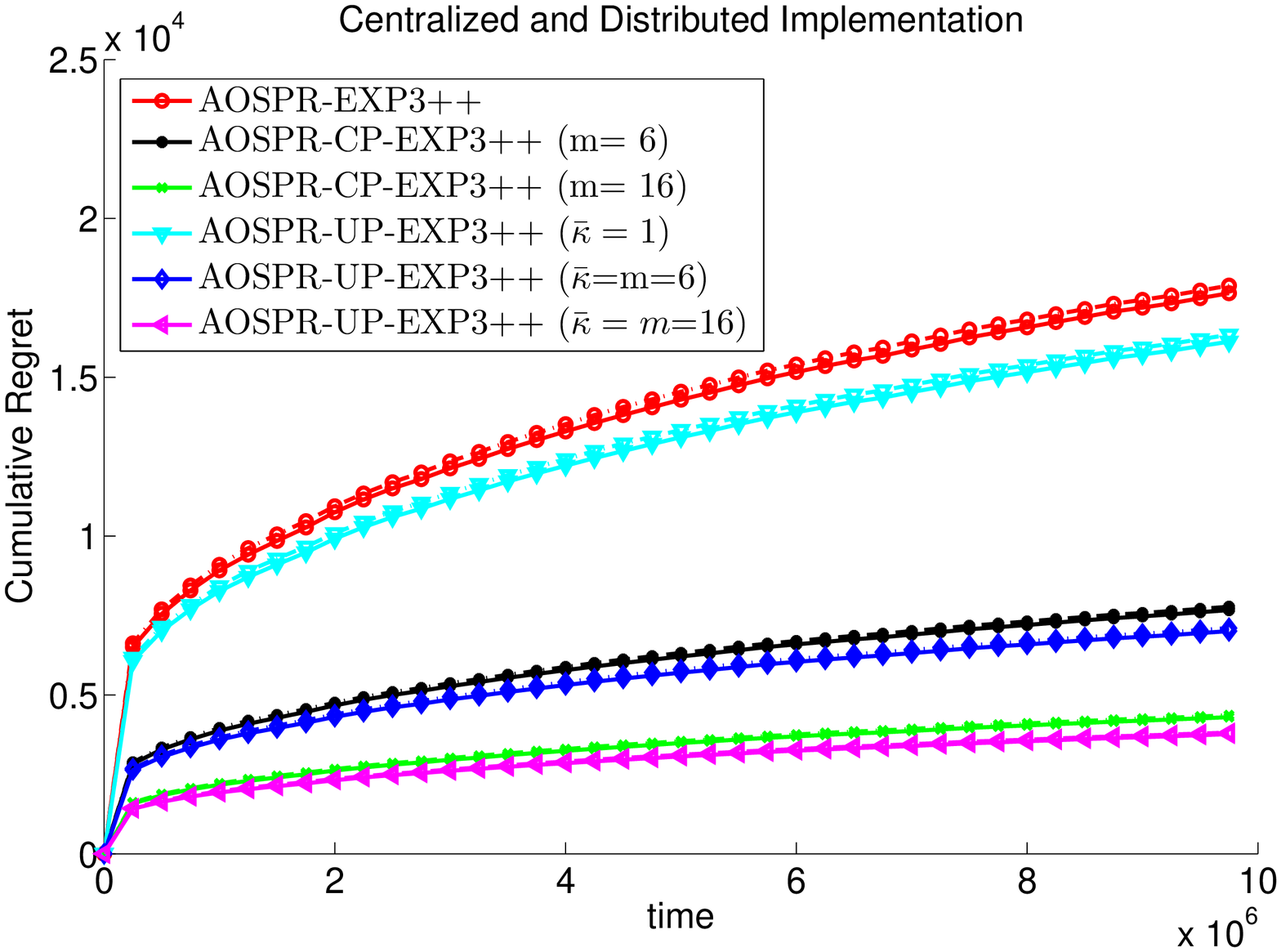}
  \caption{Centr. and Distr. Implementations}
  \end{minipage}
   \begin{minipage}[t]{0.33\textwidth}
\includegraphics[scale=.253]{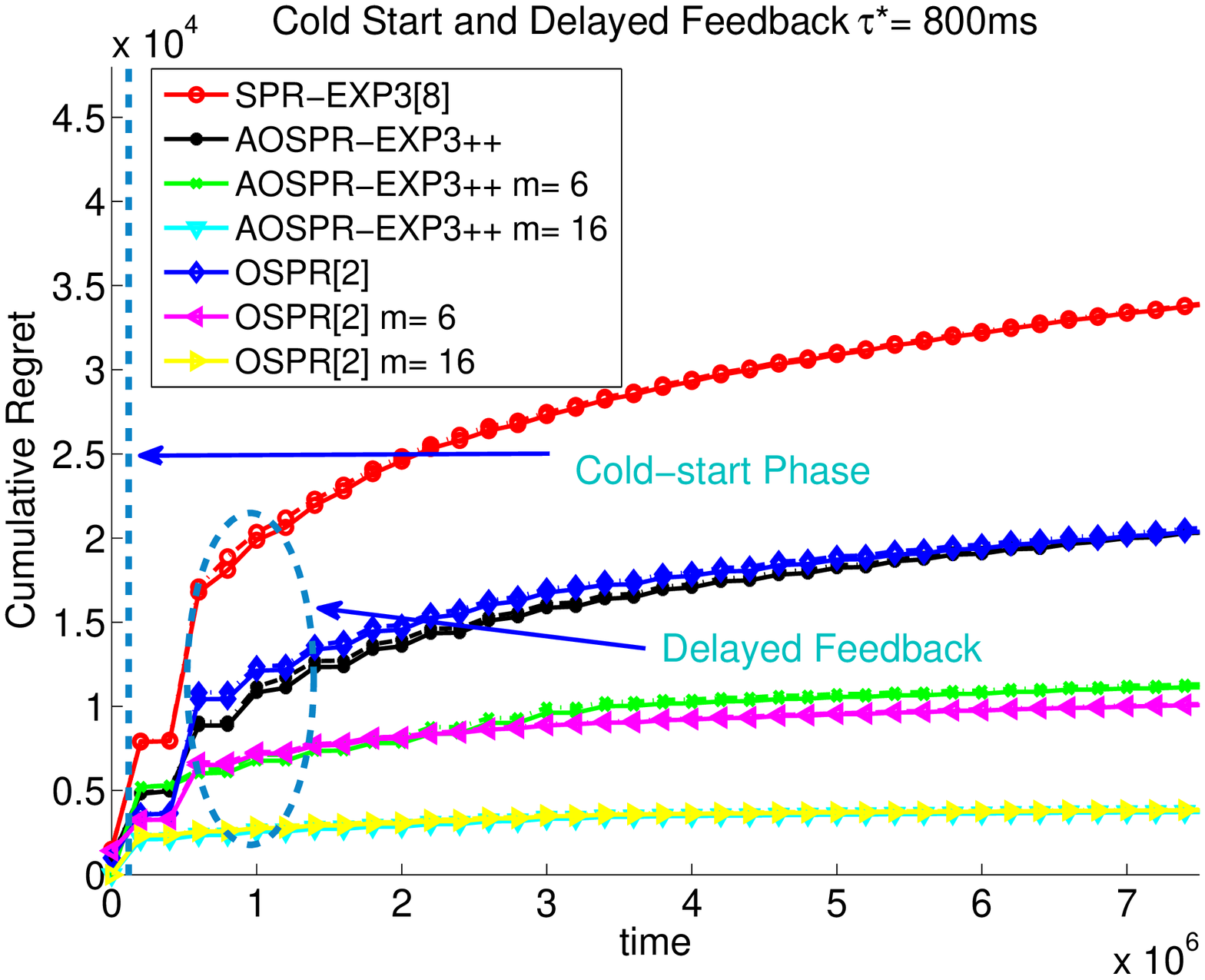}
  \caption{Cold Start and Delayed Feedback}
  \end{minipage}
  \end{figure*}

All computations of collected datasets were conducted on an off-the-shelf desktop with dual
$6$-core Intel i7 CPUs clocked at $2.66$Ghz.  We  make ten repetitions of each experiment to reduce the potential performance
bias. The \emph{solid} lines in the graphs represent  the mean performance over the experiments and the \emph{dashed} lines represents the mean plus on
standard deviation (std) over the ten repetitions of the corresponding experiments.  To show the advantage of our AOPSR-EXP3++ algorithms, we need to
compare the performance of ours to other existing MAB based algorithms. They include:
the EXP3 based SPR algorithm in \cite{Routing07}, which is named as ``SPR-EXP3"; the Upper-Confidence-Bound (UCB) based
online SPR algorithm ``OSPR" in \cite{Info_Rout13} and their variations.  We set all versions of our AOPSR-EXP3++ algorithms
parameterized by ${\xi _t}(e) = \frac{{\ln (t{{\hat \Delta }_t}{{(e)}^2})}}{{32t{{\hat \Delta }_t}{{(e)}^2}}}$, where ${{{\hat \Delta }_t}(e)}$ is the
empirical estimate of ${{{ \Delta }_t}(e)}$ defined in (\ref{eq:EstD1}).


In our first group of experiments  in the  stochastic regime (environment) as shown in Fig. 1, it is clear to see that AOPSR-EXP3++
 enjoys almost the same (cumulative) regrets as OSPR \cite{Info_Rout13}  and has much lower regrets over time than the
 adversarial SPR-EXP3 \cite{Routing07}. We also see the significantly regrets reduction when accelerated learning ($m=6, 16$)
  is employed for both OSPR and AOPSR-EXP3++.

In our second group of experiments in the moderately contaminated stochastic environment, there are several contaminated
time slots as labeled in Fig. 3.
In this case, the contamination is not fully adversarial, but drawn
from a different stochastic model. Despite the  corrupted rounds the AOPSR-EXP3++
algorithm successfully returns to the stochastic operation
mode and achieves better results than SPR-EXP3 \cite{Routing07}.  With light contaminations, the performance of OSPR in \cite{Info_Rout13} is
 comparable to  AOPSR-EXP3++,
although it is not applicable here due to  the i.i.d.  assumption of OSPR.

We conducted the third group of  experiments in the
adversarial regimes.  We studied the oblivious adversary case in Fig. 4. Due to the strong  interference effect on each link and
the arbitrarily changing feature of the jamming behavior, all algorithms experience very high accumulated
regrets.  It can be find that our AOPSR-EXP3++
algorithm will have close and slightly worst learning  performance when compared to SPR-EXP3 \cite{Routing07}, which confirms our theoretical
analysis. Note that we do not implement stochastic MAB algorithms such as ``OSPR" in \cite{Info_Rout13}, since it is inapplicable in this regime.
Moreover, we studied the adaptive adversary case in Fig. 5. Compared with Fig. 4, the learning performance is much worse, which results in
 close to linear (but still sublinear) regret values, especially when the memory size of  $\Theta$ is large.  From the collected data,
 we see a 252\% increase in the network delay under the adaptive adversary with $\Theta=4$ when compared to the
 oblivious adversary conditions. The value becomes 845\% when $\Theta=20$, which shows the adaptive attacker is very hard to defend.
%


The centralized and distributed implementation of our cooperative learning AOSPR-EXP3++ algorithms are presented in Fig. 6. The sensitivity of
the   deviation of observed values $m$ and $n$ in stochastic and adversarial regimes are presented in Fig. 8. It is
obvious to see the effects of $m_{\Delta}$ and $n_{\Delta}$ on the regret of AOSPR-EXP3++ in the stochastic regime is
much smaller compared to the counterpart in the adversarial regimes. On  average, we see a deviation of regret
about 12\% in the stochastic regime for the values of $m_{\Delta}$ and $n_{\Delta}$ show in Fig. 8, while the
deviation of regret is about 126\% in the adversarial regime. This indicated that our algorithm is more
sensitive to the attacked  environments than benign  environments.

We conduct the ``Cold-Start" and delayed feedback version of all algorithms in Fig. 7 in a general
unknown environment  that consists of data randomly mixed in all four regimes.  The
``Cold-Start" phase takes about $3\sim 20$ packets delivery timslots. Although it is hard to see the first 20 rounds
on the plot, their effect on all the algorithms is clearly
visible. For delayed feedback problem, we see a ``quick jump" of regret for adversarial MAB algorithms (e.g., SPR-EXP3\cite{Routing07}) at  initial rounds that confirms its multiplicative  effect to $\tau^*$, while the relative
small regret increase is seen for stochastic MAB algorithms (e.g., OSPR \cite{Info_Rout13}  and AOSPR-EXP3++) that confirms its additive  effect to  $\tau^*$.

\begin{figure}
\centering
\includegraphics[scale=.22]{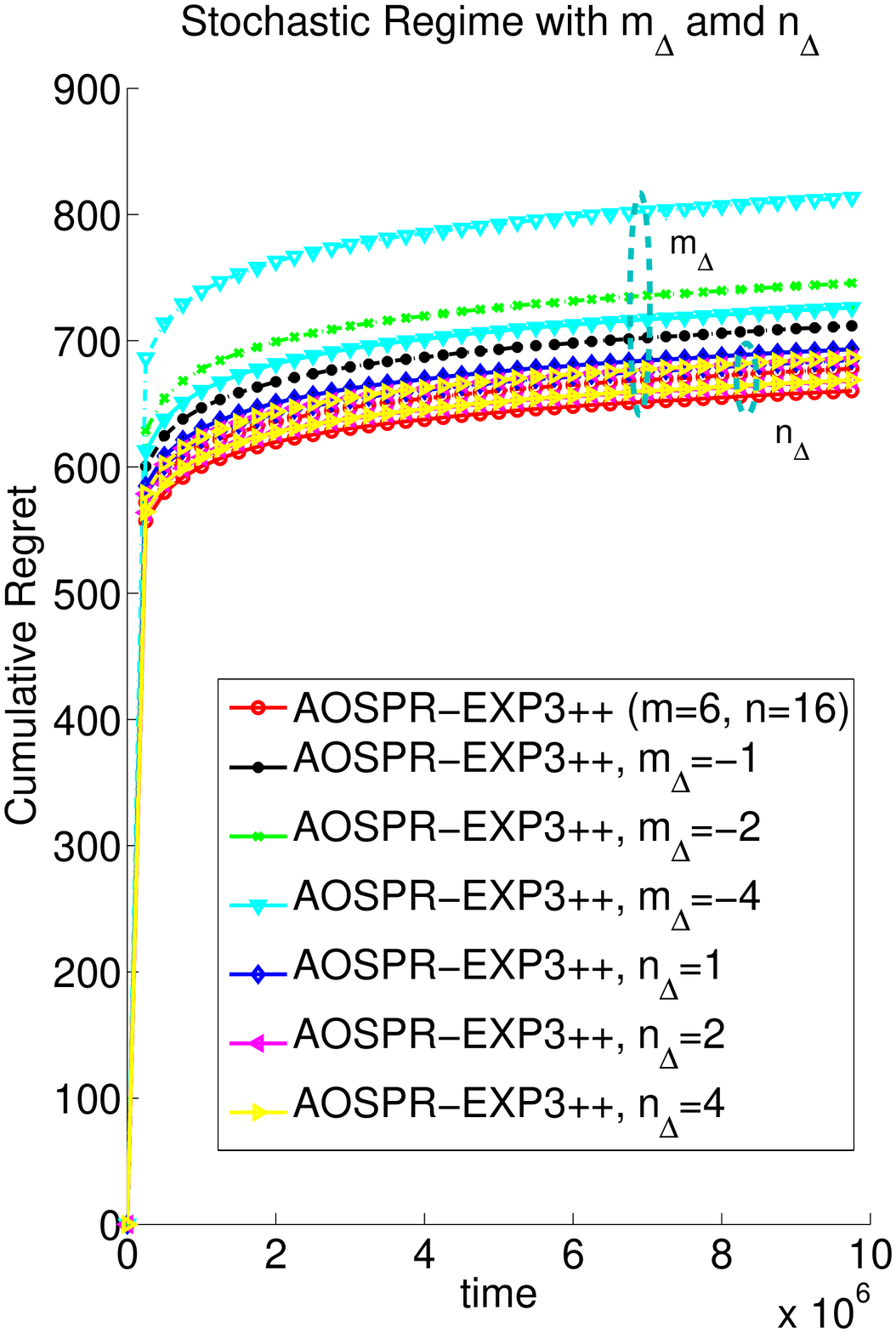}
\includegraphics[scale=.22]{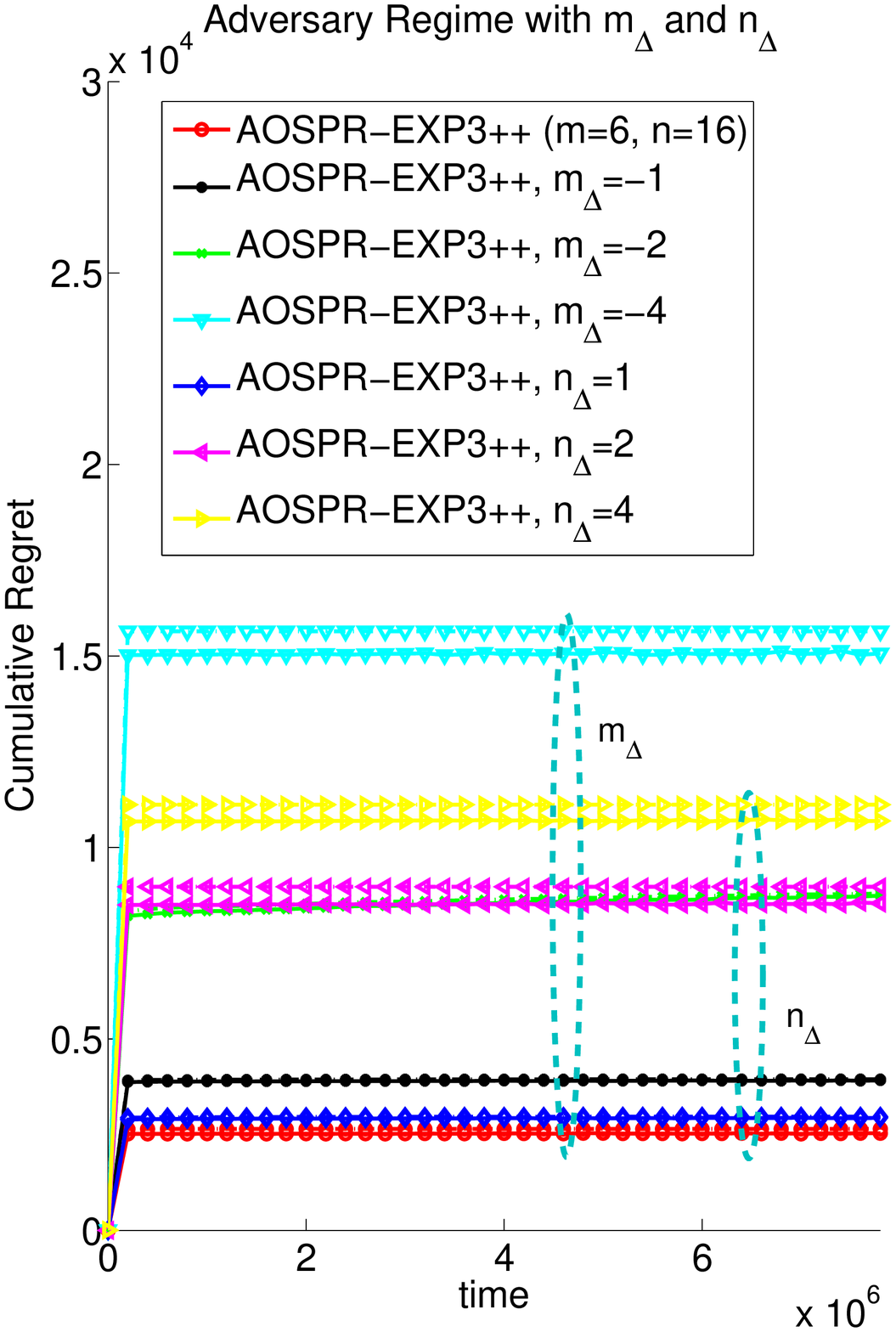}
\caption{Sensitivity of  $m$ and $n$ in Adver. and Stoc. Regimes}
\label{fig:digraph}
\end{figure}

We also compared the averaged received data packets delay  with different network
sizes as shown in Fig. 9 for the
 mixed stochastic and adversarial regime under different number of links after a relative long period of learning rounds $n=7*10^7$. We can find that with the increasing of the network size, the learning performance
of our AOSPR-EXP3++ is approaching the state-of-the-art  algorithms OSPR \cite{Info_Rout13} and  SPR-EXP3\cite{Routing07} in the
stochastic and adversarial regimes, which indicates its superior flexibility  in large scale network deployments.

\begin{table*}\centering
\caption{Computation Time Comparisons of Algorithm 1 and Algorithm 3}
\begin{tabular}{|l|l|l|l|l|l|l|l|}
\hline

\multicolumn{8}{|c|}{$(n, k)$} \\

\hline
 Alg. Ver. \emph{vs} Comp. Time (micro seconds)  
  & $(12, 4)$ & $(24, 4)$ & $(48, 6)$ & $(48, 12)$ & $(64, 6)$  & $(64, 12)$ & $(64, 24)$ \\
\hline
AOPSR-EXP3++:¡¡Algorithm1          & 46.1610 & 267.3351 & 819.7124 & 2622.1341 & 11087.0957 & 222376.0135 & 1868341.2324  \\
AOPSR-EXP3++¡¡ Algorithm3         & 14.5137 & 29.1341 & 61.6366 & 157.6732 &  258.3622 & 456.1143 &  790.5101  \\
\hline
\end{tabular}
\end{table*}

With the increasing of the network size, we find that the learning performance
of our AOSPR-EXP3++ is approaching the state-of-the-art algorithms OSPR \cite{Info_Rout13} and  SPR-EXP3\cite{Routing07} in the
stochastic and adversarial regimes, which indicates its superior flexibility  in large scale network deployments.  Comparing the
 values of average received data packets delays of our AOSPR-EXP3++ to that of the classic algorithm SPR-EXP3, we see a $65.3\%$ improvements of the EE in average
 under different set of links $n=4,8,16,32,64$ under oblivious jamming attack, and a $124.8\%$ improvements under adaptive jamming attack. Moreover,
 to reach the same value of delay for the  AOSPR-EXP3++ algorithm, the SPR-EXP3 takes a total of $n=12.8*10^7$ learning rounds. This
 indicates a $81.5\%$ improvement in the learning period of our proposed AOSPR-EXP3++ algorithm.

Moreover, we test the performance of the  computational efficiency of our algorithms,  where the computational time is compared in Table 1. From the results, we know that
the computationally  efficient version of the AOPSR-EXP3++ algorithm, i.e., Algorithm 3, takes about several hundreds of
micro-seconds on average, while the original algorithm takes about several hundreds of seconds, which is prohibitive in practical
implementations.

\begin{figure}
\centering
\includegraphics[scale=.22]{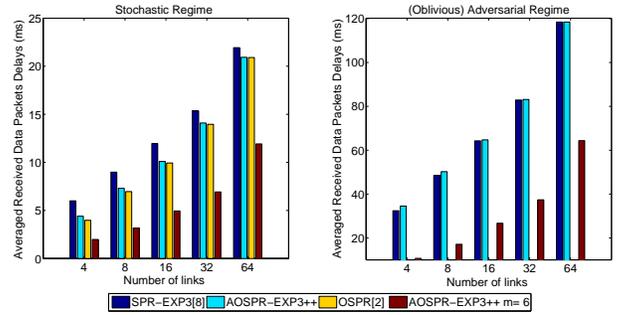}
\caption{Delay Performance with Different Network Size}
\label{fig:digraph}
\end{figure}

 \section{Proofs of Regrets in Different Regimes for the Signal-Source AOSPR}
 We prove the theorems of the performance results in Section III in the order they were presented.
  \subsection{The Adversarial Regimes}
 The proof of Theorem 1 borrows some of the analysis of EXP3 of the loss model in  \cite{Bubeck12}. However, the introduction
 of the new mixing   exploration parameter and the truth of link dependency as a special type of combinatorial MAB
 problem in the loss model  makes the proof a non-trivial task, and we prove it for the first time.

 \emph{Proof of Theorem 1.}
  \begin{IEEEproof}
Note first that the following equalities can be easily verified:
${\mathbb{E}_{\textbf{i} \sim {\rho _t}}}{\tilde{\ell}_t}(\textbf{i}) = {\ell_t}({\textbf{I}_t}),{\mathbb{E}_{{\tilde{\ell}_t} \sim {\rho _t}}}{\ell_t}(\textbf{i}) = {\ell_t}(\textbf{i}),{\mathbb{E}_{\textbf{i} \sim {\rho _t}}}{\tilde{\ell}_t}{(\textbf{i})^2} = \frac{{{\ell_t}{{({\textbf{I}_t})}^2}}}{{{\rho _t}({\textbf{I}_t})}}$ and ${\mathbb{E}_{{\textbf{I}_t} \sim {\rho _t}}}\frac{1}{{{\rho _t}({\textbf{I}_t})}} = N$.


Then, we can immediately rewrite $R(t)$ and have
\begin{IEEEeqnarray*}{l}
R(t)  =\mathbb{E}_t \left[\sum\limits_{s = 1}^t {{\mathbb{E}_{\textbf{i} \sim {p_s}}}{\tilde{\ell}_s}(\textbf{i})}  -
\sum\limits_{s = 1}^t {{\mathbb{E}_{{I_s} \sim {p_s}}}{\tilde{\ell}_s}(\textbf{i})} \right].
\end{IEEEeqnarray*}

The key step here is to consider the expectation of the cumulative losses ${\tilde{\ell}_t}(\textbf{i})$ in the sense of distribution
$\textbf{i} \sim {\rho _t}$. Let ${\varepsilon_t}(\textbf{i})=\sum\nolimits_{e \in \textbf{i}} {{\varepsilon _t}(e)}$.  However, because of the mixing terms of $\rho _t$, we need to introduce a few more notations. Let $u =
( {\underbrace {\sum\nolimits_{e \in 1}
{{\varepsilon _t}(e)},...,\sum\nolimits_{e \in \textbf{i}}
{{\varepsilon _t}(e)},...,\sum\nolimits_{e \in |\mathcal{C}|}
 {{\varepsilon _t}(e)}}_{\textbf{i} \in \mathcal{C}},\underbrace {0,...,0}_{\textbf{i}
\notin \mathcal{C}}} )$ be the distribution over all the strategies. Let ${\omega_{t-1}} = \frac{{{\rho _t} - u}}{{1 - \sum\nolimits_{e} {{\varepsilon _t}(e)} }}$ be the distribution
induced by AOSPR-EXP3++ at the time $t$ without mixing. Then we have:
\begin{IEEEeqnarray*}{l}
\!\!\begin{array}{l}
{\mathbb{E}_{\textbf{i} \sim {p_s}}}{{\tilde \ell}_s}(\textbf{i}) = ( {1 - \sum\nolimits_{e} {{\varepsilon _s}(e)} } ){\mathbb{E}_{\textbf{i} \sim {\omega_{s-1}}}}{{\tilde \ell}_s}(\textbf{i}) + {\varepsilon _s}(\textbf{i}){\mathbb{E}_{\textbf{i} \sim u}}{{\tilde \ell}_s}(\textbf{i})\\
\quad\quad \quad \quad \ \  = ( {1 - \sum\nolimits_{e} {{\varepsilon _s}(e)} } )(\frac{1}{{{\eta_s}}}\ln {\mathbb{E}_{\textbf{i} \sim {\omega_{s-1}}}}\exp ( - {\eta_s}({{\tilde \ell}_s}(\textbf{i}) \\
\quad\quad \quad \quad \quad \ - {\mathbb{E}_{\textbf{j} \sim {\omega_{s-1}}}} \tilde \ell_t(\textbf{j}))))\\
 \quad\quad \quad \quad \quad \ - \frac{( {1 - \sum\nolimits_{e} {{\varepsilon _s}(e)} } )}{{{\eta_s}}}\ln
 {\mathbb{E}_{\textbf{i} \sim {\omega_{s-1}}}}\exp ( - {\eta_s}{{\tilde \ell}_s}(\textbf{i}))) \\
\quad\quad \quad \quad \quad \  + {\mathbb{E}_{\textbf{i} \sim u}}{{\tilde \ell}_t}(\textbf{i}).
\end{array}\IEEEyesnumber \label{eq:AppenAS}
\end{IEEEeqnarray*}

Recall that for all the strategies, we have distribution
$\omega_{t-1} = (\omega_{t-1}(1),...,\omega_{t-1}(N))$ with
\begin{IEEEeqnarray*}{l}
{\omega_{t-1}}(\textbf{i}) = \frac{{\exp ( - \eta_t {{\tilde L}_{t - 1}}(\textbf{i}))}}{{\sum\nolimits_{j = 1}^N {\exp ( - \eta_t {{\tilde L}_{t - 1}}(\textbf{j}))} }},
\IEEEyesnumber \label{eq:Apart2}
\end{IEEEeqnarray*}
 and  for all the links, we have distribution  $\omega_{t-1,e} = (\omega_{t-1,e} (1),...,\omega_{t-1,e} (n))$
\begin{IEEEeqnarray*}{l}
{\omega_{t-1,e}}(e') = \frac{{\sum\nolimits_{\textbf{i}:e' \in \textbf{i}}\exp ( - \eta_t {{\tilde L}_{t
- 1}}(\textbf{i}))}}{{\sum\nolimits_{j = 1}^N {\exp ( - \eta_t {{\tilde L}_{t - 1}}(\textbf{j}))} }}.\IEEEyesnumber \label{eq:Apsart2}
\end{IEEEeqnarray*}

In the second step, we use the inequalities $lnx \le x-1$ and $exp(-x) -1 +x \le x^2/2$, for all $x \ge 0$, and the fact that take expectations over $\textbf{j}
\sim {\omega_{s-1}}$ and over $\textbf{i}
\sim {\omega_{s-1}}$ are equivalent, to obtain:
\begin{IEEEeqnarray*}{l}
\begin{array}{l}
\ln {\mathbb{E}_{\textbf{i} \sim {\omega_{s-1}}}}\exp ( - {\eta_s}({{\tilde \ell}_s}(\textbf{i}) - {\mathbb{E}_{\textbf{j}
\sim {\omega_{s-1}}}}{{\tilde \ell}_s}(\textbf{j})))\\
\quad\quad \quad = \ln {\mathbb{E}_{\textbf{i} \sim {\omega_{s-1}}}}\exp ( - {\eta_s}{{\tilde \ell}_s}(\textbf{i})) +
 {\eta_s}{\mathbb{E}_{\textbf{i} \sim {\omega_{s-1}}}}{{\tilde \ell}_s}(\textbf{i})\\
\quad\quad \quad \le {\mathbb{E}_{\textbf{i} \sim {\omega_{s-1}}}}( {\exp ( - {\eta_s}{{\tilde \ell}_s}(\textbf{i})) - 1 + {\eta_s}{{\tilde \ell}_s}(\textbf{i})} )\\
\quad\quad \quad \le {\mathbb{E}_{\textbf{i} \sim {\omega_{s-1}}}}\frac{{\eta_s^2{{\tilde \ell}_s}{{(\textbf{i})}^2}}}{2}.
\end{array}\IEEEyesnumber \label{eq:Apart1S}
\end{IEEEeqnarray*}
Take expectations over all random strategies of losses ${{\tilde \ell}_s}{(\textbf{i})^2}$, we have
\begin{IEEEeqnarray*}{l}
\begin{array}{l}
{\mathbb{E}_t}\left[{\mathbb{E}_{\textbf{i} \sim {w_s}}}{{\tilde \ell}_s}{(\textbf{i})^2} \right] =
{\mathbb{E}_t}\left[\sum\limits_{\textbf{i} = 1}^N {{\omega_{s-1}}(\textbf{i}){{\tilde \ell}_s}{{(\textbf{i})}^2}} \right]\\
= {\mathbb{E}_t}\!\!\left[\sum\limits_{\textbf{i} = 1}^N {{\omega_{s-1}}(\textbf{i}){{(\sum\limits_{e \in \textbf{i}} {{{\tilde \ell}_s}(e)} )}^2}}
  \right]
 \le  {\mathbb{E}_t}\!\!\left[ \sum\limits_{\textbf{i} = 1}^N {{\omega_{s-1}}(\textbf{i}){k}\!\!\sum\limits_{e \in \textbf{i}}
 {{{\tilde \ell}_s}{{(e)}^2}} }\right]
 \\= \!{\mathbb{E}_t}k \left[\! \sum\limits_{e = 1}^n {{{\tilde \ell}_s}{{(e)}^2}} \!\!\!\!\sum\limits_{\textbf{i}
  \in {\mathcal{P}}:e \in \textbf{i}} \!\!\!{{\omega_{s-1}}(\textbf{i})} \right]\! = \!
 {k}{\mathbb{E}_t}\!\! \left[\!\sum\limits_{e' = 1}^n \!{{{\tilde \ell}_s}{{(e')}^2}{\omega_{s-1,e}}(e')} \right]\\
 \end{array}
\end{IEEEeqnarray*}
\begin{IEEEeqnarray*}{l}
\begin{array}{l}
 = {k}{\mathbb{E}_s}\left[ {\sum\limits_{e' = 1}^n {{{\left( {\frac{{{l_t}(e')}}{{{{{\tilde \rho }_s}}(e')}}{\mathds{1}_s}(e')} \right)}^2}} {\omega _{s-1,e}}(e')} \right]\\
 \le {k}{\mathbb{E}_s}\left[ {\sum\limits_{e' = 1}^n {\frac{{{\omega _{s-1,e}}(e')}}{{{{{\tilde \rho }_s}}{{(e')}^2}}}{\mathds{1}_s}(e')} } \right]
  = {k}\sum\limits_{e' = 1}^n {\frac{{{\omega _{s-1,e}}(e')}}{{{{{\tilde \rho }_s}}(e')}}} \\
 = {k}\sum\limits_{e' = 1}^n {\frac{{{\omega _{s-1,e}}(e')}}{{\left( {1 - \sum\nolimits_{e} {{\varepsilon _s}(e)} } \right)
{\omega _{s-1,e}}(e')
 + {{\sum\nolimits_{e \in \textbf{i}} {{\varepsilon _s}(e)} }}{{}} \left|
  {\left\{ {\textbf{i} \in \mathcal{C}:e \in \textbf{i}} \right\}} \right|}}}
 \le 2k n,
\end{array}\IEEEyesnumber \label{eq:Apart1sb}
\end{IEEEeqnarray*}
where the last inequality follows the fact that $( {1 - \sum\nolimits_{e} {{\varepsilon _t}(e)} }) \ge \frac{1}{2}$ by the definition of
${{\varepsilon _t}(e)}$.

In the third step, note that ${{\tilde L}_0}(\textbf{i}) = 0$. Let ${\Phi _t}(\eta ) = \frac{1}{\eta }\ln \frac{1}{N}\sum\nolimits_{\textbf{i} = 1}^N
{\exp ( - \eta {{\tilde L}_t}(\textbf{i}))}$ and ${\Phi _0}(\eta )=0 $. The second term in (\ref{eq:AppenAS})
 can be bounded by using the same technique in \cite{Bubeck12} (page 26-28).
 Let us substitute  inequality (\ref{eq:Apart1sb}) into (\ref{eq:Apart1S}), and then substitute (\ref{eq:Apart1S}) into equation  (\ref{eq:AppenAS}) and sum over $t$ and take expectation over all random strategies of losses up to time $t$, we obtain
\begin{IEEEeqnarray*}{l}
\begin{array}{l}
\hspace{-.3cm}{\mathbb{E}_t}\left[ \sum\limits_{s = 1}^t {\mathbb{E}_{\textbf{i} \sim {p_s}}}{{\tilde \ell}_s}(\textbf{i}) \right]
 \le k n \!\sum\limits_{s = 1}^t \eta_s + \frac{{\ln N}}{\eta_t } + \!\! \sum\limits_{s = 1}^t  \!{\mathbb{E}_{\textbf{i} \sim u}}{{\tilde \ell}_s}(\textbf{i}) \\
\hspace{1.65cm} + {\mathbb{E}_t}\left[ \sum\limits_{s = 1}^{t - 1} {{\Phi _s}({\eta _{s + 1}}) - {\Phi _s}({\eta _s})}\right] +
\sum\limits_{s = 1}^t {{\mathbb{E}_{{I_s} \sim {p_s}}}{\tilde{\ell}_s}(\textbf{i})}.
\end{array}
\end{IEEEeqnarray*}

Then, we get \begin{IEEEeqnarray*}{l}
R(t)  =\mathbb{E}_t \sum\limits_{s = 1}^t {{\mathbb{E}_{\textbf{i} \sim {p_s}}}{\tilde{\ell}_s}(\textbf{i})}  - \mathbb{E}_t
\sum\limits_{s = 1}^t {{\mathbb{E}_{{I_s} \sim {p_s}}}{\tilde{\ell}_s}(\textbf{i})}\\
\hspace{.7cm}\le k n \!\sum\limits_{s = 1}^t \eta_s + \frac{{\ln N}}{\eta_t }
+  \sum\limits_{s = 1}^t  \!{\mathbb{E}_{\textbf{i} \sim u}}{{\tilde \ell}_s}(\textbf{i})\\
\hspace{.7cm}\mathop  \le \limits^{(a)} k n \!\sum\limits_{s = 1}^t \eta_s + \frac{{\ln N}}{\eta_t }
+ k \sum\limits_{s = 1}^t {\sum\limits_{e = 1}^n {{\varepsilon _s}(e)} } \\
\end{IEEEeqnarray*}
\begin{IEEEeqnarray*}{l}
\hspace{.7cm}\mathop  \le \limits^{(b)} 2k n \!\sum\limits_{s = 1}^t \eta_s + \frac{{\ln N}}{\eta_t } \\
\hspace{.7cm}\mathop  \le \limits^{(c)} 2k n \!\sum\limits_{s = 1}^t \eta_s + k\frac{{\ln n}}{\eta_t }.
\end{IEEEeqnarray*}
Note that, the inequality $(a)$ holds by setting ${{\tilde \ell}_s}(\textbf{i})=k, \forall \textbf{i}, s$, and the
upper bound is $k \sum\nolimits_{\textbf{i} \in C} {{\sum\nolimits_{e \in \textbf{i}} {{\varepsilon _t}(e)} }}=
k \sum\nolimits_{s = 1}^t {\sum\nolimits_{e = 1}^n {{\varepsilon _s}(e)} }$. The inequality $(b)$ holds, because of,  for every time slot $t$,
 $\eta_t \ge {\varepsilon _t}(e)$. The
 inequality $(c)$ is due to the fact that $N \le n^{k}$. Setting $\eta_t=\beta_t$, we prove the theorem.
  \end{IEEEproof}

 \emph{Proof of Theorem 2.}
  \begin{IEEEproof}
To defend against the $\theta$-memory-bounded adaptive adversary, we need to adopt the idea of the mini-batch protocol proposed in \cite{Arora12}.
We define a new algorithm by wrapping AOSPR-EXP3++ with a mini-batching loop \cite{Dekel11}. We specify a batch
size $\tau$ and name the new algorithm AOSPR-EXP3++$_\tau$. The idea is to group the overall
time slots $1,...,t$ into consecutive and disjoint mini-batches of size $\tau$. It can be viewed that one signal mini-batch
as a round (time slot) and use the average loss suffered during that mini-batch to feed the original AOSPR-EXP3++. Note that our
new algorithm does not need to know $m$, which only appears as a constant as shown in Theorem 2. So our new
AOSPR-EXP3++$_\tau$ algorithm still runs in an adaptive way without any prior about the
environment. If we set the batch $\tau= {(4{k}\sqrt {n\ln n} )^{ - \frac{1}{3}}}{t^{^{\frac{1}{3}}}}$ in Theorem 2 of \cite{Arora12},
we can get the regret upper bound in our Theorem 2.
  \end{IEEEproof}

\subsection{The Stochastic Regime}
 Our proofs are based on the following form of Bernstein's inequality with minor improvement as shown in \cite{Seldin14}.

\textbf{Lemma 1.} (Bernstein's inequality for martingales). Let $X_1,...,X_m$ be martingale difference sequence with
respect to filtration $\mathcal{F}=(\mathcal{F}_\textbf{i})_{1 \le k \le m}$ and let $Y_k = \sum\nolimits_{j = 1}^k {{X_j}}$ be the
associated martingale. Assume that there exist positive numbers $\nu$ and $c$, such that $X_j \le c$ for all $j$ with probability
$1$ and $\sum\nolimits_{k = 1}^m {\mathbb{E}\left[ {{{\left( {{X_k}} \right)}^2}|{\mathcal{F}_{k - 1}}} \right]}  \le \nu$ with probability 1.
\begin{IEEEeqnarray*}{l}
\mathbb{P}[{Y_m} > \sqrt {2\nu b}  + \frac{{cb}}{3}] \le {e^{ - b}}.
\end{IEEEeqnarray*}

We also need to use the following technical lemma, where the proof can be found in \cite{Seldin14}.

\textbf{Lemma 2. } For any $c > 0$, we have $\sum\nolimits_{t = 0}^\infty  {{e^{ - c\sqrt t }}}  = O\left( {\frac{2}{{{c^2}}}} \right)$.

To obtain the tight regret performance for AOSPR-EXP3++, we need to study and estimate the
number of times each of link is selected up to time $t$, {i}.e., $N_t(e)$. We summarize it in the following lemma.

\textbf{Lemma 3. } Let $\left\{ {{{\underline \varepsilon }_t}(e)} \right\}_{t = 1}^\infty$ be non-increasing deterministic sequences, such that
${{\underline \varepsilon }_t}(e) \le {{ \varepsilon }_t}(e)$ with probability $1$ and ${{\underline \varepsilon }_t}(e) \le {{ \varepsilon }_t}(e^*)$
 for all $t$ and $e$. Define $\nu_t(e)= \sum\nolimits_{s = {1}}^t  \frac{1}{{{k \underline \varepsilon  }_s}(e)} $, and define the event $\mathcal{E}^e_t$
\begin{IEEEeqnarray*}{l}
 {t\Delta (e) - ( {{{\tilde L}_t}(e) - {{\tilde L}_t}({e^*})} )} \\
\hspace{1.3cm} \le {\sqrt {2({\nu _t}(e) + {\nu _t}({e^*})){b_t}}  + \frac{{{(1/k+ 0.25)} {b_t}}}{{3k  {\underline \varepsilon _t}({e^*})}}}\hspace{.3cm}
(\mathcal{E}^e_t).
\end{IEEEeqnarray*}
Then for any positive sequence $b_1, b_2,...,$ and any $t^* \ge 2$ the number of times link $e$ is played by AOSPR-EXP3++ up to
round $t$ is bounded as:
\begin{IEEEeqnarray*}{l}
\begin{array}{l}
\mathbb{E}[{N_t}(e)] \le \left( {{t^*} - 1} \right) + \sum\limits_{s = {t^*}}^t {{e^{ - {b_s}}}}  + k \sum\limits_{s = {t^*}}^t {{\varepsilon _s}(e){\mathds{1}_{\{ \mathcal{E}^e_t\} }}} \\
    \hspace{1.6cm}               + \sum\limits_{s = {t^*}}^t {{e^{ - {\eta _s}{h_{s - 1}}(e)}}},
\end{array} \IEEEyesnumber \label{eq:AHigh}
\end{IEEEeqnarray*}
where
\begin{IEEEeqnarray*}{l}
\begin{array}{l}
{h_t}(e) = t\Delta (e) - \sqrt {2t{b_t}\left( {\frac{1}{{{k}{{\underline \varepsilon  }_t}(e)}} + \frac{1}{{{k}{{\underline \varepsilon  }_t}({e^*})}}} \right)}  - \frac{{(\frac{1}{4} + \frac{1}{k}){b_t}}}{{3{{\underline \varepsilon  }_t}({e^*})}}.
\end{array}
\end{IEEEeqnarray*}

\begin{IEEEproof}
Note that the elements of the martingale difference sequence $\{ {\Delta (e)
- ({{\tilde \ell}_t}(e) - {{\tilde \ell}_t}({e^*}))} \}_{t = 1}^\infty$ by $\max \{ \Delta (e) +{{\tilde \ell}_t}({e^*}) \}= {\frac{1}{{{k}{{\underline \varepsilon  }_t}({e^*})}}} +1$. Since ${{{\underline \varepsilon  }_t}({e^*})} \le {{{ \varepsilon  }_t}({e^*})} \le 1/(2n) \le 1/4$, we can simplify
the upper bound by using ${\frac{1}{{{{k\underline \varepsilon   }_t}({e^*})}}} +1 \le
    \frac{{(\frac{1}{4} + \frac{1}{k})}}{{{{\underline \varepsilon   }_t}({e^*})}}$.

  We further note that
   \begin{displaymath}
\begin{array}{l}
\sum\limits_{s = 1}^t {{\mathbb{E}_s}\left[{{(\Delta (e) - ({{\tilde \ell}_s}(e) - {{\tilde \ell}_s}({e^*})))}^2}\right]} \\
 \hspace{1.1cm}  \le \sum\limits_{s = 1}^t {{\mathbb{E}_s}\left[{{({{\tilde \ell}_s}(e) - {{\tilde \ell}_s}({e^*}))}^2}\right]} \\
 \hspace{1.1cm}  = \sum\limits_{s = 1}^t {\left( {{\mathbb{E}_s}\left[({{\tilde \ell}_s}{{(e)}^2}\right] + {E_s}\left[({{\tilde \ell}_s}
 {{({e^*})}^2}\right]} \right)} \\
 \hspace{1.1cm}  \le \sum\limits_{s = 1}^t \left( {\frac{1}{{{q_s}(e)}} + \frac{1}{{{q_s}({e^*})}}} \right) \\
 \hspace{1.1cm} \mathop \le \limits^{(a)} \sum\limits_{s = 1}^t {\left( {\frac{1}{{k{\varepsilon  _s}(e)}} +
  \frac{1}{{k{\varepsilon  _s}({e^*})}}} \right)} \\
 \hspace{1.1cm}  \le \sum\limits_{s = 1}^t {\left( {\frac{1}{{k{{\underline \varepsilon   }_s}(e)}} +
 \frac{1}{{k{{\underline \varepsilon   }_s}({e^*})}}} \right)}  = {\nu _t}(e) + {\nu _t}({e^*})
\end{array}
\end{displaymath}
with probability $1$.  The above inequality (a) is due to the fact that ${{\tilde \rho }_t}(e) \ge \sum\nolimits_{e \in \textbf{i}} {{\varepsilon _t}(e)} \left| {\left\{ {\textbf{i} \in \mathcal{C}:e \in \textbf{i}} \right\}} \right|$. Since each $e$ only belongs to one of the covering strategies $\textbf{i} \in \mathcal{C}$, $\left| {\left\{ {\textbf{i} \in \mathcal{C}:e \in \textbf{i}} \right\}} \right|$ equals to 1 at time slot $t$ if link $e$ is selected. Thus, ${{\tilde \rho }_t}(e) \ge \sum\nolimits_{e \in \textbf{i}} {{\varepsilon _t}(e)}= k{\varepsilon_t}(e)$.

Let $\mathcal{\bar E}_t^e$ denote the complementary of event
 $\mathcal{E}_t^e$. Then by the Bernstein's inequality $\mathbb{P}[\mathcal{\bar E}_t^e] \le e^{-b_t}$. The number of
 times the link $e$ is selected up to round $t$ is bounded as:
\begin{IEEEeqnarray*}{l}
 \begin{array}{l}
 \mathbb{E}[{N_t}(e)] = \sum\limits_{s = 1}^t {\mathbb{P}[A_s= e]} \\
 \hspace{4em} = \sum\limits_{s = 1}^t {\mathbb{P}[A_s= e|\mathcal{E}_{s - 1}^e]P[\mathcal{E}_{s - 1}^e]} \\
  \hspace{5em}+ \mathbb{P}[A_s= e|\overline {\mathcal{E}_{s - 1}^e} ]P[\overline {\mathcal{E}_{s - 1}^e} ]\\
 \hspace{4em} \le \sum\limits_{s = 1}^t {\mathbb{P}[A_s= e|\mathcal{E}_{s - 1}^e]} {\mathds{1}_{\{ \mathcal{E}_{s - 1}^e\} }} +
  \mathbb{P}[\overline {\mathcal{E}_{s - 1}^S} ]\\
  \hspace{4em} \le \sum\limits_{s = 1}^t {\mathbb{P}[A_s= e|\mathcal{E}_{s - 1}^e]} {\mathds{1}_{\{ \mathcal{E}_{s - 1}^e\} }} + {e^{ - {b_{s - 1}}}}.
\end{array}
\end{IEEEeqnarray*}

We further upper bound $ {\mathbb{P}[A_s= e|\mathcal{E}_{s - 1}^e]} {\mathds{1}_{\{ \mathcal{E}_{s - 1}^e\} }} $ as follows:
    \begin{displaymath}
\begin{array}{l}
\mathbb{P} {[A_s= e|{\cal \mathcal{E}}_{s - 1}^e]} {\mathds{1}_{\{ {\cal \mathcal{E}}_{s - 1}^e\} }}
= {{{\tilde \rho }_s}}(e){\mathds{1}_{\{ {\cal \mathcal{E}}_{s - 1}^e\} }}\\
  \hspace{4em} \le ({\omega_{s-1}}(e) + k{\varepsilon_s}(e)){\mathds{1}_{\{ {\cal \mathcal{E}}_{s - 1}^e\} }}\\
 \hspace{4em} =({k\varepsilon  _s}(e) + \frac{{\sum\nolimits_{\textbf{i}:e \in \textbf{i}} {{w_{s - 1}}\left( \textbf{i} \right)} }}{{{W_{s - 1}}}}){\mathds{1}_{\{ {\cal \mathcal{E}}_{s - 1}^e\} }}\\
    \hspace{4em} =({k\varepsilon  _s}(e) + \frac{{\sum\nolimits_{\textbf{i}:e \in \textbf{i}}^{} {{e^{ - {\eta _s}{ \tilde L_{s - 1}}(\textbf{i})}}} }}{{\sum\nolimits_{\textbf{i} = 1}^N {{e^{ - {\eta _s}{ \tilde L_{s - 1}}(\textbf{i})}}} }}){\mathds{1}_{\{ {\cal \mathcal{E}}_{s - 1}^e\} }}\\
  \hspace{4em}\mathop \le \limits^{(a)} ({k\varepsilon  _s}(e) + {e^{ - {\eta _s}\left( {{{\tilde L}_{s - 1}}(\textbf{i}) - {{\tilde L}_{s - 1}}({\textbf{i}^*})} \right)}}){\mathds{1}_{\{ {\cal \mathcal{E}}_{s - 1}^e\} }}\\
    \hspace{4em}\mathop \le \limits^{(b)} ({k\varepsilon  _s}(e) + {e^{ - {\eta _t}\left( {{{\tilde L}_{s - 1}}(e) - {{\tilde L}_{s - 1}}({e^*})} \right)}}){\mathds{1}_{\{ {\cal \mathcal{E}}_{s - 1}^e\} }}\\
   \hspace{4em}\mathop \le \limits^{(c)}  k{\varepsilon  _s}(e){\mathds{1}_{\{ {\cal \mathcal{E}}_{s - 1}^e\} }} + {e^{ - {\eta _s}{h_{s - 1}}(e)}}.
\end{array}
  \end{displaymath}
The above inequality (a) is due to the fact that link
$e$ only belongs to one chosen path $\textbf{i}$ in $t-1$, inequality (b) is because the  cumulative
regret of each path is great than the cumulative regret of each link that belongs to the
  path, and the last inequality (c) we used the fact that $\frac{t}{{{{\underline \varepsilon   }_t}(e)}}$ is a non-increasing
 sequence ${\upsilon _t}(e) \le \frac{t}{{{{k \underline \varepsilon   }_t}(e)}}$. Substitution of this result back into
 the computation of $ \mathbb{E}[{N_t}(e)]$ completes the proof. \end{IEEEproof}

\emph{Proof of Theorem 3.}
\begin{IEEEproof}
The proof is based on Lemma 3. Let $b_t =ln(t \Delta(e)^2)$ and ${{{\underline \varepsilon  }_t}(e)}
={{{ \varepsilon  }_t}(e)}$. For any $c \ge 18$ and any $t \ge t^*$, where $t^*$ is the minimal integer for which
 ${t^*} \ge \frac{{4{c^2}n \ln {{({t^*}\Delta {{(e)}^2})}^2}}}{{\Delta {{(e)}^4}\ln (n)}}$, we have
     \begin{displaymath}
\begin{array}{l}
{h_t}(e) = t\Delta (e) - \sqrt {2t{b_t}\left( {\frac{1}{{k{\varepsilon _t}(e)}} + \frac{1}{{k{\varepsilon _t}({e^*})}}} \right)}  - \frac{{\left( {\frac{1}{4} + \frac{1}{k}} \right){b_t}}}{{3{\varepsilon _t}({e^*})}}\\
   \hspace{2.43em} \ge t\Delta (e) - 2\sqrt {\frac{{t{b_t}}}{{k{\varepsilon _t}(e)}}}  - \frac{{\left( {\frac{1}{4} + \frac{1}{k}} \right){b_t}}}{{3{\varepsilon _t}(e)}}\\
  \hspace{2.43em} = t\Delta (e)(1 - \frac{2}{{\sqrt {k c} }} - \frac{{\left( {\frac{1}{4} + \frac{1}{k}} \right)}}{{3c}})\\
  \hspace{2.43em} \mathop  \ge \limits^{(a)} t\Delta (e)(1 - \frac{2}{{\sqrt c }} - \frac{{1.25}}{{3c}}) \ge \frac{1}{2}t\Delta (e).
\end{array}
  \end{displaymath}
The above inequality (a) is due to the fact that $(1 - \frac{2}{{\sqrt {k c} }} - \frac{{\left( { \frac{1}{4}+\frac{1}{k}} \right)}}{{3c}})$ is
an increasing function with respect to $k (k \ge 1)$. Plus, as indicated in work \cite{Branislav2015}, by a bit more sophisticated  bounding $c$ can be made almost as small as 2 in our case. By substitution of the lower bound on $h_t(e)$ into Lemma 3, we have
     \begin{displaymath}
\begin{array}{l}
\!\!\!\mathbb{E}[{N_t}(e)] \le {t^*} + \frac{{\ln (t)}}{{\Delta {{(e)}^2}}} + k   \frac{{c\ln {{(t)}^2}}}{{\Delta {{(e)}^2}}}  +  \sum\limits_{s = 1}^t \!\!\left(\!{{e^{ - \frac{{\Delta (e)}}{4}\sqrt {\frac{{(s - 1)ln(n)}}{n}} }}}\!\right)\\
 \hspace{3.5em} \le k\frac{{c\ln {{(t)}^2}}}{{\Delta {{(e)}^2}}} + \frac{{\ln (t)}}{{\Delta {{(e)}^2}}} + O(\frac{{{n
 }}}{{\Delta {{(e)}^2}}}) + {t^*},
\end{array}
  \end{displaymath}
  where lemma 3 is used to bound the sum of the exponents. In addition, please
  note that $t^*$ is of the order $O(\frac{{k n}}{{\Delta {{(e)}^4}\ln (n)}})$.
\end{IEEEproof}

\emph{Proof of Theorem 4.}
\begin{proof} The proof is based on the similar idea of Theorem 2 and Lemma 3. Note that
by our definition ${{\hat \Delta }_t}(e) \le 1$ and the sequence ${\underline \varepsilon _t}(e) = {\underline \varepsilon  _t} =
\min \{ \frac{1}{{2n}},{\beta _t},\frac{{c\ln {{(t)}^2}}}{t}\} $ satisfies the condition of Lemma 3. Note that when ${\beta _t} \ge
\frac{{c\ln {{(t)}^2}}}{t}\}$, {i}.e., for $t$ large enough such that
$
t \ge \frac{{4{c^2}\ln {{(t)}^4}n }}{{\ln (n)}}
$, we have ${\underline \varepsilon  _t}=\frac{{c\ln {{(t)}^2}}}{t}$. Let $b_t=ln(t)$ and let $t^*$ be
large enough, so that for all $t \ge t^*$ we have $t \ge \frac{{4{c^2}\ln {{(t)}^4}n }}{{\ln (n)}}$ and
$t \ge e^{\frac{1}{\Delta(e)^2}}$. With these parameters and conditions on hand, we are going to bound the
rest of the three terms in the bound on $\mathbb{E}[N_t(e)]$ in Lemma 3. The upper bound of
 $\sum\nolimits_{s = {t^*}}^t {{e^{ - {b_s}}}} $ is easy to obtain. For bounding
 $k\sum\nolimits_{s = {t^*}}^t {{\varepsilon _s}(e){\mathds{1}_{\{ \mathcal{E}_{s - 1}^e\} }}}$,  we note that $\mathcal{E}_{t}^e$ holds and
 for $c\ge 18$ we have
      \begin{displaymath}
\begin{array}{l}
{{\hat \Delta }_t}(e) \ge \frac{1}{t}({{\tilde L}_t}(e)-
\mathop {\max }\limits_{e'} ({{\tilde L}_t}(e'))  ) \ge \frac{1}{t}( {{\tilde L}_t}(e)-{{\tilde L}_t}({e^*}))\\
 \hspace{2.43em} \ge \frac{1}{t}{h_t}(e) = \frac{1}{t}\left( {t\Delta (e) - 2\sqrt {\frac{{t{b_t}}}{{k{{\underline \varepsilon  }_t}}}}  - \frac{{(\frac{1}{4} + \frac{1}{k}){b_t}}}{{3{{\underline \varepsilon  }_t}}}} \right)\\
  \end{array}
\end{displaymath}
   \begin{displaymath}
\begin{array}{l}
 \hspace{2.43em} = \frac{1}{t}\left( {t\Delta (e) - \frac{{2t}}{{\sqrt {c k \ln (t)} }} - \frac{{(\frac{1}{4} + \frac{1}{k})t}}{{3c\ln (t)}}} \right)\\
 \hspace{2.43em} \mathop  \ge \limits^{(a)} \frac{1}{t}\left( {t\Delta (e) - \frac{{2t}}{{\sqrt {c\ln (t)} }} - \frac{{1.25t}}{{3c\ln (t)}}} \right)\\
\hspace{2.43em} \mathop  \ge \limits^{(b)} \Delta (e)\left( {1 - \frac{2}{{\sqrt c }} - \frac{{1.25}}{{3c}}} \right) \ge \frac{1}{2}\Delta (e),
\end{array}
  \end{displaymath}
  where the inequality (a) is due to the fact that $\frac{1}{t}( t\Delta (e) - \frac{2t}{\sqrt {c k \ln (t)} } -
  \frac{(\frac{1}{4} + \frac{1}{k})t}{3c\ln (t)} )$ is
an increasing function with respect to $k (k \ge 1)$ and the inequality (b) due to the fact that for $t \ge t^*$ we have $\sqrt {ln(t)}
\ge 1/\Delta(e).$ Thus,
\begin{displaymath}
{\varepsilon _n}(e){\mathds{1}_{\{ \mathcal{E}_{n - 1}^e\} }}
\le \frac{{c{{\left( {\ln t} \right)}^2}}}{{t{{\hat \Delta }_t}{{(e)}^2}}} \le
 \frac{{4{c^2}{{\left( {\ln t} \right)}^2}}}{{t\Delta {{(e)}^2}}}
 \end{displaymath}
 and $k\sum\nolimits_{s = {t^*}}^t {{\varepsilon _s}(e){\mathds{1}_{\{ \mathcal{E}_{n - 1}^e\} }}} = O\left(
 {\frac{{k\ln {{\left( t \right)}^3}}}{{\Delta {{(e)}^2}}}} \right)$. Finally, for the last term in Lemma 3, we have
 already get $h_t(e) \ge \frac{1}{2}\Delta(e)$ for $t \ge t^*$ as an intermediate step in the calculation of bound
 on ${{{\hat \Delta }_t}(e)}$. Therefore, the last term
 is bounded in a order of $O(\frac{{{n
 }}}{{\Delta {{(e)}^2}}})$. Use all these results together we obtain the results of the theorem. Note that the
 results holds for any $\eta_t \ge \beta_t$.
 \end{proof}

\subsection{Mixed Adversarial and Stochastic Regime}
\emph{Proof of Theorem 5.}
\begin{proof}
The proof of the regret performance in the mixed adversarial and stochastic regime is simply a combination of the performance of
the AOSPR-EXP3++$^\emph{AVG}$ algorithm in adversarial and stochastic regimes. It is very straightforward from Theorem 1 and Theorem
3.
 \end{proof}
\emph{Proof of Theorem 6.}
\begin{proof}
Similar as above, the proof is very straightforward from Theorem 2 and Theorem 3.
 \end{proof}

\subsection{Contaminated Stochastic Regime}
\emph{Proof of Theorem 7.}
\begin{proof}
The key idea of proving the regret bound under  moderately contaminated stochastic  regime
 relies on how to estimate the performance loss by taking into account the contaminated pairs. Let $\mathds{1}
 _{t,e}^\star$ denote the indicator functions of the occurrence of contamination at location $(t,e)$, {i}.e.,
 $\mathds{1} _{t,e}^\star$  takes value $1$ if contamination occurs and $0$ otherwise.
 Let $m_t(e)= \mathds{1}  _{t,e}^\star \tilde \ell_t(e) + (1-\mathds{1}  _{t,e}^\star)\mu(e)$.  If either base arm $e$
 was contaminated on round $t$ then $m_t(e)$ is adversarially assigned a value of loss that is
 arbitrarily affected by some adversary, otherwise we use the expected loss. Let  ${Z_t}(e) = \sum\nolimits_{s = 1}^t {{m_t}(e)}$
 then $\left( {{Z_t}({e}) - {Z_t}(e^*)} \right) - \left( {{{\tilde L}_t}({e}) - {{\tilde L}_t}(e^*)} \right)$ is a martingale.
 After
 $\tau$ steps, for $t \ge \tau$,
  \begin{displaymath}
\begin{array}{l}
\left( {{Z_t}({e}) - {Z_t}(e^*)} \right) \ge t\min \{ \mathds{1}  _{t,e}^\star ,\mathds{1}  _{t,e^*}^\star \} ({\tilde \ell_t}(e) - {\tilde \ell_t}({e^*}))\\
\hspace{6em} + t\min \{ 1 - \mathds{1}  _{t,e}^\star ,1 - \mathds{1}  _{t,e^*}^\star \} (\mu ({e}) - \mu (e^*))\\
 \hspace{3.6em} \ge  - \zeta t\Delta (e) + (t - \zeta t\Delta (e))\Delta (e) \ge (1-2\zeta){t\Delta (e)}.
\end{array}
  \end{displaymath}

Define the event $\mathcal{Z}_t^e$:
  \begin{displaymath}
(1-2\zeta)t\Delta (e) - \left( {{{\tilde L}_t}({e}) - {{\tilde L}_t}(e^*)} \right)
\le 2\sqrt {{\nu _t}{b_t}}  + \frac{{\left( {\frac{1}{4} + \frac{1}{k}} \right){b_t}}}{{3{{\underline \varepsilon  }_t}}},
  \end{displaymath}
where ${\underline \varepsilon  }_t$ is defined in the proof of Theorem 3 and $\nu _t = \sum\nolimits_{s = 1}^t
 {\frac{1}{{{{k\underline \varepsilon  }_t}}}}$. Then by Bernstein's inequality
  $\mathbb{P}[\mathcal{Z}_t^e] \le e^{-b_t}$. The remanning proof is identical to the proof of Theorem 3.

  For the regret performance in the moderately contaminated stochastic regime, according to our definition with the attacking strength
  $\zeta \in [0,1/4]$, we only need to replace  $\Delta(e)$  by $\Delta(e)/2$ in Theorem 5.
\end{proof}

\section{Proof of Regret for Accelerated AOSPR Algorithm}
  We prove the theorems of the performance results in Section IV in the order they were presented.
\subsection{Accelerated Learning in Adversarial Regime}
The proof the Theorem 8 requires the following Lemma from Lemma 7 \cite{Seldin14lim}. We restate it for completeness.

\textbf{Lemma 4.}  For any probability distribution $\tilde\omega$ on $\{1,...,n\}$   and any $m \in [1,n]$:
\begin{IEEEeqnarray*}{l}
\sum\limits_{e = 1}^n {\frac{{\tilde \omega (e)(n - 1)}}{{\tilde \omega (e)(n - m) + m - 1}} \le \frac{n}{m}} .
\end{IEEEeqnarray*}

 \emph{Proof of Theorem 8.}
  \begin{IEEEproof}
Note first that the following equalities can be easily verified:
${\mathbb{E}_{\textbf{i} \sim {\varrho _t}}}{\tilde{\ell}_t}(\textbf{i})
= {\ell_t}({\textbf{I}_t}),{\mathbb{E}_{{\tilde{\ell}_t} \sim {\varrho _t}}}{\ell_t}(\textbf{i}) = {\ell_t}(\textbf{i}),{\mathbb{E}_{\textbf{i} \sim {\varrho _t}}}{\tilde{\ell}_t}{(\textbf{i})^2} = \frac{{{\ell_t}{{({\textbf{I}_t})}^2}}}{{{\varrho _t}({\textbf{I}_t})}}$ and ${\mathbb{E}_{{\textbf{I}_t} \sim {\varrho _t}}}\frac{1}{{{\varrho _t}({\textbf{I}_t})}} = N$.



Then, we can immediately rewrite $R(t)$ and have
\begin{IEEEeqnarray*}{l}
R(t)  =\mathbb{E}_t \left[\sum\limits_{s = 1}^t {{\mathbb{E}_{\textbf{i} \sim {\rho_s}}}{\tilde{\ell}_s}(\textbf{i})}
- \sum\limits_{s = 1}^t {{\mathbb{E}_{{\textbf{I}_s} \sim {\rho_s}}}{\tilde{\ell}_s}(\textbf{i})} \right].
\end{IEEEeqnarray*}

The key step here is to consider the expectation of the cumulative losses ${\tilde{\ell}_t}(\textbf{i})$ in the sense of distribution
$\textbf{i} \sim {\varrho _t}$. Let ${\varepsilon_t}(\textbf{i})=\sum\nolimits_{e \in \textbf{i}} {{\varepsilon _t}(e)}$.  However,
because of the mixing terms of $\varrho _t$, we need to introduce a few more notations. Let $\varphi_s =
( {\underbrace {\sum\nolimits_{e \in 1}
{{\varepsilon _t}(e)},...,\sum\nolimits_{e \in \textbf{i}}
{{\varepsilon _t}(e)},...,\sum\nolimits_{e \in |\mathcal{C}|}
 {{\varepsilon _t}(e)}}_{\textbf{i} \in \mathcal{C}},\underbrace {0,...,0}_{\textbf{i}
\notin \mathcal{C}}} )$ be the distribution over all the strategies. Let ${\omega_{t-1}} = \frac{{{\rho _t} - u}}{{1 - \sum\nolimits_{e} {{\varepsilon _t}(e)} }}$ be the distribution
induced by AOSPR-EXP3++ at the time $t$ without mixing. Then we have:
\begin{IEEEeqnarray*}{l}
\!\!\begin{array}{l}
\!\!{\mathbb{E}_{\textbf{i} \sim {\rho_s}}}{{\tilde \ell}_s}(\textbf{i}) = ( {1 - \sum\nolimits_{e} {{\varepsilon _s}(e)} } )
{\mathbb{E}_{\textbf{i} \sim {\omega_{s-1}}}}{{\tilde \ell}_s}(\textbf{i}) +
\varepsilon _s(\textbf{i}){\mathbb{E}_{\textbf{i}\sim u}{\tilde \ell}_s(\textbf{i})}  \\
\ \quad \quad \quad \ \  = ( {1 - \sum\nolimits_{e} {{\varepsilon _s}(e)} } )(\frac{1}{{{\eta_s}}}\ln {\mathbb{E}_{\textbf{i} \sim {\omega_{s-1}}}}\exp ( - {\eta_s}({{\tilde \ell}_s}(\textbf{i}) \\
\quad\quad \quad \quad \quad \ - {\mathbb{E}_{\textbf{j} \sim {\omega_{s-1}}}} \tilde \ell_t(\textbf{j}))))\\
 \quad\quad \quad \quad \quad \ - \frac{( {1 - \sum\nolimits_{e} {{\varepsilon _s}(e)} } )}{{{\eta_s}}}\ln {\mathbb{E}_{\textbf{i} \sim {\omega_{s-1}}}}\exp ( - {\eta_s}{{\tilde \ell}_s}(\textbf{i}))) \\
\quad\quad \quad \quad \quad \  + \varepsilon _s(\textbf{i}){\mathbb{E}_{\textbf{i}\sim u}{\tilde \ell}_s(\textbf{i})}.
\end{array}\IEEEyesnumber \label{eq:AppenA}
\end{IEEEeqnarray*}

In the second step, by similar arguments as in the proof of Theorem 1, we have:
\begin{IEEEeqnarray*}{l}
\begin{array}{l}
\ln {\mathbb{E}_{\textbf{i} \sim {\omega_{s-1}}}}\exp ( - {\eta_s}({{\tilde \ell}_s}(\textbf{i}) - {\mathbb{E}_{\textbf{j} \sim {\omega_{s-1}}}}
{{\tilde \ell}_s}(\textbf{j})))\\
\quad\quad \quad = \ln {\mathbb{E}_{\textbf{i} \sim {\omega_{s-1}}}}\exp ( - {\eta_s}{{\tilde \ell}_s}(\textbf{i})) +
{\eta_s}{\mathbb{E}_{\textbf{j} \sim {\omega_{s-1}}}}{{\tilde \ell}_s}(\textbf{j})\\
\quad\quad \quad \le {\mathbb{E}_{\textbf{i} \sim {\omega_{s-1}}}}( {\exp ( - {\eta_s}{{\tilde \ell}_s}(\textbf{i})) - 1 + {\eta_s}{{\tilde \ell}_s}(\textbf{i})} )\\
\quad\quad \quad \le {\mathbb{E}_{\textbf{i} \sim {\omega_{s-1}}}}\frac{{\eta_s^2{{\tilde \ell}_s}{{(\textbf{i})}^2}}}{2}.
\end{array}\IEEEyesnumber \label{eq:Apart1}
\end{IEEEeqnarray*}
 Take expectations over all random strategies of losses ${{\tilde \ell}_s}{(\textbf{i})^2}$, we have
\begin{IEEEeqnarray*}{l}
\begin{array}{l}
 \!\!\!{\mathbb{E}_t}\left[{\mathbb{E}_{\textbf{i} \sim {\omega_{s-1}}}}{{\tilde \ell}_s}{(\textbf{i})^2} \right] =
{\mathbb{E}_t}\left[\sum\limits_{\textbf{i} = 1}^N {{\omega_{s-1}}(\textbf{i}){{\tilde \ell}_s}{{(\textbf{i})}^2}} \right]\\
 \!\!\!\!= {\mathbb{E}_t}\!\!\left[\sum\limits_{\textbf{i} = 1}^N {{\omega_{s-1}}(\textbf{i}){{(\sum\limits_{e \in \textbf{i}}
  {{{\tilde \ell}_s}(e)} )}^2}}
  \right]
  \le {\mathbb{E}_t}\!\!\left[ \sum\limits_{\textbf{i} = 1}^N {{\omega_{s-1}}(\textbf{i}){k}\!\!\sum\limits_{e
   \in \textbf{i}} {{{\tilde \ell}_s}{{(e)}^2}} }\right]
 \\
  \!\!\!\!= \!{\mathbb{E}_t}k \left[\! \sum\limits_{e = 1}^n {{{\tilde \ell}_s}{{(e)}^2}} \!\!\!\!\sum\limits_{\textbf{i} \in {\mathcal{P}}:e \in \textbf{i}} \!\!\!{{\omega_{s-1}}(\textbf{i})} \right]\! = \!
 {k}{\mathbb{E}_t}\!\! \left[\!\sum\limits_{e' = 1}^n \!{{{\tilde \ell}_s}{{(e')}^2}{\omega_{s-1,e}}(e')} \right]\\
 \!\!\!\! = {k}{\mathbb{E}_t}\left[ {\sum\limits_{e = 1}^n {{{\left( {\frac{{{l_s}(e)}}{{{{{\tilde \varrho }_s}}(e)}}
 {\mathds{1}_s}(e)} \right)}^2}} {\tilde \omega _{s-1}}(e)} \right]\\
 \!\!\!\! \le {k}{\mathbb{E}_t}\left[ {\sum\limits_{e = 1}^n {\frac{{{\tilde \omega _{s-1}}(e)}}{{{{{\tilde \varrho }_s}}{{(e)}^2}}}{\mathds{1}_s}(e
 )} } \right] = {k}\sum\limits_{e = 1}^n {\frac{{{\tilde \omega _{s-1}}(e)}}{{{{{\tilde \varrho }_s}}(e)}}} \\
 \!\!\!\!= {k}\!\!\sum\limits_{e = 1}^n\!\! {\frac{{{\tilde \omega _{s-1}}(e)}}{{
{\tilde\rho _s}(e)
 + (1-{\tilde\rho _s}(e))\frac{{{m_s} - 1}}{{{n} - 1}}}}} \mathop  \le \limits^{(a)}  {k}\!\!\sum\limits_{e = 1}^n\!\! {\frac{{2\tilde\rho _s}(e)}{{
{\tilde\rho _t}(e)
 + (1-{\tilde\rho _s}(e))\frac{{{m_s} - 1}}{{{n} - 1}}}}} \\
 \mathop  \le \limits^{(b)}  2k \frac{n}{m},
\end{array}\IEEEyesnumber \label{eq:Apart1s}
\end{IEEEeqnarray*}
where the above inequality $(a)$ follows the fact that $( {1 - \sum\nolimits_{e} {{\varepsilon _t}(e)} }) \ge \frac{1}{2}$ by the definition of
${{\varepsilon _t}(e)}$ and the equality (\ref{eq:Alg1p2}) and the above inequality $(b)$ follows the Lemma 4.  Note that ${\varphi_{s-1}}(e)= \sum\nolimits_{e \in \textbf{i}} {{\varepsilon _t}(e)} \left| {\left\{ {\textbf{i} \in \mathcal{C}:e
\in \textbf{i}} \right\}} \right|, \forall e \in [1,n]$ Take expectations over all random strategies of losses
 ${{\tilde \ell}_s}{(\textbf{i})}$ with respective
to distribution $u$, we have
\begin{IEEEeqnarray*}{l}
\begin{array}{l}
{\mathbb{E}_t}\left[{\mathbb{E}_{\textbf{i} \sim {\varphi_s}}}{{\tilde \ell}_s}{(\textbf{i})} \right] =
{\mathbb{E}_t}\left[\sum\limits_{\textbf{i} = 1}^N {{\varphi_s}(\textbf{i}){{\tilde \ell}_s}{{(\textbf{i})}}} \right]\\
= {\mathbb{E}_t}\!\!\left[\sum\limits_{\textbf{i} = 1}^N {{\varphi_s}(\textbf{i}){{(\sum\limits_{e \in \textbf{i}} {{{\tilde \ell}_s}(e)} )}}}
  \right]    \le  {\mathbb{E}_t}\!\!\left[ \sum\limits_{\textbf{i} = 1}^N {{\varphi_s}(\textbf{i}) (
 \sum\limits_{e \in \textbf{i}} {{{\tilde \ell}_s}{{(e)})}} }\right]
 \\= \!{\mathbb{E}_t}  \left[\! \sum\limits_{e = 1}^n {{{\tilde \ell}_s}{{(e)}}} \!\!\!\!\sum\limits_{\textbf{i} \in {\mathcal{P}}:e \in
  \textbf{i}} \!\!\!{{\varphi_s}(\textbf{i})} \right]\! = \!
 { }{\mathbb{E}_t}\!\! \left[\!\sum\limits_{e' = 1}^n \!{{{\tilde \ell}_s}{{(e')}}{\varphi_{s}}(e')} \right]\\
\le {k}{\mathbb{E}_t}\left[ {\sum\limits_{e' = 1}^n {\frac{{\varphi_{s}}(e')}{{{{{\tilde \rho }_s}}{{(e')}
}}}{\mathds{1}_s}(e')} } \right] = {k}\sum\limits_{e' = 1}^n {\frac{{\varphi_{s}}(e')}{{{{{\tilde \rho }_s}}(e')}}} \\
 = {k}\sum\limits_{e' = 1}^n {\frac{{\varphi_{s}}(e')}{{\tilde\rho _s}(e)
 + (1-{\tilde\rho _s}(e))\frac{{{m_s} - 1}}{{{n} - 1}}}}\\
  \mathop  \le \limits^{(a)}  {k}\sum\limits_{e' = 1}^n {\frac{{\tilde\rho _s}(e)}{{\tilde\rho _s}(e)
 + (1-{\tilde\rho _s}(e))\frac{{{m_s} - 1}}{{{n} - 1}}}}
 \le 2k \frac{n}{m},
\end{array}\IEEEyesnumber \label{eq:Apart4s}
\end{IEEEeqnarray*}
where the above inequality $(a)$ is because  ${\tilde\rho _t}(e) \ge {\varphi_{s-1}}(e)$.

\vspace{-.26cm}
In the third step, note that ${{\tilde L}_0}(\textbf{i}) = 0$. Let
${\Phi _t}(\eta ) = \frac{1}{\eta }\ln \frac{1}{N}\sum\nolimits_
{\textbf{i} = 1}^N {\exp ( - \eta {{\tilde L}_t}(\textbf{i}))}$ and ${\Phi _0}(\eta )=0 $. The second term in (\ref{eq:AppenA}) can be
bounded by using the same technique in \cite{Bubeck12} (page 26-28).  Let us substitute  inequality (\ref{eq:Apart1s}) into (\ref{eq:Apart1}), and then substitute (\ref{eq:Apart1}) into equation  (\ref{eq:AppenA}) and sum over $t$ and take expectation over all random strategies of losses up to time $t$, we obtain
\begin{IEEEeqnarray*}{l}
\begin{array}{l}
\hspace{-.3cm}{\mathbb{E}_t}\left[ \sum\limits_{s = 1}^t {\mathbb{E}_{\textbf{i} \sim {\rho_s}}}{{\tilde \ell}_s}(\textbf{i}) \right]
 \le k n \!\sum\limits_{s = 1}^t \eta_s + \frac{{\ln N}}{\eta_t } + \!\! \sum\limits_{s = 1}^t  \!{\mathbb{E}_{\textbf{i} \sim \varphi_s}}{{\tilde \ell}_s}(\textbf{i}) \\
\hspace{2.1cm} + {\mathbb{E}_t}\left[ \sum\limits_{s = 1}^{t - 1} {{\Phi _s}({\eta _{s + 1}}) - {\Phi _s}({\eta _s})}\right] + {
\mathbb{E}_t}\sum\limits_{s = 1}^t {{{\tilde \ell}_s}(\textbf{i})}.
\end{array}
\end{IEEEeqnarray*}
Then, we get
\begin{IEEEeqnarray*}{l}
R(t)  =\mathbb{E}_t \sum\limits_{s = 1}^t {{\mathbb{E}_{\textbf{i} \sim {p_s}}}{\tilde{\ell}_s}(\textbf{i})}  -
\mathbb{E}_t \sum\limits_{s = 1}^t {{\mathbb{E}_{{\textbf{I}_s} \sim {\varphi_s}}}{\tilde{\ell}_s}(\textbf{i})}\\
\hspace{.7cm}\le k \frac{n}{m} \!\sum\limits_{s = 1}^t \eta_s + \frac{{\ln N}}{\eta_t }
+  \sum\limits_{s = 1}^t  \!{\mathbb{E}_{\textbf{i} \sim \varphi_s}}{{\tilde \ell}_s}(\textbf{i})\\
\hspace{.7cm}\mathop  \le \limits^{(a)} k \frac{n}{m} \!\sum\limits_{s = 1}^t \eta_s + \frac{{\ln N}}{\eta_t }
+ k \frac{n}{m} \sum\limits_{s = 1}^t {\sum\limits_{e = 1}^n {{\varepsilon _s}(e)} } \\
\end{IEEEeqnarray*}
\begin{IEEEeqnarray*}{l}
\hspace{.7cm}\mathop  \le \limits^{(b)} 2k \frac{n}{m} \!\sum\limits_{s = 1}^t \eta_s + \frac{{\ln N}}{\eta_t } \\
\hspace{.7cm}\mathop  \le \limits^{(c)} 2k \frac{n}{m} \!\sum\limits_{s = 1}^t \eta_s + k\frac{{\ln n}}{\eta_t }.
\IEEEyesnumber \label{eq:ApartAc}
\end{IEEEeqnarray*}
Note that, the inequality $(a)$ holds according to (\ref{eq:Apart4s}). The inequality $(b)$ holds is because of,  for every time slot $t$,
 $\eta_t \ge {\varepsilon _t}(e)$. The
 inequality $(c)$ is due to the fact that $N \le n^{k}$. Setting $\eta_t=b_t$, we prove the theorem.
  \end{IEEEproof}

 \emph{Proof of Theorem 9.}
 \begin{IEEEproof}
 The proof of Theorem 9 for adaptive adversary is based on Theorem 8, and use the same idea as in the proof
 of Theorem 2. Here, If we set the batch $\tau= {(4{k}\sqrt {\frac{n}{m}\ln n} )^{ - \frac{1}{3}}}{t^{^{\frac{1}{3}}}}$ in Theorem 2 of \cite{Arora12},
we can get the regret upper bound in our Theorem 9.
 \end{IEEEproof}

\subsection{Accelerated AOSPR Algorithm in The Stochastic Regime}

To obtain the tight regret performance for AOSPR-MP-EXP3++, we need to study and estimate the
number of times each of link is selected up to time $t$, {i}.e., $N_t(e)$. We summarize it in the following lemma.

\textbf{Lemma 5. } In the multipath probing case, let $\left\{ {{{\underline \varepsilon }_t}(e)} \right\}_{t = 1}^\infty$ be non-increasing deterministic sequences, such that
${{\underline \varepsilon }_t}(e) \le {{ \varepsilon }_t}(e)$ with probability $1$ and ${{\underline \varepsilon }_t}(e) \le {{ \varepsilon }_t}(e^*)$
 for all $t$ and $e$. Define $\nu_t(e)= \sum\nolimits_{s = {1}}^t  \frac{1}{{{k \underline \varepsilon  }_s}(e)} $, and define the event $\Xi^e_t$
\begin{IEEEeqnarray*}{l}
 {mt\Delta (e) - ( {{{\tilde L}_{t}}(e^*) - {{\tilde L}_{t}}({e})} )} \\
\hspace{.3cm} \le {\sqrt {2({\nu _{t}}(e) + {\nu _t}({e^*})){b_{t}}}  + \frac{{{(1/k+ 0.25)} {b_{t}}}}{{3k
{\underline \varepsilon _{t}}({e^*})}}}\hspace{.3cm}
(\Xi^e_{t}).
\end{IEEEeqnarray*}
Then for any positive sequence $b_1, b_2,...,$ and any $t^* \ge 2$ the number of times link $e$ is played by AOSPR-EXP3++ up to
round $t$ is bounded as:
\begin{IEEEeqnarray*}{l}
\begin{array}{l}
\mathbb{E}[{N_t}(e)] \le \left( {{t^*} - 1} \right) + \sum\limits_{s = {t^*}}^{t} {{e^{ - {b_s}}}}  +
 k \sum\limits_{s = {t^*}}^{t} {{\varepsilon _s}(e){\mathds{1}_{\{ \Xi_{t}^e\} }}} \\
    \hspace{1.6cm}               + \sum\limits_{s = {t^*}}^{t} {{e^{ - {\eta _s}{\hslash_{s - 1}}(e)}}},
\end{array}
\end{IEEEeqnarray*}
where
\begin{IEEEeqnarray*}{l}
\begin{array}{l}
{\hslash_t}(e) = {mt}\Delta (e) - \sqrt {2{mt}{b_t}\left( {\frac{1}{{{k}{{\underline \varepsilon
 }_t}(e)}} + \frac{1}{{{k}{{\underline \varepsilon  }_t}({e^*})}}} \right)}  - \frac{{(\frac{1}{4} +
 \frac{1}{k}){b_t}}}{{3{{\underline \varepsilon  }_t}({e^*})}}.
\end{array}
\end{IEEEeqnarray*}

\begin{IEEEproof}
Note that AOSPR-MP-EXP3++ probes $M_t$ paths rather than $1$ path each time slot $t$. Let $\# \left\{  \cdot  \right\}$ stands for the number of elements
 in the set $\left\{  \cdot  \right\}$. Hence,
 \begin{IEEEeqnarray*}{l}
\begin{array}{l}
\mathbb{E}[{N_t}(e)] = \mathbb{E}[\# \left\{ {1 \le s \le t:{A_s} = e,\mathcal{E}_t^e} \right\} + \\
\hspace{4cm}\# \left\{ {1 \le s \le t:{A_s} = e,\overline {\mathcal{E}_t^e} } \right\}],
\end{array}
\end{IEEEeqnarray*}
where $A_s$ denotes the action of link selection at time slot $s$.
By the following  simple trick, we have
 \begin{IEEEeqnarray*}{l}
\begin{array}{l}
\!\!\!\!\!\!\!\!\mathbb{E}[{N_t}(e)] = \mathbb{E}[\# \left\{ {1 \le s \le t:{A_s} = e,\mathcal{E}_t^e} \right\}] + \\
\hspace{3.1cm}\mathbb{E}[\# \left\{ {1 \le s \le t:{A_s} = e,\overline {\mathcal{E}_t^e} } \right\}]] \\
 \hspace{.4cm}\le \mathbb{E}[ \sum\limits_{s = 1}^t \mathds{1}_{ \left\{ {1 \le s \le t:{A_s} = e} \right\}}\mathbb{P}[\# \{\mathcal{E}_t^e\}]] + \\
 \hspace{3.1cm}\mathbb{E}[ \sum\limits_{s = 1}^t \mathds{1}_{ \left\{ {1 \le s \le t:{A_s} = e} \right\}}\mathbb{P}[\# \{\overline {\mathcal{E}_t^e}\}]] \\
 \end{array}\IEEEyesnumber
\end{IEEEeqnarray*}
  \begin{IEEEeqnarray*}{l}
\begin{array}{l}
 \hspace{.4cm}\le \mathbb{E}[\sum\limits_{s = 1}^t  \mathds{1}_{ \left\{ {1 \le s \le t:{A_s} = e} \right\}}\mathbb{P}[\Xi _{mt}^e]] + \\
\hspace{3.1cm}\mathbb{E}[ \sum\limits_{s = 1}^t \mathds{1}_{ \left\{ {1 \le s \le t:{A_s} = e} \right\}}\mathbb{P}[\Xi _{mt}^e]].
\end{array}\IEEEyesnumber \label{eq:Apart23s}
\end{IEEEeqnarray*}




Note that the elements of the martingale difference sequence in the  $\{ {\Delta (e)
- ({{\tilde \ell}_t}(e) - {{\tilde \ell}_t}({e^*}))} \}_{t = 1}^\infty$ by $\max \{ \Delta (e) +{{\tilde \ell}_t}({e^*}) \}= {\frac{1}{{{k}{{\underline \varepsilon  }_t}({e^*})}}} +1$. Since ${{{\underline \varepsilon  }_t}({e^*})} \le {{{ \varepsilon  }_t}({e^*})} \le 1/(2n) \le 1/4$, we can simplify
the upper bound by using ${\frac{1}{{{{k\underline \varepsilon   }_t}({e^*})}}} +1 \le
    \frac{{(\frac{1}{4} + \frac{1}{k})}}{{{{\underline \varepsilon   }_t}({e^*})}}$.

We further note that
   \begin{displaymath}
\begin{array}{l}
{\mathbb{E}_s}  \left\{  \# \left\{
\sum\limits_{s = 1}^{t} {\left[{{(\Delta (e) - ({{\tilde \ell}_s}(e) - {{\tilde \ell}_s}({e^*})))}^2}\right]} \right\} \right\}\\
 \hspace{1.1cm}  \mathop \le \limits^{(a)}   {\mathbb{E}_s}  \left\{ m
\sum\limits_{s = 1}^{t} {\left[{{(\Delta (e) - ({{\tilde \ell}_s}(e) - {{\tilde \ell}_s}({e^*})))}^2}\right]}\right\} \\
 \hspace{1.1cm}  \le m\sum\limits_{s = 1}^{t}  {{\mathbb{E}_s}\left[{{({{\tilde \ell}_s}(e) - {{\tilde \ell}_s}({e^*}))}^2}\right]} \\
  \hspace{1.1cm}  = m\sum\limits_{s = 1}^{t}  {\left( {{\mathbb{E}_s}\left[({{\tilde \ell}_s}{{(e)}^2}\right] + {E_s}\left[({{\tilde \ell}_s}{{({e^*})}^2}\right]} \right)} \\
  \end{array}
\end{displaymath}
   \begin{displaymath}
\begin{array}{l}
 \hspace{1.1cm}  \le m\sum\limits_{s = 1}^{t}  \left( {\frac{1}{{{\tilde \varrho_s}(e)}} + \frac{1}{{{\tilde \varrho_s}({e^*})}}} \right) \\
 \hspace{1.1cm} \mathop \le \limits^{(b)} m\sum\limits_{s = 1}^{t}  {\left( {\frac{1}{{k{\varepsilon  _s}(e)}} + \frac{1}{{k{\varepsilon  _s}({e^*})}}} \right)} \\
 \hspace{1.1cm}  \le m\sum\limits_{s = 1}^{t} {\left( {\frac{1}{{k{{\underline \varepsilon   }_s}(e)}} +
 \frac{1}{{k{{\underline \varepsilon   }_s}({e^*})}}} \right)} = m{\nu _t}(e) + m{\nu _t}({e^*})
\end{array}
\end{displaymath}
with probability $1$. The above inequality $(a)$ is because the number of probes for each link
$e$  at time slot $s$ is at most $m$ times, so does the accumulated value of the variance ${{(\Delta (e) - ({{\tilde \ell}_s}(e)
- {{\tilde \ell}_s}({e^*})))}^2}$. The above inequality (b) is due to the fact
that ${{\tilde \varrho }_t}(e) \ge
{{\tilde \rho }_t}(e) \ge \sum\nolimits_{e \in \textbf{i}}
 {{\varepsilon _t}(e)} \left| {\left\{ {\textbf{i} \in \mathcal{C}:e \in \textbf{i}} \right\}} \right|$. Since each $e$ only belongs to one of the covering strategies $\textbf{i} \in \mathcal{C}$, $\left| {\left\{ {\textbf{i} \in \mathcal{C}:e \in \textbf{i}} \right\}} \right|$ equals to 1 at time slot $t$ if link $e$ is selected. Thus, ${{\tilde \rho }_t}(e) \ge \sum\nolimits_{e \in \textbf{i}} {{\varepsilon _t}(e)}= k{\varepsilon_t}(e)$.

Let $\mathcal{\bar E}_t^e$ denote the complementary of event
 $\mathcal{E}_t^e$. Then by the Bernstein's inequality $\mathbb{P}[\mathcal{\bar E}_t^e] \le e^{-b_t}$. According to (\ref{eq:Apart23s}), the number of
 times the link $e$ is selected up to round $t$ is bounded as:
\begin{IEEEeqnarray*}{l}
 \begin{array}{l}
  \mathbb{E}[{N_t}(e)] \le  \sum\limits_{s = 1}^t {\mathbb{P}[A_s= e|\Xi_{s - 1}^e]P[\Xi_{s - 1}^e]} \\
  \hspace{5em}+ \mathbb{P}[A_s= e|\overline {\Xi_{s - 1}^e} ]P[\overline {\Xi_{s - 1}^e} ]\\
 \hspace{4em} \le \sum\limits_{s = 1}^t {\mathbb{P}[A_s= e|\Xi_{s - 1}^e]} {\mathds{1}_{\{ \Xi_{s - 1}^e\} }} +
  \mathbb{P}[\overline {\Xi_{s - 1}^S} ]\\
  \hspace{4em} \le \sum\limits_{s = 1}^t {\mathbb{P}[A_s= e|\Xi_{s - 1}^e]} {\mathds{1}_{\{ \Xi_{s - 1}^e\} }} + {e^{ - {b_{s - 1}}}}.
\end{array}
\end{IEEEeqnarray*}
We further upper bound $ {\mathbb{P}[A_s= e|\Xi_{s - 1}^e]} {\mathds{1}_{\{ \Xi_{s - 1}^e\} }} $ as follows:
    \begin{displaymath}
\begin{array}{l}
\mathbb{P} {[A_s= e|{ \Xi}_{s - 1}^e]} {\mathds{1}_{\{ { \Xi}_{s - 1}^e\} }}
= {{{\tilde \rho }_s}}(e){\mathds{1}_{\{ { \Xi}_{s - 1}^e\} }}\\
  \hspace{4em} \le ({\omega_{s-1}}(e) + k{\varepsilon_s}(e)){\mathds{1}_{\{ { \Xi}_{s - 1}^e\} }}\\
 \hspace{4em} =({k\varepsilon  _s}(e) + \frac{{\sum\nolimits_{\textbf{i}:e \in \textbf{i}} {{w_{s - 1}}\left( \textbf{i} \right)} }}
 {{{W_{s - 1}}}}){\mathds{1}_{\{ {\Xi}_{s - 1}^e\} }}\\
    \hspace{4em} =({k\varepsilon  _s}(e) + \frac{{\sum\nolimits_{\textbf{i}:e \in \textbf{i}}^{} {{e^{ - {\eta _s}{
    \tilde L_{s - 1}}(\textbf{i})}}} }}{{\sum\nolimits_{\textbf{i} = 1}^N {{e^{ - {\eta _s}{ \tilde L_{s - 1}}(\textbf{i})}}} }})
    {\mathds{1}_{\{ { \Xi}_{s - 1}^e\} }}\\
  \hspace{4em}\mathop \le \limits^{(a)} ({k\varepsilon  _s}(e) +
   {e^{ - {\eta _s}\left( {{{\tilde L}_{s - 1}}(\textbf{i}) -
    {{\tilde L}_{s - 1}}({\textbf{i}^*})} \right)}}){\mathds{1}_{\{ { \Xi}_{s - 1}^e\} }}\\
    \hspace{4em}\mathop \le \limits^{(b)} ({k\varepsilon  _s}(e) + {e^{ - {\eta _s}\left( {{{\tilde L}_{s - 1}}(e) - {{\tilde L}_{s - 1}}
    ({e^*})} \right)}}){\mathds{1}_{\{ {\Xi}_{s - 1}^e\} }}\\
   \hspace{4em}\mathop \le \limits^{(c)}  k{\varepsilon  _s}(e){\mathds{1}_{\{ {\Xi}_{s - 1}^e\} }} + {e^{ - {\eta _s}{\hslash_{s - 1}}(e)}}.
\end{array}
  \end{displaymath}
The above inequality (a) is due to the fact that link
$e$ only belongs to one chosen path $\textbf{i}$ in $t-1$, inequality (b) is because the  cumulative
regret of each path is great than the cumulative regret of each link that belongs to the
  path, and the last inequality (c) we used the fact that $\frac{t}{{{{\underline \varepsilon   }_t}(e)}}$ is a non-increasing
 sequence ${\upsilon _t}(e) \le \frac{t}{{{{k \underline \varepsilon   }_t}(e)}}$. Substitution of this result back into
 the computation of $ \mathbb{E}[{N_t}(e)]$ completes the proof. \end{IEEEproof}

 \emph{Proof of Theorem 10.}
\begin{IEEEproof}
The proof is based on Lemma 3. Let $b_t =ln(t \Delta(e)^2)$ and ${{{\underline \varepsilon  }_t}(e)}
={{{ \varepsilon  }_t}(e)}$. For any $c \ge 18$ and any $t \ge t^*$, where $t^*$ is the minimal integer for which
 ${t^*} \ge \frac{{4{c^2}n \ln {{({t^*}\Delta {{(e)}^2})}^2}}}{{m^2\Delta {{(e)}^4}\ln (n)}}$, we have
     \begin{displaymath}
\begin{array}{l}
{\hslash_t}(e) = mt\Delta (e) - \sqrt {2mt{b_t}\left( {\frac{1}{{k{\varepsilon _t}(e)}} + \frac{1}{{k{\varepsilon _t}({e^*})}}} \right)}  - \frac{{\left( {\frac{1}{4} + \frac{1}{k}} \right){b_t}}}{{3{\varepsilon _t}({e^*})}}\\
   \hspace{2.43em} \ge mt\Delta (e) - 2\sqrt {\frac{{mt{b_t}}}{{k{\varepsilon _t}(e)}}}  - \frac{{\left( {\frac{1}{4} + \frac{1}{k}} \right){b_t}}}{{3{\varepsilon
   _t}(e)}}\\
     \hspace{2.43em} = mt\Delta (e)(1 - \frac{2}{{\sqrt {k c} }} - \frac{{\left( {\frac{1}{4} + \frac{1}{k}} \right)}}{{3c}})\\
  \end{array}
\end{displaymath}
   \begin{displaymath}
\begin{array}{l}
  \hspace{2.43em} \mathop  \ge \limits^{(a)}m t\Delta (e)(1 - \frac{2}{{\sqrt c }} - \frac{{1.25}}{{3c}}) \ge \frac{1}{2}mt\Delta (e),
\end{array}
  \end{displaymath}
where ${\varepsilon _t}(e) = \frac{{c\ln (t\Delta {{(e)}^2})}}{{tm\Delta {{(e)}^2}}}$.  By substitution of the lower bound on $h_t(e)$ into Lemma 3, we have
\begin{IEEEeqnarray*}{l}
\begin{array}{l}
\!\!\!\!\!\mathbb{E}[{N_t}(e)] \le {t^*} \!+  \! \frac{{\ln (t)}}{{\Delta {{(e)}^2}}} \!+ \! k \frac{{c\ln {{(t)}^2}}}{{m\Delta {{(e)}^2}}}  +  \!\!
\sum\limits_{s = 1}^t \!(\!{{e^{ - \frac{{m\Delta (e)}}{4}\sqrt {\frac{{(s - 1)ln(n)}}{n}} }}}\!)\\
 \hspace{3.2em} \le k\frac{{c\ln {{(t)}^2}}}{{m\Delta {{(e)}^2}}} + \frac{{\ln (t)}}{{\Delta {{(e)}^2}}} + O(\frac{{{n
 }}}{{m^2\Delta {{(e)}^2}}}) + {t^*},
\end{array}\IEEEyesnumber \label{eq:Apart10}
\end{IEEEeqnarray*}
  where lemma 3 is used to bound the sum of the exponents. In addition, please
  note that $t^*$ is of the order $O(\frac{{k n}}{{m^2\Delta {{(e)}^4}\ln (n)}})$.
\end{IEEEproof}

 \emph{Proof of Theorem 11-Theorem 13.}
 The proofs of Theorem 11-Theorem 13 use similar idea as in the proof of Theorem 14. We omitted here for brevity.

 \emph{Proof of Theorem 14.}
 \begin{IEEEproof}
For the  AOSPR-CP-EXP3++ algorithm, multiple source-destination pairs are coordinated to avoid probing
the overlapping path as little as possible, where now the statistically
collected link-level probing rate $m'_t$  is no less than the $m_t$ at each time slot. Thus, the actual
link probability  ${\tilde \varrho _t}(e)$ is no less than the one in (\ref{eq:Lim2}). Following the same line of
analysis, the regret upper bounds in Theorem 8-13 hold for  the  AOSPR-CP-EXP3++ algorithm. \end{IEEEproof}

 \emph{Proof of Theorem 15.}
 \begin{IEEEproof}
The proof of Theorem 15 also relies on the Theorem 8-13. Moreover, it requires  the construction of a linear program.
 Let $C_{es} (e=1,...,n, s=1,...,S)$ be the indicator that link $i$ is covered
 by the paths of the source-destination pair $s$, ${E_s} \buildrel \Delta \over = \left\{ {e \in E:{C_{es}} = 1} \right\}$ be
 the subset of links constructing path $s$ and ${k_s} \buildrel \Delta
 \over = \sum\nolimits_{e = 1}^n {{C_{es}}}$ the size of this subset. Consider a
  source-destination pair $s$. The key point is to bound the minimum link
   sample size ${\min _{e \in {E_s'}}}{z_e}(t)$ for general set of $C_{es}$. It is obvious that $\sum\nolimits_e {{z_{es}}(t)}  \ge t$
   for all $s=1,...,S$. In the worst case, we have $\sum\nolimits_e {{z_{es}}(t)}  = t$. In the Step 4 in the Algorithm 2, it iteratively
   solves the following integer linear programm (LP).
\begin{IEEEeqnarray*}{l}
\begin{array}{l}
\max  \quad \kappa \\
s.t.  \  \sum\limits_{s = 1}^S {{z_{es}}(t){C_{es}}}  \ge \kappa ,e = 1,...,n,\\
  \quad\quad     \sum\limits_{i = 1}^N {{z_{es}}(t)}  \le t,s = 1,...,S,\\
   \quad\quad \quad  \   {z_{es}}(t) \in \mathds{N} ,\forall i,s.
\end{array}\IEEEyesnumber \label{eq:Apart30}
\end{IEEEeqnarray*}
The aim of this LP
 is that  distributing the $t$ probing of each source-destination pair $s$ to evenly cover the links to maximize
 the minimum link sample size $\sum\nolimits_{s = 1}^S {{z_{es}}(t){C_{es}}}$. Particulary, we consider the minimum
 link sample size for source-destination pair $s'$, i.e., ${\min _{e \in {E_s'}}}\sum\nolimits_{s = 1}^S {{z_{es}}(t){C_{es}}}$.
 Denote the maximum value of the LP (\ref{eq:Apart30}) by $\kappa^*$. Note that ${z_{es}}(t) = \left\lfloor {t/{k_s}} \right\rfloor {C_{es}}$
  is a feasible solution to (\ref{eq:Apart30}). Thus, ${\min _{e \in {E_s}}}{z_e}(t) \ge {\kappa ^*} \ge {\min _{e \in {E_s}}}\sum\nolimits_{s = 1}^S {\left\lfloor {t/{k_s}} \right\rfloor {C_{es}} \buildrel \Delta \over = } \kappa \left( t \right)$.

Actually, normalized $\kappa \left( t \right)$ to $\bar \kappa= \sum\nolimits_{t = 1}^t\kappa \left( t \right)/t$, which is the average probing
rate up to time slot $t$.
The AOSPR-CP-EXP3++ Algorithm needs to use in the link probability calculation of (\ref{eq:Lim2}).  Under the complete overlap of paths over
 the entire network, i.e., ${{C_{es}}}\equiv 1$ and $k_s \equiv n$, we have $\bar \kappa= S=m$.  Following the same line of analysis, the regret upper
  bounds in Theorem 8-Theorem 13 hold for  the AOSPR-CP-EXP3++ algorithm in the multi-source
accelerated learning case by replacing $m$ with $S$. In the absence of any overlap, i.e., $\sum\nolimits_{s = 1}^S {{C_{es}}} \equiv 1$,
we have the probing rate $\bar \kappa= 1$. This correspond to the single source-destination case, and the  now
Theorem 1-Theorem 7 hold for the AOSPR-CP-EXP3++ algorithm. \end{IEEEproof}

 \emph{Proof of Theorem 17.}
 \begin{IEEEproof}
 To analysis the deviation of regret $R(t)$ to $m_{\Delta}$ and $n_{\Delta}$ in the adversarial regime,
 we need to focus on the following function
 $f\left( {m,n, {\tilde\varrho _t}(e)} \right) = \sum\nolimits_{e = 1}^n {\frac{{{\tilde\varrho _t}(e)}}{{{\tilde\varrho _t}(e) + \left( {1 - {\tilde\varrho _t}(e)} \right)\frac{{m - 1}}{{n - 1}}}}} $ subject to  $\sum\nolimits_{e = 1}^n  {\tilde\varrho _t}(e) =1$. The corresponding Lagrangian
 is:
 \begin{displaymath}
 \begin{array}{l}
 \mathcal{L}\left( {m,n, {\tilde\varrho _t}(e)} \right) = \sum\nolimits_{e = 1}^n {\frac{{{\tilde\varrho _t}(e)}}{{{\rho _t}(e) + \left( {1 - {\tilde\varrho _t}(e)} \right)\frac{{m - 1}}{{n - 1}}}}}\\
 \hspace{2.3cm} + \lambda \left( {1 - \sum\nolimits_{e = 1}^n {{\tilde\varrho _t}(e)} } \right).
 \end{array}
  \end{displaymath}
  As shown in \cite{Seldin14lim}, ${\tilde\varrho _t}(e) = \frac{1}{n}, \forall e \in [1,n]$ is the only maximizer of  $f\left( {m,n, {\tilde\varrho _t}(e)} \right)$.

 At first, take the first derivative of the Lagrangian with respect to $m$ we get
  \begin{displaymath}
 \begin{array}{l}
\!\!\!{\left. {\frac{{\partial L\left( {m,n,{\tilde\varrho _t}(e)} \right)}}{{\partial m}}} \right|_{{\tilde\varrho _t}(e) = \frac{1}{n}}} =  \sum\limits_{e = 1}^n {\frac{{{- \tilde\varrho _t}(e)\left( {1 - {\tilde\varrho _t}(e)} \right)\frac{1}{{n - 1}}}}{{{{\left( {{\tilde\varrho _t}(e) + \left( {1 - {\tilde\varrho _t}(e)} \right)\frac{{m - 1}}{{n - 1}}} \right)}^2}}}}  =  - \frac{{{n^2}}}{{{m^2}}}.
  \end{array}
  \end{displaymath}
We can make the first order approximation, i.e., $f\left( {m + {m_\Delta },n,{\tilde\varrho _t}(e)} \right) \simeq f\left( {m,n,{\tilde\varrho _t}(e)} \right) + {m_\Delta }{\left. {\frac{{\partial L\left( {m,n,{\tilde\varrho _t}(e)} \right)}}{{\partial m}}} \right|_{{\tilde\varrho _t}(e) = \frac{1}{n}}}= f\left( {m,n,{\tilde\varrho _t}(e)} \right)
-\frac{n^2}{m^2}m_{\Delta}.$
Then, according to (\ref{eq:Apart1s}) and  (\ref{eq:Apart4s}). We have the deviated version of regret in (\ref{eq:ApartAc})
as
  \begin{displaymath}
   \begin{array}{l}
R(t) \le 2k\left( {\frac{n}{m} - \frac{{{n^2}{m_\Delta }}}{{{m^2}}}} \right)\sum\limits_{s = 1}^t {{\eta _s}}  + k\frac{{\ln n}}{{{\eta _t}}} \\
 \hspace{.75cm}  \le 4k\sqrt {t\left( {\frac{n}{m} - \frac{{{n^2}{m_\Delta }}}{{{m^2}}}} \right)\ln n}.
 \end{array}
  \end{displaymath}
  Make the first order approximation of the upper bound of $R(t)$, i.e., $\bar R(t)$,  around $\frac{n}{m} $ we get the $R_{m_{\Delta}}(t)$ is $\frac{1}{2}{m_\Delta }\frac{n}{m}\bar R(t)$. Use similar approach we get the result for adaptive jammer is $\frac{1}{3}{m_\Delta }\frac{n}{m}\bar R(t)$. Combine the two, we prove the
  part (a) of the Theorem 17.

To prove the part (b) of the theorem, let us view the proof of the upper bound of $R(t)$ in the stochastic regime  in  (\ref{eq:Apart10}).
Take a first order approximation of $m_{\Delta}$ on the leading term $\frac{1}{m}$ (as a function), we easily get the  $R_{m_{\Delta}}(t)=\frac{1}{2}\frac{m_{\Delta}}{m} \bar R(t)$. Similarly, the results hold in the contaminated stochastic regimes.

The result (c) in the mixed adversarial and stochastic regime straightforward, which is just a combination of the results in adversarial and
stochastic regimes.

Secondly, take the first derivative of the Lagrangian with respect to $n$ we get
  \begin{displaymath}
   \begin{array}{l}
\!\!\!{\left. {\frac{{\partial L\left( {m,n,{\tilde\varrho _t}(e)} \right)}}{{\partial n}}} \right|_{{\tilde\varrho _t}(e) = \frac{1}{n}}} \!=\! \sum\limits_{e = 1}^n {\frac{{{\tilde\varrho _t}(e)\left( {1 - {\tilde\varrho _t}(e)} \right)\frac{{m - 1}}{{{{\left( {n - 1} \right)}^2}}}}}{{{{\left( {{\tilde\varrho _t}(e) + \left( {1 - {\tilde\varrho _t}(e)} \right)\frac{{m - 1}}{{n - 1}}} \right)}^2}}}}  = \frac{{{n^2}}}{{{m^2}}}\frac{{m - 1}}{{n - 1}}.
  \end{array}
  \end{displaymath}
Let us take the first order approximation, i.e., $f\left( {m,n + {n_\Delta },{\tilde\varrho _t}(e)} \right) \simeq f\left( {m,n,{\tilde\varrho _t}(e)} \right) + {n_\Delta }{\left. {\frac{{\partial L\left( {m,n,{\tilde\varrho _t}(e)} \right)}}{{\partial n}}} \right|_{{\tilde\varrho _t}(e) = \frac{1}{n}}}= f\left( {m,n,{\tilde\varrho _t}(e)} \right) + n_\Delta  \frac{{{n^2}}}{{{m^2}}}\frac{{m - 1}}{{n - 1}}.$

Then, according to (\ref{eq:Apart1s}) and  (\ref{eq:Apart4s}). We have the deviated version of regret in (\ref{eq:ApartAc})
as
  \begin{displaymath}
   \begin{array}{l}
R(t) \le 2k\left( {\frac{n}{m} + {n_\Delta }\frac{{{n^2}}}{{{m^2}}} \frac{{m - 1}}{{n - 1}}} \right)\sum\limits_{s = 1}^t {{\eta _s}}  + k\frac{{\ln n}}{{{\eta _t}}} \\
 \hspace{.7cm}  \le 4k\sqrt {t\left( {\frac{n}{m} + {n_\Delta }\frac{{{n^2}}}{{{m^2}}} \frac{{m - 1}}{{n - 1}}} \right)\ln n}.
 \end{array}
  \end{displaymath}
  Make the first order approximation of the upper bound of $R(t)$, i.e., $\bar R(t)$,  around $\frac{n}{m} $ we get the $R_{m_{\Delta}}(t)$ is $\frac{1}{2}{n_\Delta }\frac{n}{m}\frac{{m - 1}}{{n - 1}} \bar R(t) \simeq \frac{1}{2} {n_\Delta } \bar R(t)$. Use similar approach we get the result for adaptive jammer is $\frac{1}{3}{n_\Delta }\bar R(t)$. Combine the two, we prove the
  part (d) of the Theorem 17.

To prove the part (e) of the theorem, let us view the proof of the upper bound of $R(t)$ in the stochastic regime  in  (\ref{eq:Apart10}).
Take a first order approximation of $n_{\Delta}$ on the leading regret term. Since there is no
estimated value of $n$ in (\ref{eq:Lim2}),  we easily get the  $R_{m_{\Delta}}(t)=0$. Similarly, the results hold in the contaminated stochastic regimes.

The result (f) in the mixed adversarial and stochastic regime straightforward, which is just a combination of the results in adversarial and
stochastic regimes. \end{IEEEproof}

 \emph{Proof of Theorem 18.}
 \begin{IEEEproof}
 The delayed regret upper bounds results of Theorem 18 comes from the general results for adversarial and stochastic MABs in the respective
  Theorem 1 and Theorem 6 in \cite{Joulani-ICML13}.  The regret upper bound under delayed feedback in the
  adversarial regime is proved by a simple Black-Box transformation in a non-delayed oblivious MAB environment, which is a
  general result. For the stochastic regimes (contaminated regimes, etc.),  we need to study the following
  high probability bounds  (\ref{eq:AHigh})
\begin{IEEEeqnarray*}{l}
\begin{array}{l}
\mathbb{E}[{N_t}(e)] \le \left( {{t^*} - 1} \right) + \sum\limits_{\tau = {t^*}}^t {{e^{ - {b_\tau}}}}  + k \sum\limits_{\tau = {t^*}}^t {{\varepsilon _\tau}(e){\mathds{1}_{\{ \mathcal{E}_t^e\} }}} \\
    \hspace{1.6cm}               + \sum\limits_{\tau = {t^*}}^t {{e^{ - {\eta _\tau}{h_{\tau - 1}}(e)}}},
\end{array}
\end{IEEEeqnarray*}
again. In the delayed-feedback setting, if we use upper confidence bounds $\mathcal{E}_{s(t)}^e$ instead of $\mathcal{E}_t^e$,
where $s(t)$ was defined
to be the number of rewards of link $e$ observed up to and including time instant $t$. In the same way as above we can write
\begin{IEEEeqnarray*}{l}
\begin{array}{l}
\!\!\mathbb{E}[{N_t}(e)] \le \left( {{t^*} - 1} \right) + \sum\limits_{\tau = {t^*}}^t {{e^{ - {b_\tau}}}}  + k \sum\limits_{\tau = {t^*}}^t {{\varepsilon _\tau}(e){\mathds{1}_{\{ \mathcal{E}_{s(t)}^e\} }}} \\
    \hspace{1.3cm}               + \sum\limits_{\tau = {t^*}}^t {{e^{ - {\eta _\tau}{h_{\tau - 1}}(e)}}}.
\end{array}\IEEEyesnumber \label{eq:ApaF1}
\end{IEEEeqnarray*}
Since ${N_{t-1}}(e)= \tau^ * + {S_{t-1}}(e)$, we get
\begin{IEEEeqnarray*}{l}
\begin{array}{l}
\!\!\!\mathbb{E}[{N_t}(e)] \le \tau^ * + \left( {{t^*} - 1} \right) + \! \! \sum\limits_{\tau = {t^*}}^t {{e^{ - {b_\tau}}}}  + \! k \!\!  \sum\limits_{\tau = {t^*}}^t {{\varepsilon _\tau}(e){\mathds{1}_{\{ \mathcal{E}_{s(t)}^e\} }}} \\
    \hspace{1.3cm}               + \sum\limits_{\tau = {t^*}}^t {{e^{ - {\eta _\tau}{h_{\tau - 1}}(e)}}}.
\end{array}\IEEEyesnumber \label{eq:ApaF2}
\end{IEEEeqnarray*}
Now the same concentration inequalities used to bound (\ref{eq:ApaF1}) in the analysis of the non-delayed setting can be used
to upper bound the expected value of the sum in (\ref{eq:ApaF2}). By the same technique, the result holds for other
stochastic regimes.
  \end{IEEEproof}

\section{Conclusion and Future Works}
In this paper, we propose the first adaptive online SPR algorithm, which can automatically detect the
feature of the environment and achieve almost optimal learning performance in all different regimes.
We have conducted extensive experiments to verify the flexibility of our algorithm  and have seen  performance improvements over classic approaches. We also considered many practical implementation issues to make our algorithm more useful and
 computationally efficient in practice. Our algorithm can be especially useful for  sensor,  ad hoc and military networks in  dynamic  environments.
 In the near future, we plan to  extend our model to mobile networks and networks with node failure and inaccessibility to gain more
 insight into the learnability of the online SPR algorithm.

\end{document}